\documentclass[sigconf, nonacm]{acmart}

\usepackage[ruled,vlined,linesnumbered]{algorithm2e}
\usepackage{amsmath,amsfonts}
\usepackage{graphics}
\usepackage{graphicx}
\usepackage{subfigure}
\usepackage{textcomp}
\usepackage{xcolor}

\usepackage{hyperref} 

\newcommand{\Paragraph}[1]{\smallskip\noindent{\bf #1.}}

\newcommand{\Removed}[1]{}
\newcommand{\NewlyAdded}[1]{#1}

\newcommand{\techreport}[1]{#1}
\newcommand{\nottechreport}[1]{}

\definecolor{mybg}{rgb}{0.82,0.91,0.84}

\newcommand\vldbdoi{XX.XX/XXX.XX}

\newcommand\vldbavailabilityurl{}
\newcommand\vldbpagestyle{plain} 

\definecolor{myblue}{rgb}{0,0,0.7}
\definecolor{mypurple}{rgb}{0.5,0,0.5}
\newcommand{\revise}[1]{#1}     

\begin{document}
\title{Density-optimized Intersection-free Mapping and Matrix
Multiplication for Join-Project Operations (extended version)}

\author{Zichun Huang, Shimin Chen}
\authornote{Shimin Chen is the corresponding author.}
\affiliation{
        \institution{SKL of Computer Architecture, ICT, CAS}
        \institution{University of Chinese Academy of Sciences}
}

\begin{abstract}

A Join-Project operation is a join operation followed by a duplicate
eliminating projection operation.  It is used in a large variety of
applications, including entity matching, set analytics, and graph
analytics.  Previous work proposes a hybrid design that exploits the
classical solution (i.e., join and deduplication), and \revise{MM (matrix
multiplication)} to process the sparse and the dense portions of
the input data, respectively.  However, we observe three problems in
the state-of-the-art solution: 1) The outputs of the sparse and dense
portions overlap, requiring an extra deduplication step; 2) Its
table-to-matrix transformation makes an over-simplified assumption of 
the attribute values; and 3) There is a mismatch between the employed
MM in BLAS packages and the characteristics of the Join-Project
operation.

In this paper, we propose DIM$^3$, an optimized algorithm for the
Join-Project operation.  To address 1), we propose an
intersection-free partition method to completely remove the final
deduplication step. For 2), we develop an optimized design for mapping
attribute values to natural numbers.  For 3), we propose DenseEC and
SparseBMM algorithms to exploit the structure of Join-Project for
better efficiency.  Moreover, we extend DIM$^3$ to consider partial
result caching and support Join-$op$ queries, including Join-Aggregate
and \revise{MJP (Multi-way Joins with Projection)}.  Experimental results using both
real-world and synthetic data sets show that DIM$^3$ outperforms
previous Join-Project solutions by a factor of
\revise{2.3$\times$-18$\times$}.  Compared to RDBMSs, DIM$^3$ achieves orders
of magnitude speedups. 

  
\end{abstract}

\maketitle

\begingroup
\renewcommand\thefootnote{}\footnote{\noindent
Our codes are available at \href{https://github.com/schencoding/JoinProject-DIM3}{https://github.com/schencoding/JoinProject-DIM3} .
}\addtocounter{footnote}{-1}\endgroup

\pagestyle{\vldbpagestyle}



\section{Introduction}
\label{sec:intro}

A Join-Project operation is a join operation followed by a duplicate
eliminating projection operation~\cite{ICDT09}. 
Given two tables $R(x,y)$ and $S(z,y)$, the Join-Project operation can
be written as follows: 
\begin{equation}
\Pi_{x,z}(R(x,y) \Join_y S(z,y)) 
\label{eqn:jp}
\end{equation}
It joins $R$ and $S$ with $y$ as the join key, then projects and
deduplicates \revise{$($}$x$,$z$\revise{$)$} tuples.  $\Pi$ denotes the duplicate
eliminating projection.

\Removed{
\emph{Example:} In the MovieLens data set~\cite{Movielens},
\emph{Rating(UserID, MovieID, Score)} table contains the ratings of
movies given by users.  To find out which users have seen and rated
the same movies, we can use the following SQL query: \\
\hspace*{0.5in}\emph{SELECT DISTINCT R.UserID, S.UserID} \\
\hspace*{0.5in}\emph{FROM Rating R, Rating S} \\
\hspace*{0.5in}\emph{WHERE R.MovieID = S.MovieID}\\ 
This is a Join-Project operation.  It joins R and S on
MovieID, then projects and deduplicates using SELECT DISTINCT.
}

\NewlyAdded{
\emph{Example}: In the HetRec2011 data set~\cite{hetrec2011},
\emph{U(userID, bookID, tagID)} table contains tags given by users to 
books that they read, and \emph{B(bookID, tagID, weight)} table records 
books and their possible tags with weights. 
The following SQL query recommends books to users based on the tags
recorded in users' reading history:
\vspace{0.03in}\\
\hspace*{0.5in}\emph{SELECT DISTINCT U.userID, B.bookID} \\
\hspace*{0.5in}\emph{FROM U, B} \\
\hspace*{0.5in}\emph{WHERE U.tagID = B.tagID} \vspace{0.03in}\\ 
This Join-Project operation joins U and B on
tagID, then projects and deduplicates using SELECT DISTINCT.
}
\revise{The results can be stored by the application for quick user-specific recommendations.}

The Join-Project operation is used in a large variety of
applications~\cite{SIGMOD20}, including entity matching, set
analytics, and graph analytics.  The above is an example of entity
matching.  Similar examples include finding users who have seen the
same movies in the MovieLens data set~\cite{Movielens}, and
discovering co-authors in the DBLP data set~\cite{DBLP}.  
Moreover, if tuple \revise{$($}$x$,$y$\revise{$)$} represents that set $x$ contains
element $y$, then the Join-Project operation using $y$ as the join key
obtains all the pairs of sets that intersect with each other. 
Furthermore, if we interpret tuple \revise{$($}$x$,$y$\revise{$)$} as an edge between
two vertices $x$ and $y$ in a graph, then the Join-Project operation
can be used to compute all pairs of vertices that are indirectly
connected.

\subsection{Previous Solutions} \label{sec:PreviousSolutions}
\label{subsec:previous}

\Paragraph{Classical Solution} 
The classical solution to compute the Join-Project in RDBMSs is to
first perform the join
operation~\cite{join1992,VLDB09Join,VLDB13Join}, then deduplicate the
projected join results using hash tables~\cite{kocberber2013meet} or
other types of indices~\cite{Btree2011,Btree2016,kraska2018case}.
The time complexity of the classical solution is
$\Theta(|R|+|S|+|OUT_J|)$, where $|R|$, $|S|$, and $|OUT_J|$ denote
the sizes of input table $R$, table $S$, and the join results before
deduplication, respectively.
This cost is reasonable when the number of duplicates is low.
However, the solution is less efficient when $|OUT_J|$ is much larger
than the size $|OUT_P|$ of the final results after deduplication.
For example, $|OUT_J|$ is 3.7x as large as $|OUT_P|$ in the HetRec2011
example, while $|OUT_J|$ is 24x larger than $|OUT_P|$ in the MovieLens
data set.
Let $|X|$, $|Y|$ and $|Z|$ denote the number of distinct values in
column $x$, $y$, and $z$ in Eqn~\ref{eqn:jp}, respectively.  Consider
the case where $|X|$=$|Y|$=$|Z|$=$n$.  $|OUT_J|$ can be
$\mathcal{O}(n^3)$ in the worst case, while $|OUT_P|$ is only
$\mathcal{O}(n^2)$.  In other words, the classical solution can spend
a lot of time generating the $|OUT_J|$ join results and then
processing them to remove a large number of duplicates.

\Removed{ A good join algorithm~\cite{join1992,VLDB09Join,VLDB13Join}
can be chosen by taking into account of factors such as the input
size, the join selectivity, and whether there are indices on the join
keys.
Deduplication of the join results can be efficiently implemented using
hash tables~\cite{kocberber2013meet} or other types of
indices~\cite{Btree2011,Btree2016,kraska2018case}.  }


\Paragraph{Matrix Multiplication} 
Alternatively, the Join-Project operation can be computed using MM.  The basic idea is to represent tables
$R(x,y)$ and $S(z,y)$ as two matrices $\mathbf{R}^{x\times y}$ and
$\mathbf{S}^{y\times z}$.  Specifically, $\mathbf{R}_{x_i,y_k}=1$ if
and only if tuple \revise{$($}$x_i$,$y_k$\revise{$)$}$\in R$ (similarly for $S$).
Then the multiplication of $\mathbf{R}^{x\times y}$ and
$\mathbf{S}^{y\times z}$ gives matrix $\mathbf{C}^{x\times z}$, where
$\mathbf{C}_{x_i,z_j}=\sum_{k=1}^{|Y|} \mathbf{R}_{x_i,y_k}
\mathbf{S}_{y_k,z_j}$.  A non-zero element $\mathbf{C}_{x_i,z_j}>0$ in
the matrix corresponds to a tuple \revise{$($}$x_i$,$z_j$\revise{$)$} in the final
output of the Join-Project operation.
Compared to the classical solution, MM performs the join and the
deduplication together.   There are efficient MM implementations in
BLAS (Basic Linear Algebra Subprograms) packages with advanced
techniques~\cite{saule2013performance,dalton2015optimizing,kurzak2009optimizing}.
Moreover, there are sub-cubic MM algorithms in theory.  The best known
is the Coppersmith-Winograd algorithm with $\mathcal{O}(n^{2.373})$
complexity~\cite{matrix2018}.

\Removed{
The complexity of dense MM is
$\mathcal{O}(|X||Y||Z|\beta^{\omega-3})$, where
$\beta=min(|X|,|Y|,|Z|)$ and $2\le \omega < 3$~\cite{SIGMOD20}.  For
square matrices, where $|X|$=$|Y|$=$|Z|$=$n$,  the complexity is
$\mathcal{O}(n^{\omega})$.  The currently best known $\omega$ is
2.373~\cite{matrix2018}.  Besides the time complexity, MM also
benefits from many advanced acceleration
techniques~\cite{saule2013performance,dalton2015optimizing,kurzak2009optimizing}.
}

\Paragraph{Hybrid Solution}
Recent studies~\cite{ICDT09,SIGMOD20} combine the classical solution
and the MM solution based on the observation that the classical
solution performs better when the data is sparse, while MM performs
better when the data is dense.
Amossen et al.~\cite{ICDT09} propose an algorithm to partition the
data into dense and sparse parts according to degrees of $x$, $y$, and
$z$.  Then dense MM is employed for the dense parts and the classical
solution is used for the remaining parts.

\begin{figure}[t]
    \centering
    \includegraphics[width=3.2in]{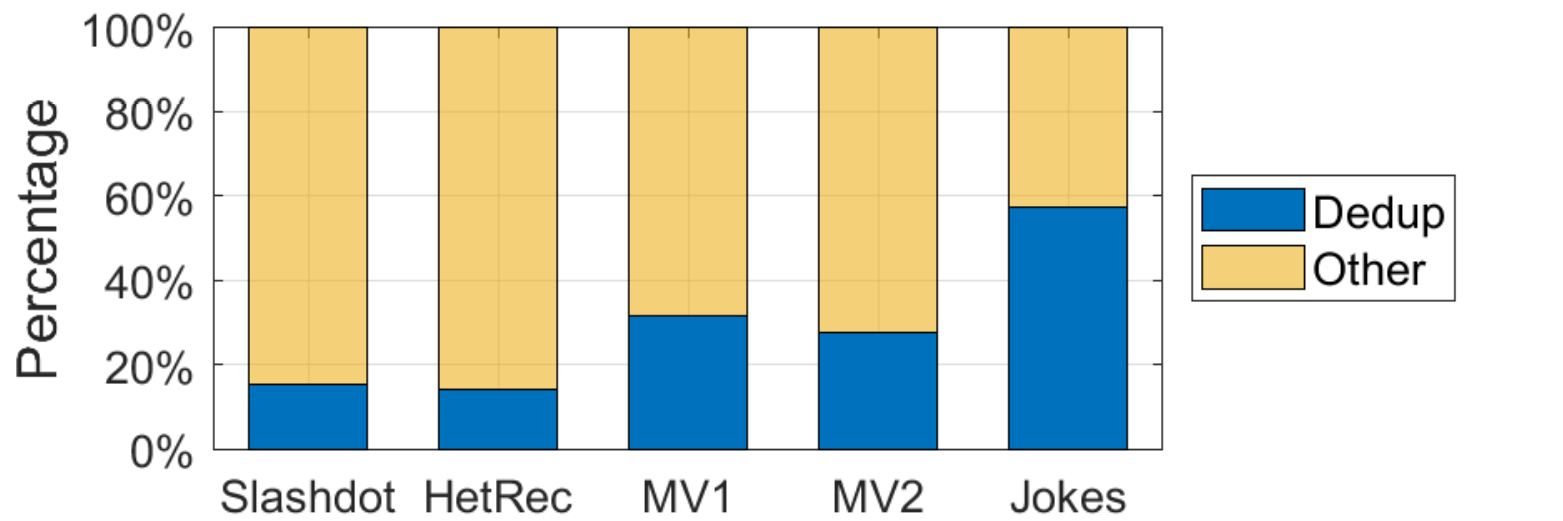}
    \vspace{-0.10in}
    \caption{\revise{Breakdown of DHK run time.}}
    \label{fig:breakdown}
    \vspace{-0.10in}
\end{figure}

Deep, Hu, and Koutris~\cite{SIGMOD20} correct errors in the cost
analysis of~\cite{ICDT09} and implement the algorithm for experimental
comparison.  We call this state-of-the-art algorithm \emph{DHK}.
%
%
DHK has been shown to significantly outperform the classical solution.
%
%
However, the following problems reduce the efficiency and practicality
of DHK:
\begin{list}{\labelitemi}{\setlength{\leftmargin}{7mm}\setlength{\itemindent}{-3mm}\setlength{\topsep}{0.5mm}\setlength{\itemsep}{0mm}\setlength{\parsep}{0.5mm}}

\item \emph{Overlapping outputs between sparse and dense parts}: The
Join-Project results computed from the sparse and from the dense parts
can overlap.  Consequently, a final deduplication step is necessary to
deduplicate the overlapping results.
Figure~\ref{fig:breakdown} shows the breakdown of the run time of DHK
for five data sets (cf.  Section~\ref{sec:exp}). 
\revise{For all these data sets, DHK partitions the
input data into dense and sparse parts.  It performs the final
deduplication step, incurring significant overhead.}

\item \emph{Over-simplified assumption for table-to-matrix
transformation}: The DHK implementation assumes that the input columns
$x$, $y$, and $z$ contain consecutive natural numbers starting from 0.
Thus, it directly uses their values as row or column ids in the
matrices.  However, in reality, database attribute values are rarely
natural numbers.  A step is missing: mapping values of input columns
to consecutive natural numbers.  
%

\item \emph{Caveats of MM Implementation}:  DHK invokes MM in BLAS
packages (e.g., Intel MKL~\cite{MKL}) as a black box.  On the one
hand, BLAS packages typically implement the $\mathcal{O}(n^{3})$ MM
algorithm.  It would be interesting to investigate sub-cubic MM
algorithms.
%
%
On the other hand, the MM invocation as a black box cannot exploit the
characteristics of the Join-Project operations for performance
improvement.

\end{list}

\subsection{Our Solution: DIM$^3$}
In this paper, we propose an efficient and practical Join-Project
algorithm, DIM$^3$ (\underline{D}ensity-optimized
\underline{I}ntersection-free \underline{M}apping and
\underline{M}atrix \underline{M}ultiplication).
We address the above problems as follows:
\begin{list}{\labelitemi}{\setlength{\leftmargin}{7mm}\setlength{\itemindent}{-3mm}\setlength{\topsep}{0.5mm}\setlength{\itemsep}{0mm}\setlength{\parsep}{0.5mm}}

\item \emph{Intersection-free partitioning}: We propose a novel
partitioning method that divides matrix $\mathbf{S}^{y\times z}$ into
subsets of rows based on the density of $z_j$ rows and then chooses
different evaluation strategies for the dense rows and the sparse
rows.  Since the results \revise{$($}$x_i$,$z_j$\revise{$)$} of the two parts have
different $z_j$s, this method is guaranteed to be intersection-free.
Hence, DIM$^3$ completely removes the final deduplication step
required by DHK.

\item \emph{Optimized mapping design}: We investigate the design of
the mapping step in DIM$^3$.  First, we identify cases (e.g., auto
increment, dictionary encoding) where columns contain roughly natural
numbers and thus the mapping step can be skipped.  Second, we exploit
the mapping of the shared join key columns to perform a semi-join like
optimization to discard tuples without matches.  Finally, we design a
cache-optimized hash-based algorithm to efficiently compute the
mappings.

\item \emph{Optimized MM algorithms}:  We obtain and evaluate an
implementation of the sub-cubic Strassen algorithm~\cite{Strassen69}
in a recent study~\cite{FMM}.  However, our results show that it
cannot beat Intel MKL (cf.  Section~\ref{sec:exp}.2).  Therefore, we
focus on $\mathcal{O}(n^{3})$ algorithms in DIM$^3$.  
For the dense rows, we observe that the computation
$\mathbf{C}_{x_i,z_j}=\sum_{k=1}^{|Y|} \mathbf{R}_{x_i,y_k}
\mathbf{S}_{y_k,z_j}$ can stop early as soon as there is a non-zero
$\mathbf{R}_{x_i,y_k} \mathbf{S}_{y_k,z_j}$.  
We design the \revise{\emph{DenseEC} (\underline{Dense} MM with
\underline{E}arly stopping \underline{C}hecking)
algorithm}.
For the sparse rows, we design the \revise{\emph{SparseBMM} (\underline{Sparse}
\underline{B}oolean \underline{MM}) algorithm} that
leverages the CSR (Compressed Sparse Row)~\cite{Book11Spmm} format of
matrix $\mathbf{S}^{y\times z}$ as a hash table on the join key $y$
with NO hash conflicts.  We also introduce a way to reduce the cost
for initializing the deduplication vector.


\end{list}

In addition to addressing the three problems, we extend the
Join-Project solution in the following two directions:
\begin{list}{\labelitemi}{\setlength{\leftmargin}{7mm}\setlength{\itemindent}{-3mm}\setlength{\topsep}{0.5mm}\setlength{\itemsep}{0mm}\setlength{\parsep}{0.5mm}}

\item \emph{Partial Result Caching}: We observe that in real-world
data sets, computing results for different $z_j$ values often take
widely different amounts of time.  This motivates us to investigate if
caching results for a subset of $z_j$ values is profitable.  We
consider caching either the original or the complement set of $x$'s
given a $z_j$.  The caching decision \revise{for $z_j$} is based on a score computed from
the computation time and the required result space.  Our experiments
show that partial result caching can effectively reduce the
Join-Project computation time with reasonable space cost for most data
sets.

\item \emph{Support for Join-Op Queries}: We discuss different types
of relational operations $op$, and investigate whether we can leverage
the Join-Project algorithm to support Join-$op$ queries.  We study two
interesting $op$s in depth:  1) For Join-Aggregate
queries~\cite{PODS20Parallel}, we show that DIM$^3$ can be applied
with simple modifications; and 2) For \Removed{multi-way joins with a
duplicate eliminating project operation} \revise{MJP (Multi-Way Joins
with Projection) queries}, a.k.a. conjunctive queries with
projection~\cite{ICDT21Enum}, we develop a dynamic programming
algorithm to find the optimal query plan that considers pushing down
deduplication operations to after the join operations.

\end{list}

\Removed{
Hu et al.~\cite{PODS20Parallel} performed a theoretical study on
parallel algorithms for Join-Aggregate queries, including line
queries, star queries and tree queries.   We show that DIM$^3$ can be
easily applied to these tasks with simple modifications.
Moreover, Deep et al.~\cite{ICDT21Enum} studied enumeration algorithms
for conjunctive queries with projection from a theoretical
perspective.  A conjunctive query contains a number of join
operations.  We find that it is straightforward to push down the
deduplication operation to replace every join with a corresponding
Join-Project operation, then apply DIM$^3$.  However, this approach
may not be optimal.  Therefore, we present a Dynamic Programming
algorithm to decide which join operations to be replaced by
Join-Project operations.
%
}

\vspace{-0.1in}

\subsection{Contributions}

The main contributions of this paper is threefold.
First, we propose DIM$^3$ with intersection-free partitioning,
optimized mapping, and DenseEC and SparseBMM algorithms to address the
three problems in  the state-of-the-art solution (cf.
Section~\ref{sec:jp}).  
\Removed{ Our contribution lies in the holistic design that combine a
number of optimization techniques that fix the weak points of previous
state-of-the-art algorithm, and exploit new opportunities to optimize
the Join-Project operation on both the dense and the sparse parts of
the data. 
Based on the analysis of our algorithm, we present the threshold
selection strategies and demonstrate the benefits of DIM$^3$ in terms
of time complexity (cf. Section~\ref{sec:threshold}).  }
\revise{Second, we propose partial result caching for the Join-Project
algorithm (cf.  Section~\ref{sec:cache}) and generalize DIM$^3$ to
support Join-$op$ queries (cf.  Section~\ref{sec:extends}).}
Third, we perform extensive experimental evaluation using both
real-world and synthetic data sets (cf. Section~\ref{sec:exp}).
Experimental results show that DIM$^3$ outperforms previous
Join-Project solutions by a factor of \revise{2.3$\times$-18$\times$}.
Compared to commercial and open-source RDBMSs, DIM$^3$ achieves orders
of magnitude speedups.

\section{DIM$^3$ for Join-Project} 
\label{sec:jp}

In this section, we first overview the DIM$^3$ algorithm in
Section~\ref{sec:overview}.  Then, we explain the components of the
algorithm in detail.  Specifically, we present the mapping phase in
Section~\ref{sec:mapping}, intersection-free partitioning in
Section~\ref{sec:partition}, SparseBMM in Section~\ref{sec:bspmm}, and
DenseEC in Section~\ref{sec:ec}.  
\nottechreport{Finally, we analyze the algorithm and describe the
strategy selection criteria in Section~\ref{subsec:threshold}.  
(An extended version of the paper provides time complexity analysis
for our solutions~\cite{dim3tr}.)
}
\techreport{We analyze the algorithm and describe the
strategy selection criteria in Section~\ref{subsec:threshold}.
Finally, we analyze the time complexities of the solutions
in Section~\ref{subsec:complexity}.}

\subsection{Overview} 
\label{sec:overview}

\begin{figure}[t]
  \centering
  \includegraphics[width=\linewidth]{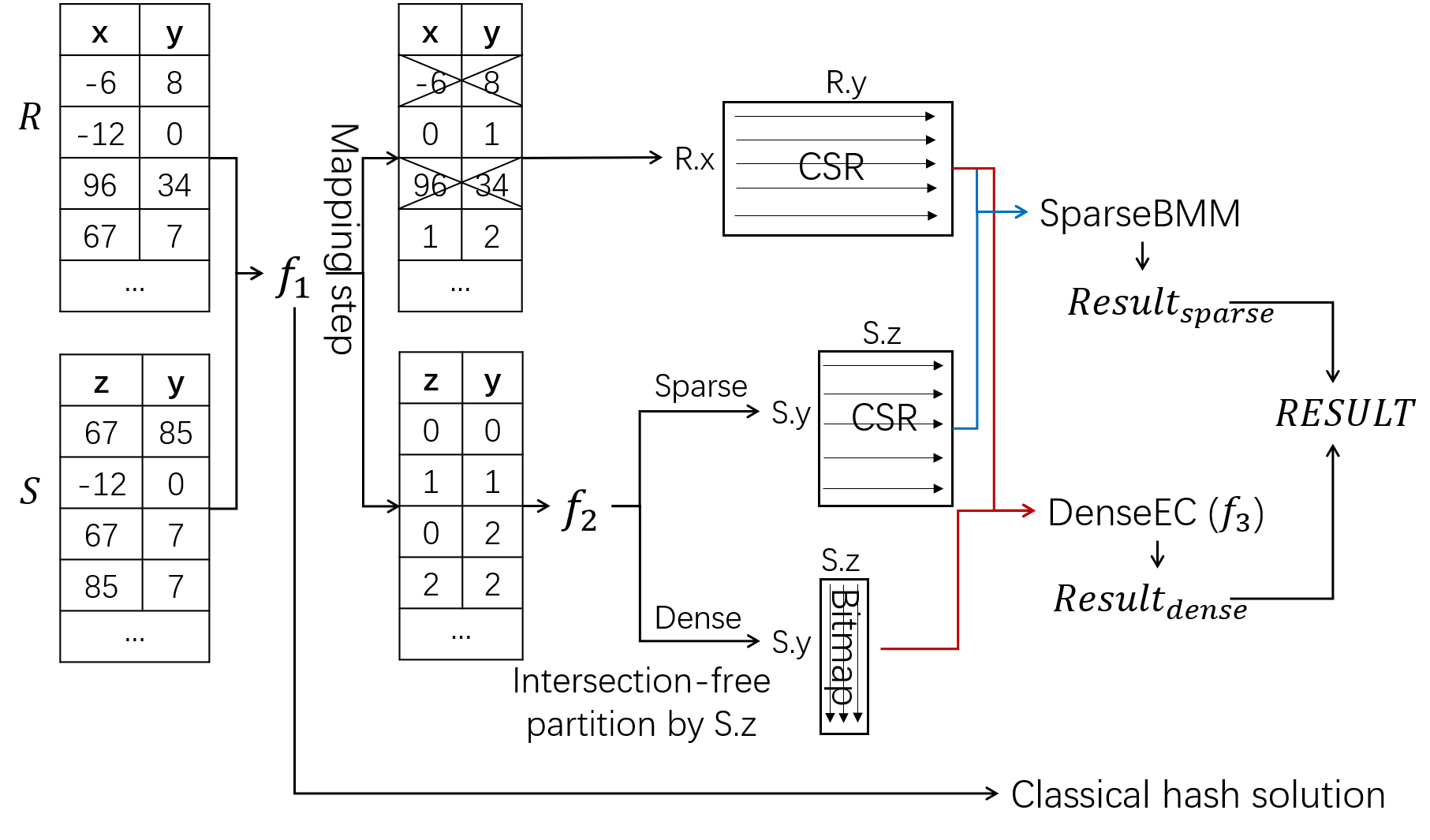}
  \vspace{-18pt}
  \caption{DIM$^3$ for Join-Project.}
  \label{fig:dim3}
  \vspace{-0.15in}
\end{figure}

The DIM$^3$ algorithm is depicted in Figure~\ref{fig:dim3} and listed
in Algorithm~\ref{alg:main}.
We perform the Join-Project operation on two tables $R(x,y)$ and
$S(z,y)$.  Without loss of generality, suppose $R$ is the larger table
and $S$ is the smaller table.

\revise{DIM$^3$ begins by selecting either the classical or the hybrid
solution for evaluating the Join-Project operation.  This allows
completely avoiding the mapping step, which can have significant cost
for highly sparse data.
DHK~\cite{SIGMOD20} uses a rule-of-thumb condition to choose the
classical solution.  In comparison, our strategy selection function
$f_1$ makes better decisions based on the estimated run times of the
classical and the hybrid solutions (cf. Section~\ref{subsec:threshold}
and~\ref{subsec:result-jp}).}

%
%


In the hybrid strategy, the first step is to map columns $x$, $y$, and
$z$ to consecutive natural numbers.  In this way, an \revise{$($}$x$, $y$\revise{$)$}
tuple in $R$ (similarly \revise{$($}$y$, $z$\revise{$)$} tuple in $S$) can be converted
to an element at row $x$ and column $y$ in matrix $\mathbf{R}^{x\times
y}$.  DHK~\cite{SIGMOD20} makes the over-simplified assumption that
the input columns contain consecutive natural numbers.  We delve into
the design of the mapping step to provide general-purpose support for
all attribute types.

The second step converts table $R$ to CSR~\cite{Book11Spmm} format,
then partitions $S$ to $S_{sparse}$ and $S_{dense}$.  We propose an
intersection-free partition method so that the Join-Project results of
$S_{sparse}$ and $S_{dense}$ can be simply combined without the final
deduplication step required by DHK~\cite{SIGMOD20}.  The partition
method uses function $f_2$ to decide which $z$ row is dense.  $f_2$
will be described in Section~\ref{subsec:threshold}.

The third step designs SparseBMM and DenseEC algorithms for processing
the sparse and dense data, respectively.  SparseBMM computes the
Join-Project $Result_{sparse}$ on $R_{CSR}$ and $S_{sparse}$.  DenseEC
multiplies $R_{CSR}$ with $S_{dense}$ to obtain $Result_{dense}$.
In DenseEC, function $f_3$ is used to choose the best method for
intersecting two bitmaps.  The computation of $f_3$ will be discussed
in Section~\ref{subsec:threshold}.

The final step is to merge $Result_{MM}$ and $Result_{EC}$ to obtain
the final results.  Because of intersection-free partitioning, there
is no need to perform an extra deduplication step.



\begin{algorithm}[t] 
  \small
  \caption{DIM$^3$}
  \label{alg:main}

  \KwIn{Table $R(x,y)$ and Table $S(z,y)$}
  \KwOut{List of result tuples $(x,z)$}

  Estimate $OUT_J$\; 
  \uIf({/* Classical is better */}){$f_1(|R|,|S|,|OUT_J|)$ $>$ 0}{ \label{alg:main:hash}
    \Return Use classical solution\;
  } 
  Mapping column values to consecutive natural numbers\;
  $R_{CSR}$ $\leftarrow$ Create CSR for $R(x,y)$\;
  Intersection-free partition $S$ by $S.z$; /* dense if
$f_2(z)$$>$0 */\label{alg:main:S}

  $S_{sparse}$ $\leftarrow$ Save the sparse part of $S$ as CSR\;
  $S_{dense}$ $\leftarrow$ Save the dense part of $S$ as Bitmap array\;
  $Result_{sparse}$ $\leftarrow$ SparseBMM($R_{CSR}$, $S_{sparse}$)\;
  $Result_{dense}$ $\leftarrow$ DenseEC($R_{CSR}$, $S_{dense}$)\; \label{alg:main:ec}
  \Return $Result_{sparse}$ $\cup$ $Result_{dense}$;
\end{algorithm}

\subsection{Mapping} 
\label{sec:mapping}


The mapping step maps columns $x$, $y$, and $z$ to consecutive natural
numbers.  This can be achieved with a baseline hash-based algorithm.
Given a column, the algorithm looks up the values of the column one by
one in a hash table.  If a value $v$ does not exist in the hash table,
it inserts ($v$, the next consecutive number) into the hash table.  As
a result, every distinct value in the column is assigned a natural
number.  We can repeat this algorithm for $x$, $y$, and $z$.

This mapping algorithm can be costly.  It creates three hash tables for
$x$, $y$, and $z$, and performs up to $2(|R|+|S|)$ hash table lookups
and/or inserts.  Compared to the hash join of $R(x,y)$ and $S(z,y)$,
which performs $|R|+|S|$ hash table accesses,  the mapping algorithm
pays twice as much cost for hash table visits.  This can be a
significant additional overhead for the Join-Project algorithm when
the join result size is not much larger than the input sizes.

In the following, we consider three opportunities to optimize the
baseline algorithm.  Then, we extend the mapping to support a wider
range of Join-Project operations.

\Paragraph{Optimization 1: Skip Mapping}
First of all, DIM$^3$ chooses the classical solution for highly sparse
data sets as shown in Figure~\ref{fig:dim3}.  This completely avoids
the mapping step.
Second, it is possible to skip mapping for columns that already
contain natural numbers.  There are two common cases in database
systems.  (i) Columns declared with auto increment (e.g.,
\emph{AUTO\_INCREMENT} in Oracle, \emph{IDENTITY} in SQLServer and
DB2, \emph{SERIAL} in PostgreSQL, \emph{AUTOINCREMENT} in MySQL)
contain consecutive natural numbers.  (ii) String columns can be
encoded by dictionary encoding~\cite{kanda2017practical} and stored as
natural numbers in database systems (e.g., SAP HANA and MonetDB).

\Removed{
For such columns, we need to consider if the column values in the join
input are sufficiently dense in light of any filtering predicates on
the base table.  Suppose a column contains values in the range [1,
$Max$], and the selectivity of the filtering predicates is $\sigma$.
Then only $\sigma Max$ values actually exist in the join input, while
the cost of Dense MM depends on $Max$.
Intuitively, the value of $\sigma$ affects the performance of
the following Join-Project computation.  
We study the impact of filtering predicates experimentally in
Section~\ref{sec:exp}.
}

\Paragraph{Optimization 2: Reduce Computation with Join Key Mapping}
The na\"ive way to map $y$ is to use $R.y \cup S.y$ as the mapping
input.  We observe that only $R.y \cap S.y$ contributes to the
equality join results.  Therefore, we can employ a semi-join like
idea, and optimize the mapping of $y$ as follows.  
Since $S$ is the smaller table, we first compute the mapping of $S.y$
using a hash table.  Then we map tuples in $R$ using the same hash
table.  If $y_k$ $\in$ $R.y$ but $y_k$ $\notin$ $S.y$, then the
corresponding $R$ tuple can be safely discarded because it does not
have any matches in $S$.
Note that we choose not to pay the cost of re-scanning $S$ to remove
$S$ tuples with $y_k$ $\in$ $S.y$ but $y_k$ $\notin$ $R.y$.
As shown in Figure \ref{fig:dim3}, the first and third $R$ tuples
are filtered out.

\Paragraph{Optimization 3: Optimize Hash Table Performance} When a
hash table is larger than the CPU cache, hash table accesses result in
expensive random memory accesses with poor CPU cache behavior.  A hash
table visit may probe multiple locations, and dereference pointers
(e.g., in the case of chained hash table), incurring significant
overhead.  Therefore, we employ the following designs to improve the
hash table performance in the mapping algorithm.
First, we estimate the hash table size for a column (e.g., based on
statistics of the number of distinct keys).  If the size exceeds the
last-level CPU cache, we employ cache partitioning.  We use the last
$k$ bits of the hash value to divide the data into $2^k$ partitions so
that the hash table of each partition fits into the last-level CPU
cache.  Then, we compute the mapping for each partition.
%
%
%
Second, we employ a linear probing hash table design to avoid pointer
dereference.  We tune the number of slots and the maximum probing
distance to reduce the cost of hash table accesses.  If no available
slots are found for a given column value, we employ a stash hash table
(Flat\_hash\_map~\cite{Flathashmap} in our implementation) to store
the overflow data.
%
%

\Paragraph{Supporting Wider Range of Join-Project Operations} We can
map not only single attribute but also multiple attributes to
consecutive numbers.  For instance, \vspace{-0.03in}
$$\Pi_{a,b,c,d,e} (R(a,b,c,d) \Join_{c,d} S(c,d,e,f)) \vspace{-0.03in}
$$ 
can be treated as \vspace{-0.08in}
$$\Pi_{x,z} (R(x,y)\Join_y S(z,y)) \vspace{-0.03in} $$ 
where $R.x= \{a,b\}$, $S.z= \{c,d,e\}$, and $R.y=S.y=\{c,d\}$.
In this way, we can support any combinations of join keys and
projection columns, including operations that contain the join key in
the output.

\subsection{Intersection-Free Partitioning} 
\label{sec:partition}

\begin{figure}[t]
  \centering
  \includegraphics[width=\linewidth]{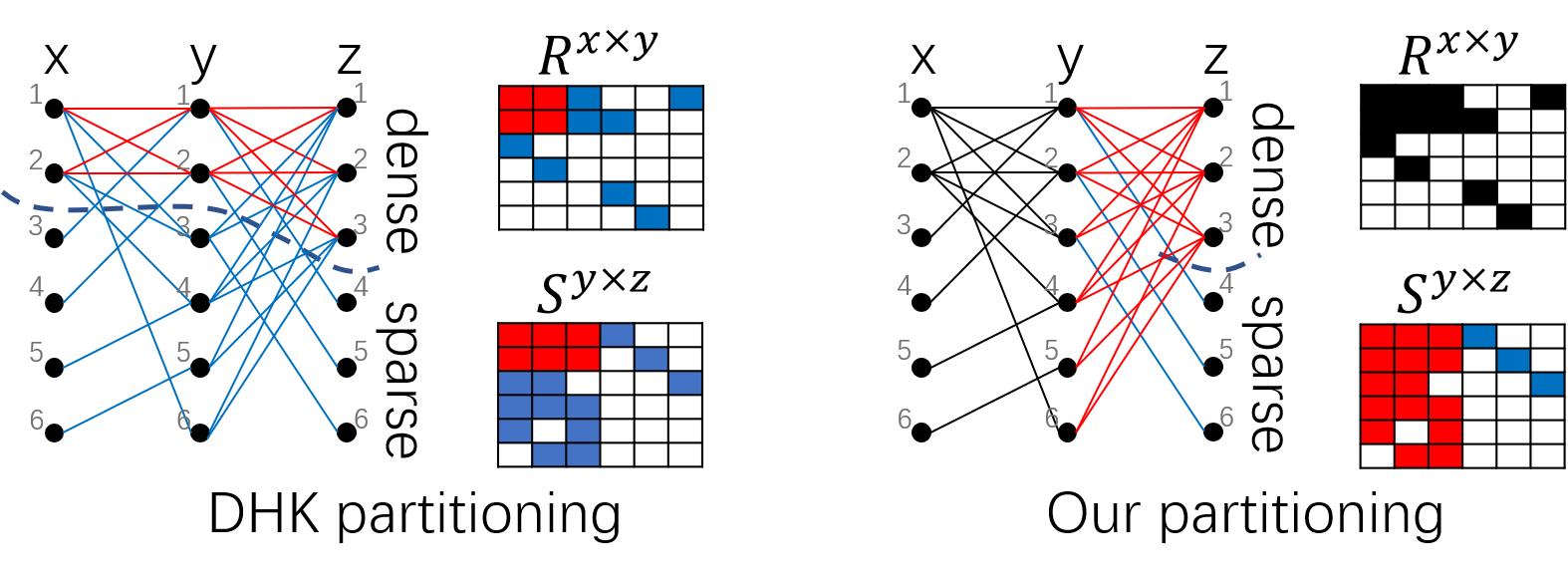}
  \vspace{-24pt}

  \caption{Comparing the partition methods in DHK~\cite{SIGMOD20} and
DIM$^3$. (A tuple in a table is displayed as an edge in graphs and an
element in matrices. Red: dense, blue: sparse)}

  \label{fig:partition}
  \vspace{-0.15in}
\end{figure}

Figure~\ref{fig:partition} compares the partition methods of
DHK~\cite{SIGMOD20} and DIM$^3$.  DHK makes separate decisions on $x$,
$y$ and $z$ according to their degrees (i.e. the number of tuples of
the same attribute value).  An \revise{$($}$x$, $y$\revise{$)$} tuple is added to the
dense part only if both its $x$ and $y$ attributes are considered as
dense (shown as red color in the figure).

In comparison, we propose an intersection-free partition
method\footnote{\revise{Previous work proposes an intersection-free
partition method in the context of query enumeration
algorithms~\cite{ICDT21Enum}.  It divides the data into two sets with
roughly equal number of join results. The method optimizes enumeration
delays rather than end-to-end query run times.  It is not directly
applicable to the Join-Project operation.}} as shown in
Figure~\ref{fig:partition}. It examines table $S$ and ensure that all
tuples with the same $S.z$ value are in the same partition.  Note that
table $R$ will not be partitioned.  Essentially, we partition matrix
$S$ by its columns.  Our partitioning method has the following
benefits.

First, the Join-Project results generated by the sparse and dense
parts do not intersect.   Given a result \revise{$($}$x$,$z$\revise{$)$}, if $z$ is
judged as dense, this result must be generated by the dense part.
Otherwise, it is generated by the sparse part.  In contrast, this
property does not hold in DHK.  As shown in
Figure~\ref{fig:partition}, the result \revise{$($}$x_2$, $z_2$\revise{$)$} is generated
both in the dense part by joining \revise{$($}$x_2$, $y_1$\revise{$)$} and \revise{$($}$y_1$,
$z_2$\revise{$)$}, and in the sparse part by joining \revise{$($}$x_2$, $y_3$\revise{$)$} and
\revise{$($}$y_3$, $z_2$\revise{$)$}.  Therefore, while DHK must deduplicate the results
from the two parts, DIM$^3$ can eliminate this final deduplication
step.

Second, DIM$^3$ may apply dense MM to more tuples.  DHK considers a
tuple as dense only if both its attributes are judged as dense.  In
comparison, DIM$^3$ makes the partitioning decision based solely on $S.z$.
Since its dense criteria tend to be more flexible, DIM$^3$ can employ
dense MM under more scenarios, as illustrated in
Figure~\ref{fig:partition}.

Third, the selection based on $z$ simplifies the Join-Project
computation.  In DHK, it is costly (spatially) to record the per-tuple
density decisions.  Therefore, DHK does not save them.  Instead, when
processing the sparse part, DHK uses the degree thresholds to
re-compute whether a tuple is dense and should be skipped.  In
comparison, DIM$^3$ avoids this complexity.  As shown in
Figure~\ref{fig:dim3}, table $S$ is divided into the sparse and the
dense parts based on $S.z$.  Hence, there is no need to re-evaluate
the density criteria any more.


Finally, the cost for computing the partition decision in DIM$^3$ is
lower compared to DHK.  DHK makes $|R|+|S|$ decisions on all input
tuples. In comparison, DIM$^3$ makes $|Z|$ decisions on $S.z$.  The
number of decisions to make is much smaller.  Consider the case where
$|X|$=$|Y|$=$|Z|$=$n$.  In the worst case, the cost is
$\mathcal{O}(n^{2})$ in DHK, but only $\mathcal{O}(n)$ in DIM$^3$.
Moreover, DHK performs binary search to determine the density
threshold, which incurs additional cost.

\subsection{SparseBMM} \label{sec:bspmm}

The classical hash-based solution is often used to process the sparse
part of the data.  We propose a \emph{SparseBMM} algorithm with two
main optimization techniques, as shown in Algorithm~\ref{alg:spmm}.

First, we observe that hash table accesses are often one main cost of
the classical hash-based computation.  Interestingly, since the column
values are mapped to natural numbers, the CSR (Compressed Sparse
Row)~\cite{Book11Spmm} format of matrix $\mathbf{S}^{y\times z}$ is
essentially a hash table on the join key $y$ with NO hash conflicts.
The original CSR structure consists of three arrays: $Val[]$, $Col[]$,
and $RowPtr[]$.  $Val[]$ and $Col[]$ contain the value and the column
index of non-zero elements in the matrix, respectively.  $RowPtr[]$
points to the row starts in $Val[]$ and $Col[]$.  In the case of
Join-Project, $Val[]$ contains all 1's and can be omitted.  Therefore,
we have two arrays $Col[]$ and $RowPtr[]$.
We employ matrix $\mathbf{S}^{y\times z}$ as the hash table.  Given
$y_k$, we locate all non-zero element $\mathbf{S}_{y_k,z_j}$ by
visiting entries $Col[RowPtr[y_k]]$ .. $Col[RowPtr[y_k$+1]-1].   In
fact, $RowPtr$ serves as the hash bucket header, and
$Col[RowPtr[y_k]]$ .. $Col[RowPtr[y_k$+1]-1] contain the hash entries
in bucket $y_k$.
In this way, we avoid the hash function computation and hash conflicts
in common hash table designs.



Second, DHK performs deduplication using a $z$-vector.  For each
$x_i$, it initializes the $z$-vector to all zeros. Then, it checks all
\revise{$($}$x_i$, $y_k$\revise{$)$}s to compute the join results.  For every result
\revise{$($}$x_i$, $z_j$\revise{$)$}, DHK increments the corresponding element in the
$z$-vector.  Hence, the non-zero elements in the $z$-vector indicate
the deduplicated Join-Project results.  However, the initialization
cost is $\mathcal{O}(|Z|)$, while the number of non-zero $z$'s can be
small for the sparse part of the data.  Consequently, the
initialization of the $z$-vector is often a main cost of the
deduplication computation.
We remove this per-$x$ initialization cost with a monotonically
increasing flag for different $x$.  As shown in
Algorithm~\ref{alg:spmm}, the $SPA$ array is the $z$-vector.  We
initialize the $SPA$ array only once before any computation.  The
$cur_y$ and $cur_z$ loops
(Line~\ref{alg:spmm:yloop}--\ref{alg:spmm:endloop}) compute the join
results for the given $cur_x$.   For a newly computed join result
\revise{$($}$cur_x$, $cur_z$\revise{$)$}, we set $SPA$[$cur_z$] to $cur_x$.  If there are
multiple duplicate \revise{$($}$cur_x$, $cur_z$\revise{$)$}, $SPA$[$cur_z$] is set only
once for the first instance.
In this way, for the next $cur_x+1$, the previous content of $SPA$ is
\emph{automatically} invalid.  This saves the cost of initializing
$SPA$
in every $cur_x$ loop iteration.

\begin{algorithm}[t] 
  \small
  \caption{SparseBMM (for sparse data).}
  \label{alg:spmm}

  \KwIn{CSR-stored $R_{CSR}$ and CSR-stored $S_{sparse}$}
  \KwOut{List of result tuples $(x,z)$}

  SPA[0..$|Z|$-1]=$-1$\;
  \For{$cur_x \leftarrow 0$ \KwTo $|X|$}{ \label{alg:spmm:forloop}
    \ForEach{$cur_y$ related to $cur_x$ in $R_{CSR}$}{ \label{alg:spmm:yloop}
      \ForEach{$cur_z$ related to $cur_y$ in $S_{sparse}$}{
        \uIf{SPA[$cur_z$]!=$cur_x$}{ \label{alg:spmm:if}
          SPA[$cur_z$]=$cur_x$\;
          Result.append(\revise{$($}$cur_x$, $cur_z$\revise{$)$})\;
        }
      }
    }
  } \label{alg:spmm:endloop}
  \Return Result\;
\end{algorithm}

We consider the time and space complexity of SparseBMM.
Line~\ref{alg:spmm:if} of Algorithm~\ref{alg:spmm} runs $|OUT_J|$
times. Thus, the time complexity is $\Theta(|R|+|S|+|OUT_J|)$.  While
this is the same as the classical hash-based solution, SparseBMM
reduces the constant factor, accelerating hash table visits and
deduplication.
Moreover, SparseBMM requires $\Theta(|R|+|S|+|Z|)$ space if the final
output is consumed by upper level operators.  While hash-based
deduplication requires $\Theta(|OUT_P|)$ space for the hash table,
SparseBMM allocates only $\Theta(|Z|)$ space for $SPA$, which is often
much smaller than $\Theta(|OUT_P|)$.



\subsection{DenseEC} \label{sec:ec}

To compute  $\mathbf{C}_{x_i,z_j}=\sum_{k=1}^{|Y|}
\mathbf{R}_{x_i,y_k} \mathbf{S}_{y_k,z_j}$, standard dense MM
enumerates all pairs of  $\mathbf{R}_{x_i,y_k}$ and
$\mathbf{S}_{y_k,z_j}$.  We observe that the computation can stop
early as soon as there is a non-zero $\mathbf{R}_{x_i,y_k}
\mathbf{S}_{y_k,z_j}$.  Therefore, we propose a DenseEC algorithm to
leverage this observation, as listed in Algorithm~\ref{alg:ec}.

In Algorithm~\ref{alg:ec}, the first two \emph{for}-loops enumerate
all the pairs of $R.x$ and $S.z$.
Line~\ref{alg:ec:R}--\ref{alg:ec:t2} use one of two methods to check
if there is any common $y$ between $Bitmap_x$ and $Bitmap_z$.  The
first method uses SIMD to compute the bit-wise \emph{AND} of
$Bitmap_x$ and $Bitmap_z$ (e.g., using \_mm256\_testz\_si256).  The
second method examines each $y$ related to $cur_x$ in $Bitmap_z$ using
a random memory access.  If any common $y$ is found, both methods stop
early.
Generally speaking, the SIMD method performs better when the $cur_x$
row has a large number of $y$.  We determine which method to use with
function $f_3$, which will be discussed in
Section~\ref{subsec:threshold}.

Apart from the time saving due to early stopping, DenseEC saves memory
space compared to dense MM.  DenseEC represents every element as a
single bit rather than a 4-byte integer or floating point value in
BLAS packages.  Moreover, a dense MM invocation would require space
to save the temporary output matrix. In comparison, DenseEC never
generates the output matrix.

\begin{algorithm}[t] 
  \small
  \caption{DenseEC (for dense data).}
  \label{alg:ec}

  \KwIn{CSR-stored $R_{CSR}$ and bitmap array $S_{dense}$}
  \KwOut{List of result tuples $(x,z)$}


  \For{$cur_x \leftarrow 0$ \KwTo $|X|$}{ \label{alg:ec:forloop}
    $Bitmap_x$ $\leftarrow$ Save the $y$ in the $cur_x$ row as bitmap\;
    \ForEach{$cur_z$ in $S_{dense}$}{ \label{alg:ec:forSz}
      \uIf(/* SIMD is better */){$f_3(m_x,m_z,|Y|)$$>$0}{ \label{alg:ec:R}
        \uIf{SIMD\_AND($Bitmap_x$,$Bitmap_z$) \revise{not all 0}}{
          Result.append(\revise{$($}$cur_x$, $cur_z$\revise{$)$})\;
        }
      }
      \uElse{
        \ForEach{$y$ related to $cur_x$ in $R_{CSR}$}{
          \uIf{the $y$th bit in $Bitmap_z$ == 1}{
            Result.append(\revise{$($}$cur_x$, $cur_z$\revise{$)$})\;
            Break\;
          }
        }
      }\label{alg:ec:t2}
    }
  }
  \Return Result\;
\end{algorithm}

\Removed{
Note that line 3--13 does not achieve the best known time complexity
of dense MM, which is $\mathcal{O}(|X||Y||Z|\beta^{\omega-3})$ with
$\omega$=2.373.  When the dimensions of the matrices are very large, a
dense MM implementation with the best time complexity will perform
better.  We consider this possibility in line 1--2 in the DenseEC
algorithm.  (However, in our experiments, we see that the main DenseEC
code in line 3--13 out-performs Intel MKL in a wide range of cases.)
}


\subsection{Evaluation Path Section Functions} \label{subsec:threshold}

We compute the three functions used in the DIM$^3$ algorithm to select
different evaluation paths.  
%
Table~\ref{tab:symbol} lists the symbols used in this subsection.
All the listed parameters can be measured in advance.

\begin{table}[b]
  \vspace{-0.1in}
  \small
  \caption{Symbols used in Section~\ref{subsec:threshold}.}
  \label{tab:symbol}
  \vspace{-12pt}
  \begin{tabular}{lp{0.8\columnwidth}}
    \hline
    Symbol & Description \\
    \hline
    $t_{seqR}$ & time for sequential memory read \\
    $t_{randR}$ & time for random memory read \\
    $t_{randRW}$ & time for random memory read-modify-write\\
    $t_{hash}$ & time of a lookup or insertion to plain hash table \\
    $t_{map}$ & time to access the optimized hash table for mapping\\
    $t_{ECs}$ & time of a non-SIMD comparison in DenseEC \\
    $t_{ECd}$ & time of a SIMD comparison in DenseEC \\ 
    \hline
\end{tabular}
\end{table}

\Removed{
\begin{figure}[h]
  \centering
  \vspace{-10pt}

  \includegraphics[width=\linewidth]{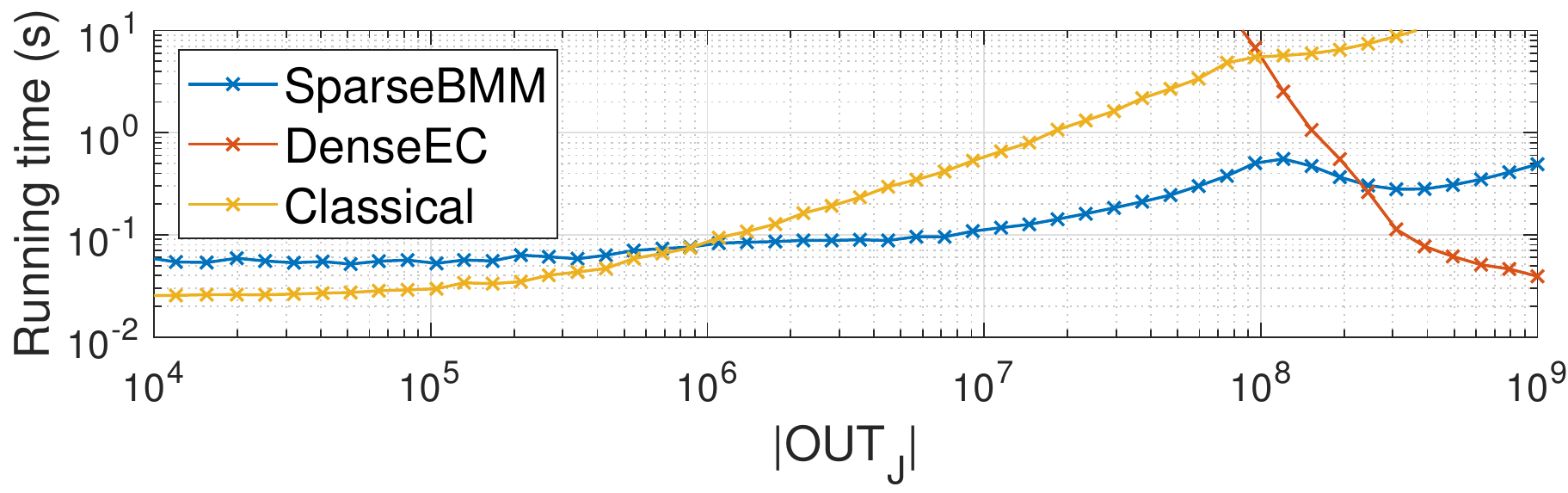}

  \vspace{-10pt}

  \caption{Comparison of three algorithms varying $|OUT_J|$
of joining two tables each having $10^6$ tuples.}

  \vspace{-6pt}

  \label{fig:algthreshold}
\end{figure}
}

\techreport{
\begin{figure*}[t]
  \centering
  \subfigcapskip=-6pt
  \subfigure[\Removed{Function} $f_1$ for strategy selection]{
    \includegraphics[width=0.3\linewidth]{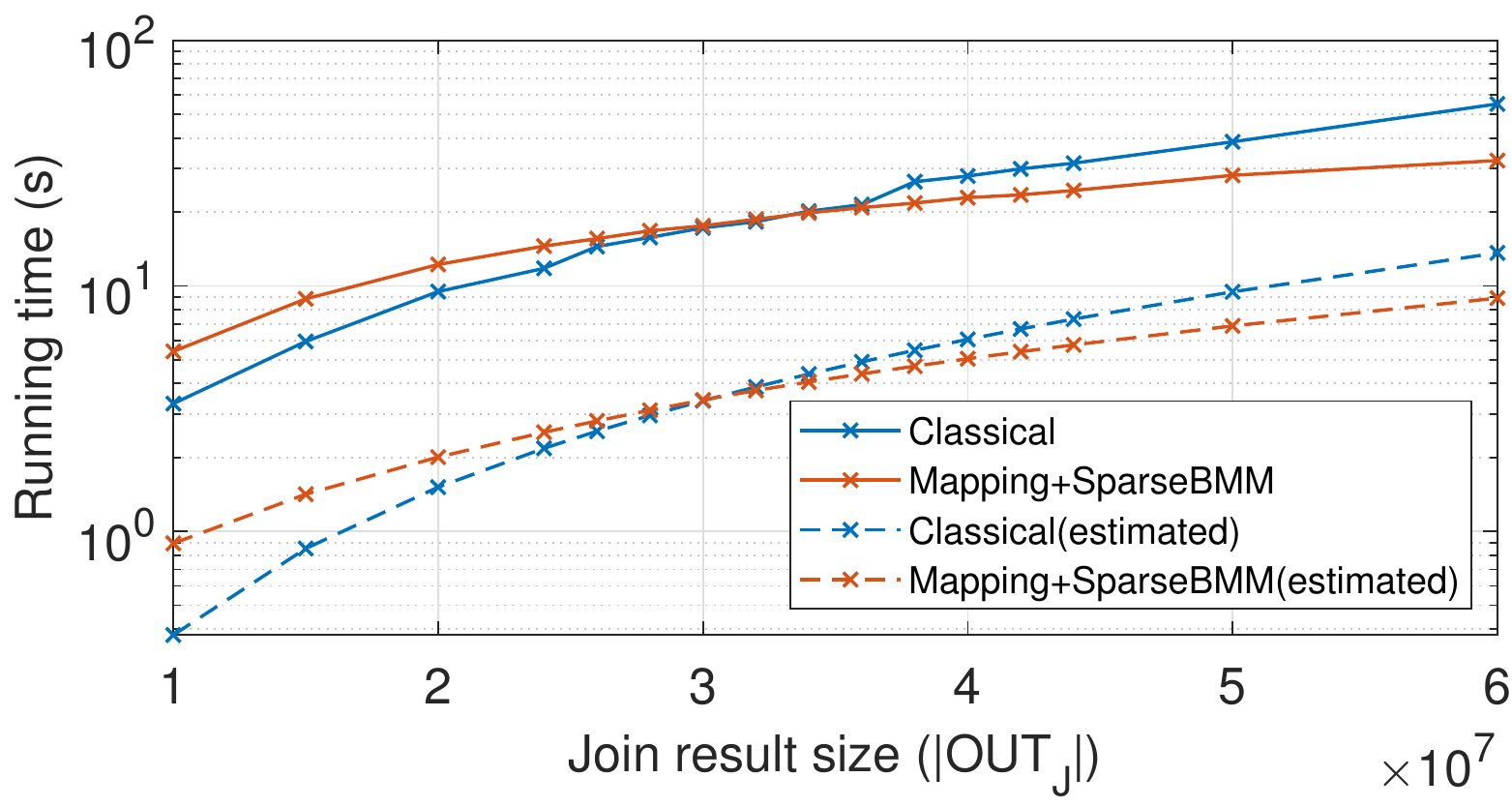}
    \label{fig:thr:alg}
  }
  \subfigure[\Removed{Function} $f_3$ for selecting bitmap comparison method]{
    \includegraphics[width=0.3\linewidth]{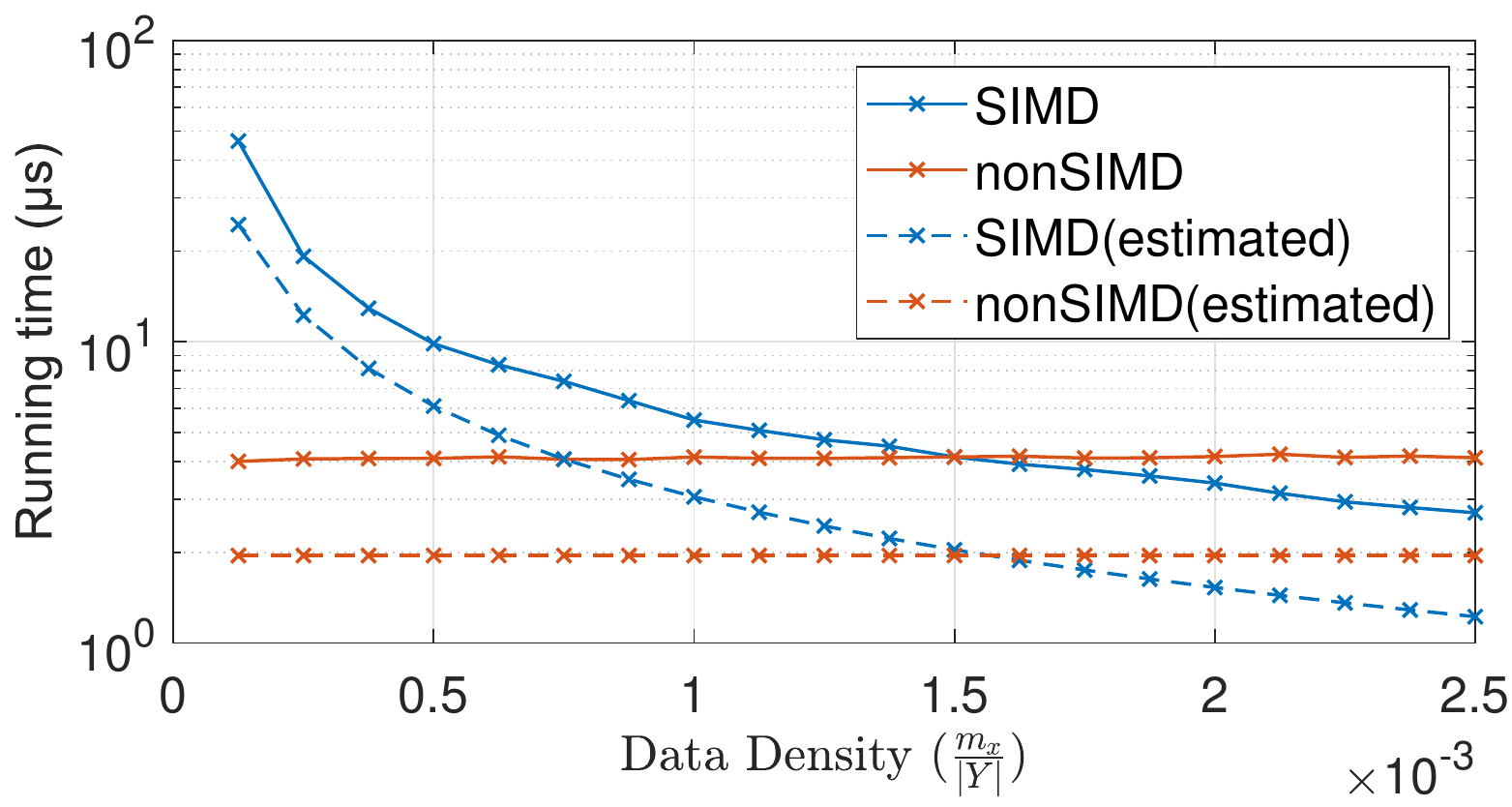}
    \label{fig:thr:r}
  }
  \subfigure[\Removed{Function} $f_2$ for deciding column density]{
    \includegraphics[width=0.3\linewidth]{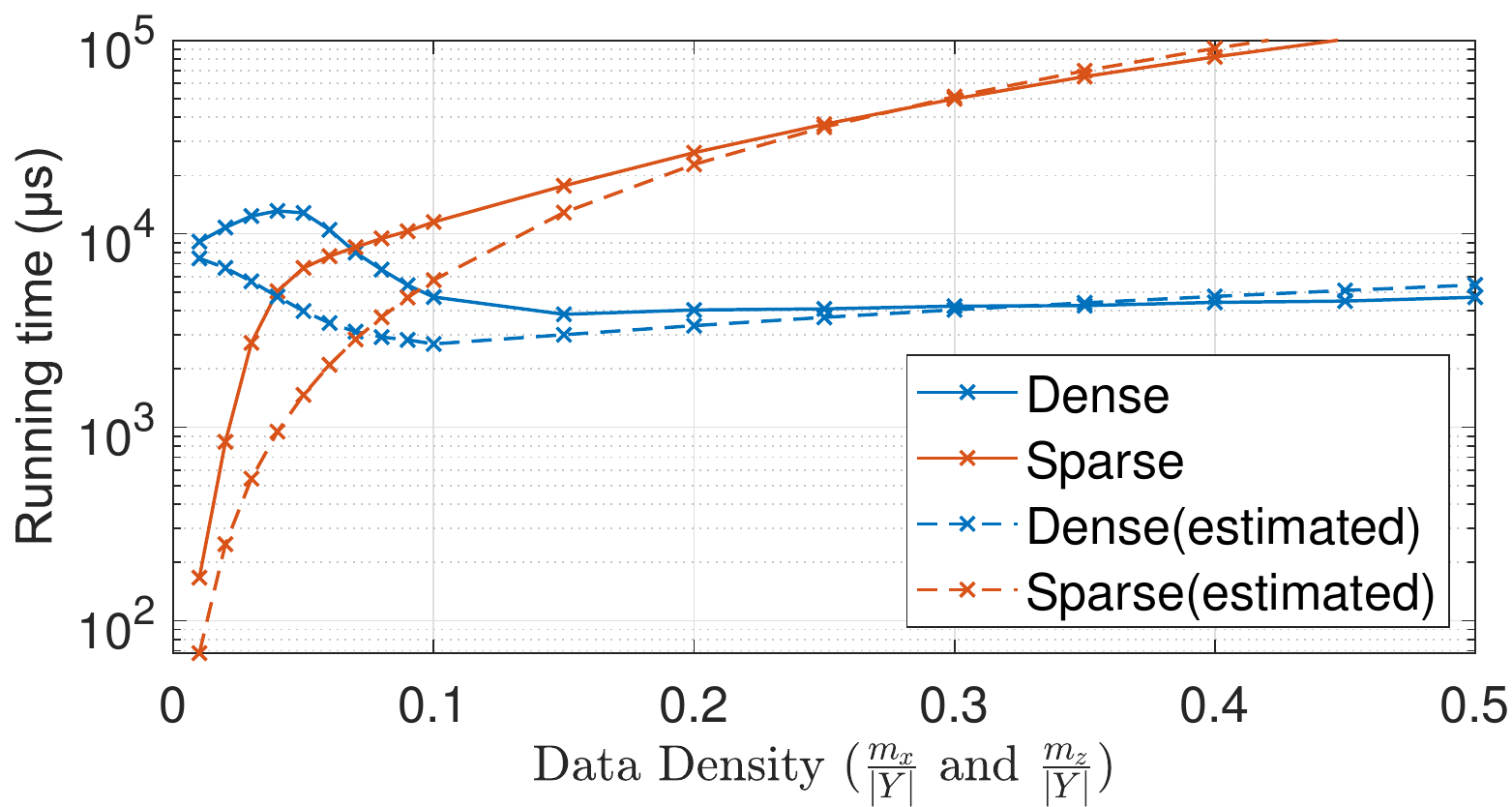}
    \label{fig:thr:s}
  }

  \vspace{-15pt}

  \caption{Effectiveness of the three functions for choosing
evaluation paths in DIM$^3$.}
  \label{fig:thr}

  \vspace{-12pt}
\end{figure*}
} 

\Removed{
Figure~\ref{fig:algthreshold} compares the classical solution,
mapping+SparseBMM, and mapping+DenseEC.  We generate $10^6$ tuples in
tables $R$ and $S$.  The values of columns $x$, $y$, and $z$ are
randomly chosen in [0, $N_{MAX}$], where $N_{MAX}$ ranges from $10^3$
to $10^8$. The data gets more sparse and $|OUT_J|$ becomes smaller as
$N_{MAX}$ increases.  The X-axis shows $|OUT_J|$, while the Y-axis
reports execution time in seconds.  Please note the logarithmic scale.
We see that the classical solution performs the best only when the
data is extremely sparse (i.e., $|OUT_J|$ is very small).  In such
cases, no $S.z$ will be judged as dense and DenseEC will not be used.
Therefore, the hybrid strategy has degenerated into mapping+SparseBMM.
}

\Paragraph{Strategy Selection ($f_1$)} 
\revise{ DHK chooses the classical solution using a rule-of-thumb
condition: $|OUT_J| \leq 20 \cdot N$, where $N=|R|=|S|$.  In
comparison, }
DIM$^3$ determines whether to use the classical solution or the hybrid
solution with function $f_1$.
As the classical solution is beneficial only when the data is sparse,
we compare mapping+SparseBMM and the classical solution to compute
$f_1$.  When $f_1 > 0$, the classical solution is faster and will be
chosen.
\revise{ \begin{align}
f_1(|R|,&|S|,|OUT_J|)=(T_{mapping}+T_{SparseBMM})-(T_{join}+T_{dedup})
\nonumber\\
& = T_{mapping} + (T_{SparseBMM}- T_{join}) - T_{dedup} \nonumber\\
& \approx 2(|R|+|S|)t_{map} + |OUT_J|t_{randRW} - |OUT_J|t_{hash} 
\label{for:alg}
\end{align} }
Here, $T_{mapping}=2(|R|+|S|)t_{map}$ from Section~\ref{sec:mapping}.
To estimate $T_{SparseBMM}- T_{join}$, we see that generating join
results with CSR in SparseBMM has smaller or similar cost compared to
hash joins.  Hence, the difference is mainly the deduplication cost of
SparseBMM.  This is $|OUT_J|* t_{randRW}$ because SparseBMM performs
a random access to the $SPA$ array per join result.  Finally,
$T_{dedup}= |OUT_J|t_{hash}$ because the classical solution performs a
hash table access for deduplicating every join output tuple.  




\Paragraph{Bitmap Comparison in DenseEC ($f_3$)}
Line~\ref{alg:ec:R} of Algorithm~\ref{alg:ec} uses function $f_3$ to
choose the SIMD or non-SIMD method for comparing $Bitmap_x$ and
$Bitmap_z$.   
The bitmaps have $|Y|$ bits.  Suppose there are $m_x$ and $m_z$ 1's in
$Bitmap_x$ and $Bitmap_z$, respectively.  




In the non-SIMD method, the comparison stops as soon as a check hits a
set bit in $Bitmap_z$.  The probability that the check hits a set bit
is $p_s=\frac{m_z}{|Y|}$.  The number of checks follows a geometric
distribution with probability $p_s$, and the method performs at most
$m_x$ checks.  Hence, the expected number of checks is calculated as:
\vspace{-0.05in}
$$
  Check_{nonsimd}= \sum_{i=1}^{m_x}i(1-p_s)^{i-1}p_s + m_x(1-p_s)^{m_x} = \tfrac{1-(1-p_s)^{m_x}}{p_s}
\vspace{-0.05in}
$$

In the SIMD method, every SIMD comparison checks 256 bits in the two
bitmaps.  When the bits at the same position in the two bitmaps are
both set, the comparison returns true and the process stops.  The
probability that an SIMD comparison returns true is
$p_d=1-(1-\frac{m_xm_z}{|Y|^2})^{256}$.  The number of SIMD checks
follows a geometric distribution with probability $p_d$, and the
method performs up to $\frac{|Y|}{256}$ checks. Hence, the expected
$Check_{simd}$= $\tfrac{1-(1-p_d)^{\frac{|Y|}{256}}}{p_d}$.


Then, we can compute $f_3$ as follows.  When $f_3 > 0$, the SIMD
method is preferred. Otherwise, the non-SIMD method is selected.
\vspace{-0.03in} \begin{equation}
  f_3(m_x,m_z,|Y|) = Check_{nonsimd}t_{ECs}-Check_{simd}t_{ECd}
\vspace{-0.03in} \end{equation}


\noindent
We can reduce the computation overhead of $f_3$ as follows.  Note that
$m_x$ and $|Y|$ are constants in the for-loop at
line~\ref{alg:ec:forSz} of Algorithm~\ref{alg:ec}.  $f_3(m_x,m_z,|Y|)$
is monotonically increasing with $m_z$. Therefore, an optimization is
to use binary search to find the threshold value of $m_{zt}$ so that
$f_3(m_x,m_{zt},|Y|)=0$.  Then, we can select the SIMD method if $m_z
> m_{zt}$.
\revise{Moreover, we find that multiple $R.x$'s can share the same
$m_x$.  Since $m_{zt}$ is determined for a given $m_x$, we can cache
the pairs of $m_x$ and $m_{zt}$, then reuse the computed $m_{zt}$ to
avoid redundant binary searches.  
In this way, the time for computing all $f_3$ thresholds is 
at most 1.15ms for the real-world data sets in
Section~\ref{sec:exp}, which is less than 0.5\% of the total run
time.}



\Paragraph{Dense vs. Sparse Partitions ($f_2$)}
At line~\ref{alg:main:S} in DIM$^3$ (Algorithm~\ref{alg:main}),
intersection-free partitioning uses function $f_2$ to determine if
$S.z$ column is dense or sparse.



For column $z$, the number of join results can be estimated as
\NewlyAdded{$\tfrac{m_{z}}{|S|}|OUT_J|$}, where $m_z$ is the number of
$S$ tuples in column $z$.  We denote this value as $OUT_{J,z}$.
We estimate the cost of processing column $z$ using either SparseBMM
and denseEC.

The cost of SparseBMM for $z$ is computed as follows:
\revise{
\vspace{-0.01in} $$
T_{sparseBMM}= \tfrac{(2|X|+|R|)t_{seqR}+2|R|t_{randR}}{|Z|}+
OUT_{J,z}( t_{seqR}+ t_{randRW})
%
\vspace{-0.01in} $$ 
}
The first component computes the cost of the two for-loops at Line
2--3 amortized to one of $|Z|$ columns.  The second component
estimates the cost of Line 4--7.
For DenseEC, we know that the cost for each pair of $x$ and $z$ from
the above.  Then we can sum this up to obtain $T_{denseEC}$=
$\sum_{i=1}^{|X|}min(Check_{nonsimd}t_{ECs},Check_{simd}t_{ECd})$.

Finally, we can compute $f_2$ as follows.  When $f_2 > 0$, we consider column
$z$ as dense.
\vspace{-0.05in} \begin{equation}
  f_2 = T_{sparseBMM} - T_{denseEC}
\vspace{-0.03in} \end{equation}

%
%
%
%
%


\techreport{

\Paragraph{Effectiveness of the Functions}
Figure~\ref{fig:thr:alg} compares the measured running time and the
estimated running time of the classical solution and
mapping+SparseBMM.  In this experiment, we generate random data with
column values in the range of 0 to $10^7$.  We vary $|R|$=$|S|$ from 5
million to 60 million.  The X-axis shows the join result size.  The
Y-axis reports the running time.  From the figure, we see that the
estimated times show consistent trends compared to the measured times.
Function $f_1$ models the intersection of the curves for the two
algorithms.  We see that the intersection points of the measured and
estimated time curves are very close, showing the effectiveness of
$f_1$ for strategy selection.

Figure~\ref{fig:thr:r} compares the measured running time and the
estimated running time of the SIMD and non-SIMD methods.  From the
figure, we see that (i) the estimated values are always smaller than
the measured values because certain runtime overheads, such as the
for-loop, are not taken into consideration; (ii) the estimated curves
show similar trends as the measured curves; and (iii) the intersection
points of the two sets of curves are very close.  

Figure~\ref{fig:thr:s} compares the measured running time and the
estimated running time of SparseBMM and DenseEC.  From the figure, we
see that (i) the estimated curves show similar trends as the measured
curves; and (ii) the intersection points of the two sets of curves are
close.  This demonstrates the effectiveness of using function $f_2$
for intersection-free partitioning.

\revise{
\subsection{Time Complexity Analysis}
\label{subsec:complexity}

Based on the above discussion in Section~\ref{subsec:threshold}, we
have the following:
\begin{align*} Complexity_{SparseBMM}&=\Theta(|R|+|S|+|OUT_J|) \\
  Complexity_{DenseEC_{nonsimd}}&=\Theta(\sum_{i=1}^{|X|}\sum_{j=1}^{|Z|}T_{Check\_nonsimd}(x_i,z_j))
\\
  Complexity_{DenseEC_{simd}}&=\Theta(\sum_{i=1}^{|X|}\sum_{j=1}^{|Z|}T_{Check\_simd}(x_i,z_j))
\\
\end{align*}
Note that $T_{Check\_nosimd}(x_i,z_j)$ and $T_{Check\_simd}(x_i,z_j)$
are the actual numbers of checks given $(x_i,z_j)$, which are
determined by the input data distribution.

\Paragraph{Worst-Case Costs} We compute the worst-case costs.  Suppose
that input tables $R$ and $S$ do not have duplicates. 
%
Then, for SparseBMM, it is clear that $|OUT_J| < |X||Y||Z|$.  For
DenseEC, early stopping never happens in the worst cases.  Hence,
$T_{Check\_nonsimd}(x_i,z_j)=\Theta(m_{x_i})$ and
$T_{Check\_simd}(x_i,z_j)=\Theta(\frac{|Y|}{SIMD\_bit\_length})$.

Therefore, the worst case time complexities are as follows:
\begin{align*}
  Complexity_{SparseBMM}&=\mathcal{O}(|X||Y||Z|) \\
  Complexity_{DenseEC_{nonsimd}}&=\mathcal{O}(|R||Z|) \\
  Complexity_{DenseEC_{simd}}&=\mathcal{O}(\frac{|X||Y||Z|}{SIMD\_bit\_length}) \\
\end{align*}

\Paragraph{Average-Case Costs} It is difficult to compute the precise
formula for average-case costs because the time complexities are
affected by the path selection decisions and the density of the input
tables.
In the following, we analyze the costs under a simplifying assumption
of uniform density distribution in the input tables.

We assume that the non-zeros in matrices $\mathbf{R}_{x,y}$ and
$\mathbf{S}_{y,z}$ are uniformly randomly distributed, and the
densities (probability) of non-zero values are $\alpha$ and $\beta$,
respectively.  Then, we have $|OUT_J|=\alpha\beta|X||Y||Z|$,
$m_x=\alpha|Y|$ and $m_z=\beta|Y|$.  We also assume that
$|OUT_J|>max(|R|,|S|)$.

Since all the rows and columns of the input data are of the same
density, $f_2$ considers the entire $S$ as dense or sparse.
Similarly, $f_3$ selects the same execution path for all pairs of
$(x,z)$.  That is, all data will be processed by the same algorithm.
We can use the expected numbers of checks ($Check_{nonsimd}$ and
$Check_{simd}$) computed in the previous subsection to estimate
$T_{Check\_nosimd}$ and $T_{Check\_simd}$.  The above formula can be
rewritten as follows:
\begin{align*} Complexity_{SparseBMM}&= \Theta(\alpha\beta|X||Y||Z|)
\\ Complexity_{DenseEC_{nonsimd}}&=
\Theta(\frac{1-(1-\beta)^{\alpha|Y|}}{\beta}|X||Z|) \\
Complexity_{DenseEC_{simd}}&=
\Theta(\frac{1-(1-\alpha\beta)^{|Y|}}{1-(1-\alpha\beta)^{SIMD\_bit\_length}}|X||Z|)
\\ \end{align*}

The classical hash solution shares the same complexity with SparseBMM
(i.e., $\Theta(\alpha\beta|X||Y||Z|)$). 

The complexity of dense MM is $\Theta(|X||Y||Z|\gamma^{\omega-3})$,
where $\gamma=min\{|X|,|Y|,|Z|\}$~\cite{SIGMOD20}. Though the best
known $\omega$ value can be 2.373~\cite{matrix2018}, these theoretical
MM algorithms are often considered
impractical~\cite{LeGall12,Pegoraro20}. And in the widely used
packages for MM (e.g. Intel MKL), $\omega=3$.

\Paragraph{Estimating the Costs of Two Special Cases}
We consider two specific cases to better understand the average-case
costs.  First, $\alpha=\beta=1$.  That is, the input matrices are
full.  In this very dense case, the time complexity of SparseBMM is
$\Theta(|X||Y||Z|)$,
while the costs of DenseEC$_{nonsimd}$ and DenseEC$_{simd}$ are both
$\Theta(|X||Z|)$.  Thanks to the early stopping technique, DenseEC is
significantly faster than the other algorithms (including Dense
MM).

Second, $\alpha=\beta=\frac{1}{|Y|}$.  This is a very sparse case. The
complexity of SparseBMM is $\Theta(\frac{|X||Z|}{|Y|})$.  No-simd will
be chosen for the very sparse case.  The complexity of
DenseEC$_{nonsimd}$ is $\Theta(|X||Z|)$. 
%
%
Therefore, SparseBMM is the best solution.

}

} 

\section{Partial Result Caching} \label{sec:cache}

\begin{figure}[t]
    \subfigure[HetRec data set.]{
      \includegraphics[width=0.45\linewidth]{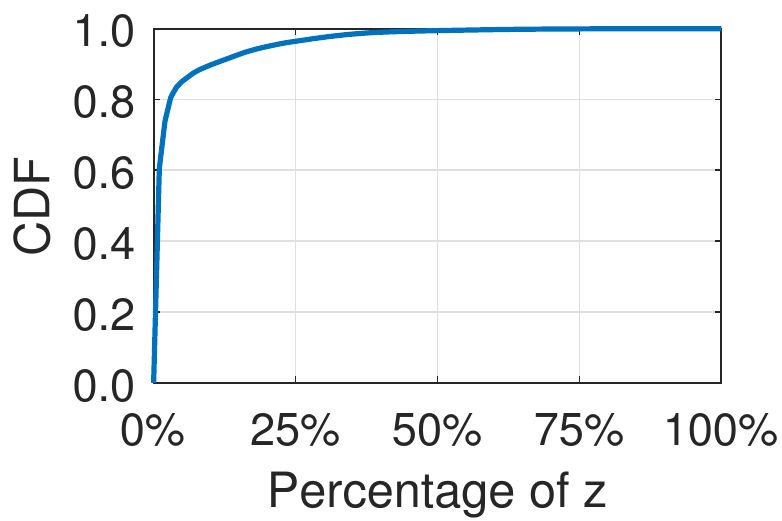}
    }
    \subfigure[MV2 data set.]{
      \includegraphics[width=0.45\linewidth]{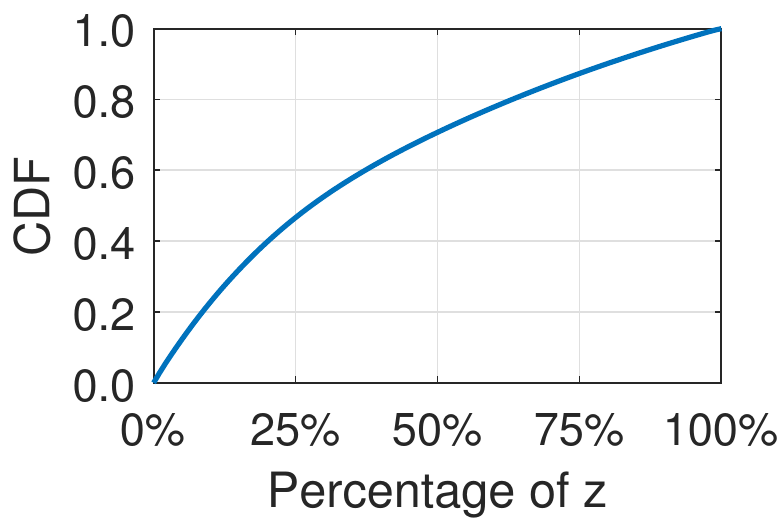}
    }
  \vspace{-0.1in}
  \caption{CDF of computation cost.}
  \label{fig:zcost}
\end{figure}

We observe that computing Join-Project results for different $z$
values can take widely different amounts of time for real-world data
sets.  Figure~\ref{fig:zcost} depicts the CDF
\revise{(Cumulative Distribution Function)} of computation cost for
two representative real-world data sets.  In HetRec, the top 1\% of
$z$ values take 60\% of the total time.  In MV2, the top 10\% of $z$
values contribute to 23\% of the total cost.  This motivates us to
study caching the results of a subset of $z$ values to improve
Join-Project performance.  Note that our technique is different from
materialized views~\cite{LarsonY85} or query result
cache~\cite{LarsonGZ04}, where the full query results are cached.

\Paragraph{DIM$^3$ with Cached Partial Results}  We choose to cache
results based on $z$ values because the intersection-free partitioning
method divides table $S$ according to the $S.z$ values.  Hence, it is
easy to integrate the cached partial results into the DIM$^3$
algorithm. 

Suppose $Z_{cached}$ is the subset of ${S.z}$, whose Join-Project
results are cached.  That is, the cached partial results are
$R_{cached}$ = $\{$\revise{$($}$x,z$\revise{$)$}$|$\revise{$($}$x,z$\revise{$)$}$\in \Pi_{x,z}(R(x,y) \Join_y S(z,y))
\land z\in Z_{cached}\}$.  In Figure~\ref{fig:dim3}, we can simply
omit any $z\in Z_{cached}$ when generating $S_{sparse}$ and
$S_{dense}$,  then keep the other steps of DIM$^3$ unchanged.
Finally, we output $R_{cached}$ in addition to the results computed by
DIM$^3$.

\Paragraph{Caching Score}  \revise{Given a caching space budget $B$
for a Join-Project query}, we rank $z$ values for caching with the
following score:
\vspace{-0.05in} $$
  CacheScore(z)= \tfrac{cost(z)}{space(z)}
\vspace{-0.05in} $$
$space(z)$ is the cache space required to store the
\revise{$($}$x$,$z$\revise{$)$} results for the given $z$.  The cache
content is \revise{$($}$z,size,x_1,x_2,...,x_k$\revise{$)$}.  When $k$
is small, $size$=$k$>0, and we cache the original results.  When
$k$>$0.5|X|$, we store the complement set of $x$'s to save space, and
set $size=k-|X|$<0.  Hence, $space(z)$=$2$+$min(k, |X|-k)$.

$cost(z)$ is the time for computing \revise{$($}$x$,$z$\revise{$)$} results for the
given $z$.  This is the benefit from caching $z$.  Since the
processing related to $z$ is distributed into many inner loops in
SparseBMM and DenseEC, we cannot directly measure $cost(z)$.  Instead,
we collect statistics and estimate the cost as follows:
\vspace{-0.05in} $$
  cost(z)=
n_{sparse}(t_{seqR}+t_{randRW})+n_{simd}t_{ECd}+n_{nonsimd}t_{ECs}
\vspace{-0.03in} $$
where $n_{sparse}$ is the number of times that $z$ is checked in the
inner loop of the SparseBMM algorithm, $n_{simd}$ and $n_{nonsimd}$
are the number of times that $z$ is encountered in the SIMD and
non-SIMD part of the inner loop of the DenseEC algorithm,
respectively.

$CacheScore(z)$ shows the benefit per unit space for caching $z$.  The
higher the $cost(z)$, the lower the $space(z)$, the more beneficial to
cache $z$.  Interestingly, a $z$ value with high $cost(z)$ may produce
a large number of results.  This actually leads to small $space(z)$
when the complement result set is saved. 
%

\Paragraph{Practical Considerations}  We discuss several practical
issues for using the partial result caching.
First, database users can enable Join-Project caching dynamically.  We
see in experiments that the statistics collection and computation for
$cost(z)$ and $space(z)$ do not incur significant overhead for
DIM$^3$.  Hence, after caching is enabled, the first run of DIM$^3$
can compute $CacheScore(z)$ and populate the partial result cache.
Then, subsequent runs of DIM$^3$ on the same tables can leverage the
cached results to improve performance.
Second, when the underlying tables are modified by
insert/delete/update, we can simply invalidate the cache and let the
next run of DIM$^3$ to re-populate the result cache.  Note that
updating the cached results (similar to the view maintenance
problem~\cite{BlakeleyLT86}) is beyond the scope of this work.
Finally, we can use the cached result to support filter predicates on
$x$ and/or $z$.  However, if there are filter predicates on $y$, we
have to run DIM$^3$ without the cached result.

\section{Support for Join-Op Query Types} \label{sec:extends}

In the above, we focus on the Join-Project operation.  Join-Project is
a special case of Join-$op$ queries, where $op$ is a common SQL
operation.  An interesting question arises: Is it possible to apply
DIM$^3$ to Join-$op$ queries in general?  We consider $op$ other
than projection in the following:
\begin{list}{\labelitemi}{\setlength{\leftmargin}{7mm}\setlength{\itemindent}{-3mm}\setlength{\topsep}{0.5mm}\setlength{\itemsep}{0mm}\setlength{\parsep}{0.5mm}}

\item \emph{Group-by Aggregation}:  The group-by operation inherently
removes duplicates. A join followed by a group-by aggregation
operation, which we call Join-Aggregate for simplicity, implicitly
performs a Join-Project operation.  We describe how to extend DIM$^3$
to support Join-Aggregate in Section~\ref{sec:ja}.

\item \emph{Join}:  Multiple join operations are followed by a
duplicate eliminating projection.  One strategy to evaluate this type
of queries is to employ Join-Project to deduplicate the results of the
last join operation.  But can we do better?  We consider how and when
to push the deduplication down in the query plan of \Removed{multi-way joins} \revise{MJP} in
Section~\ref{sec:cq}.

\item \emph{Selection or Sorting}: As Join-Selection or Join-Sorting
do not require deduplication, there is no need to employ Join-Project
for these types of queries in general.  However, in special cases
where there are too many duplicates, an alternative evaluation
strategy can be more efficient.  We compute Join-Project and keep the
duplicate count for each generated join result tuple.  Then, the
selection or sorting operation processes the much smaller,
deduplicated join result, thereby achieving better performance.
Finally, we output the correct number of duplicates based on the
per-tuple duplicate counts.

\item \emph{Intersection/Difference/Union}: 
First, the computation of intersection is similar to a join operation.
Hence, for Join-Intersection, we can employ \Removed{multi-way joins with
projection} \revise{MJP}.
Second, set difference can be evaluated as a left-outer join followed
by deduplication.  We can modify DIM$^3$ to compute Outer-Join-Project
for Join-Difference.  To support outer-joins, DIM$^3$ can be extended
with a bitmap for $x$ ($z$).  It sets a bit if the corresponding $x$
row ($z$ column) has generated join results.  In this way, the
modified DIM$^3$ can compute $x$ ($z$) with no matches for
outer-joins.
Finally, Join-Union requires the deduplication of the outputs from two
joins.  We can push the deduplication operation down 
in the query plan, and apply similar considerations as in
Section~\ref{sec:cq}.

\end{list}


\subsection{Join-Aggregate} \label{sec:ja}

Our DIM$^3$ algorithm can be applied to Join-Aggregate operations with
slight modifications. Without loss of generality, we divide
Join-Aggregate operations into two categories based on the group by
attributes: i) group-by attributes are from both tables, and ii)
group-by attributes come from one table.

\Paragraph{Group-by Attributes from Both Tables} For instance, given
tables $R(x,y,v)$ and $S(z,y,u)$, we want to compute 
\vspace{-0.02in}
$$ 
_{x,z}\mathcal{G}_{aggregate(f(R.v,S.v))}(R(x,y,v) \Join_y
S(z,y,v)) 
\vspace{-0.02in}
$$
This task is similar to the original Join-Project operation.  The main
difference is that it computes $aggregate(f(R.v,S.u))$ on the join
results with the same \revise{$($}$x$, $z$\revise{$)$} rather than deduplicating the
results.

We modify DIM$^3$ to support this task.  In SparseBMM, we change the
$SPA$ array to contain the aggregate for each $cur_z$.  $SPA$ is
initialized in each outer-loop iteration.  Line 5--7 is modified to
accumulate the aggregate for group \revise{$($}$cur_x$, $cur_z$\revise{$)$}.
In DenseEC, bitmaps cannot support the aggregates.  Therefore, we use
a plain dense MM, while computing aggregates for each pair of
\revise{$($}$cur_x$, $cur_z$\revise{$)$}.


\Paragraph{Group-by Attributes from One Table} Given tables $R(x,y)$
and $S(z,y)$, we want to compute the following query.  If the group-by
attribute is from $S$, we switch table $R$ and $S$.
\vspace{-0.02in} $$ _{x}\mathcal{G}_{aggr(z)}(R(x,y) \Join_y
S(z,y)) \vspace{-0.02in} $$
We can rewrite the query as follows:
\revise{ \vspace{-0.02in} $$ _{x}\mathcal{G}_{aggr"(z')}(R(x,y)
\Join_y (_{y}\mathcal{G}_{aggr'(z)}(S(z,y)))) \vspace{-0.02in} $$ }
We first compute the group-by aggregate on $S$ with $y$ as the
group-by key.  
\revise{If $aggr$ is $sum$, $min$, or $max$, then $aggr'$ and $aggr"$
are the same.  For $count$, $aggr'$ is $count$ and $aggr"$ is $sum$.
For $avg$, the $aggr'$ consists of both $sum$ and $count$.  Then
$aggr"$ accumulates the two components, and finally computes a
division to obtain the $avg$.}
We denote the resulting table as $SG(y,z')$, where $z'$ is the
intermediate aggregate value(s) for $y$.  Note that $|SG|=|Z|$ is
often much smaller than $|S|$.
$SG$ is highly sparse because there is no duplicate $y$'s in the
table.  Therefore, we employ the classical hash-based algorithm to
join $R$ and $SG$ then compute the final group-by aggregates.





\subsection{\revise{MJP} (Multi-Way Joins with Projection)} \label{sec:cq}

So far, we have been focusing on Join-Project on
\emph{two} tables. In this subsection, we study deduplication on the
results of joining \emph{multiple} tables.  This is also called
conjunctive queries with projection~\cite{ICDT21Enum}. For example,
the line join projection with $n$ tables is expressed as:
$$
  \Pi_{x_1,x_{n+1}}(R_1(x_1,x_2) \Join_{x_2} R_2(x_2,x_3) \Join_{x_3}
... \Join_{x_{n}} R_n(x_n,x_{n+1}))
$$
%

\begin{figure}[t]
  \centering
  \includegraphics[width=0.7\linewidth]{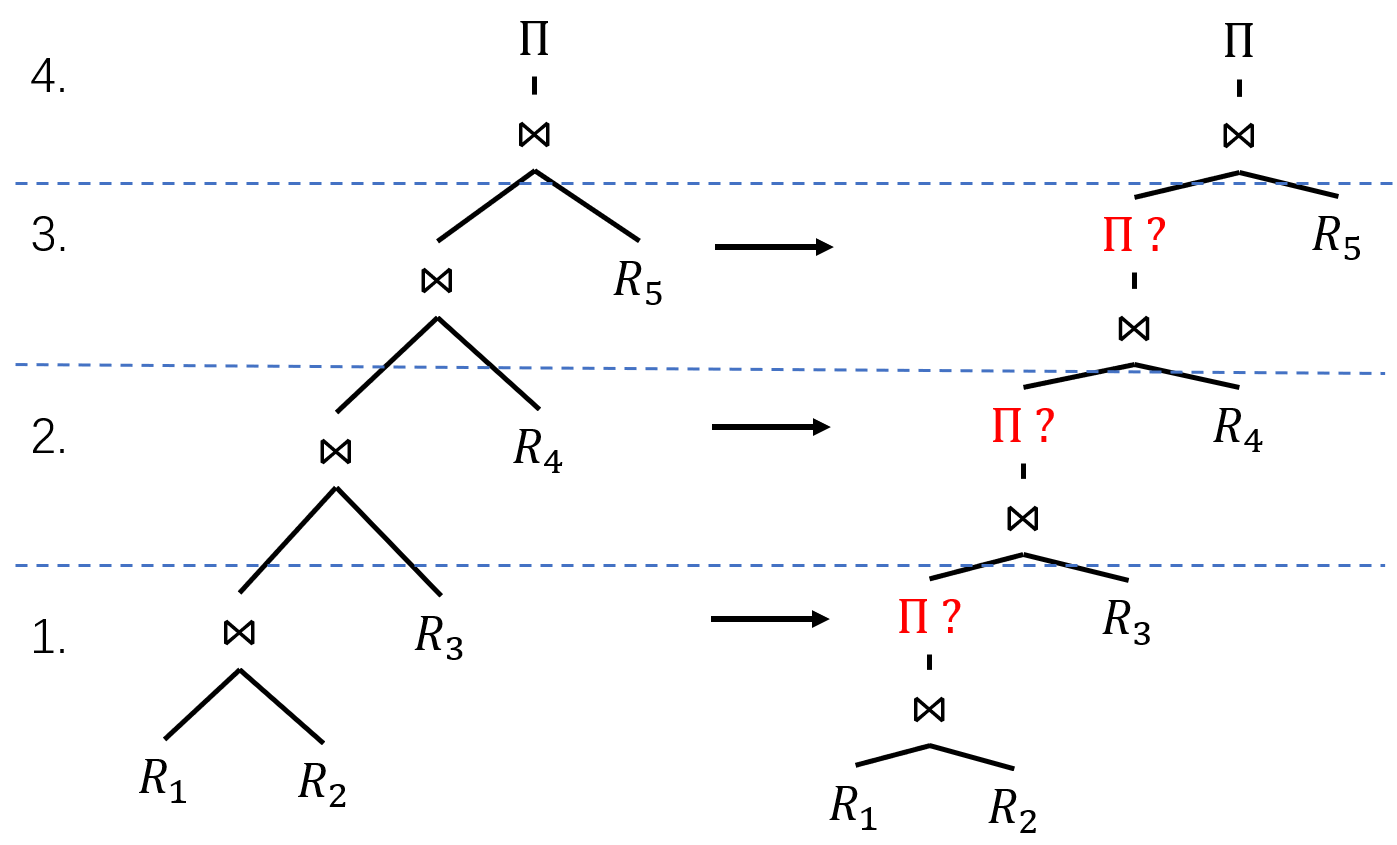}

  \vspace{-0.12in}
  \caption{The deduplication placement problem.}
  \label{fig:dp}
  \vspace{-0.2in}

\end{figure}

Figure~\ref{fig:dp} illustrates the query plan tree for evaluating a
\Removed{multi-way join query with projection} \revise{MJP query}.  The deduplication operation
($\Pi$) can be pushed down to after each join operation.  While it incurs
extra overhead, deduplication reduces the intermediate result size,
thereby reducing the cost of subsequent join operations.
There are two baseline execution plans.  The first plan computes all
the joins and then deduplicate the final join results.  The second
plan deduplicates the results immediately after each join.
However, it is easy to construct cases to show neither plan is
optimal.
Therefore, we need to judiciously place the deduplication operations
into the query plan tree.

In the following, we develop a DP (Dynamic Programming) algorithm to
find the optimal query plan.  To limit the scope of our investigation,
we make the following assumptions.
\begin{enumerate}

  \item \emph{The join order is given by the query optimizer.} We
focus on the deduplication placement problem.  For ease of
presentation, we number the tables in the join order as $R_1$, ...,
$R_k$.  The problem of optimizing both join order and deduplication
placement is beyond the scope of this work.

  \item \emph{The query plan is in the form of a left-deep
tree~\cite{schneider1990tradeoffs,ioannidis1991left}.}  Left-deep or
right-deep trees are widely used in RDBMSs to process multi-way joins.
If a solution produces left-deep trees, it is easy to modify it
to support right-deep trees. 

  \item \emph{Estimated $|OUT_{Ji}|$ and $|OUT_{Pi}|$, where
$i$=2,...,$k$, are available}.  \revise{(Please refer to recent work
on cardinality esitmation for 
details~\cite{qiu2021weighted,cai2019pessimistic,hertzschuch2021simplicity,leis2017cardinality}.)} Here,
$|OUT_{Ji}|$ = $|R_1 \Join \cdots \Join R_i|$ denotes the size of the
intermediate results after joining the first $i$ tables without
deduplication.  $|OUT_{Pi}|$ = $|\Pi(OUT_{Ji})|$ denotes the size of
the intermediate results after deduplication.  For simplicity, we
define $|OUT_{P1}|$ = $|OUT_{J1}|$ = $|R_1|$.  

  \item \emph{Adding deduplication to the $i$-th join reduces the
final join result by a factor of $\tfrac{|OUT_{Pi}|}{|OUT_{Ji}|}$}.
That is, $|\Pi(R_1 \Join \cdots \Join R_i) \Join \cdots \Join R_k|
\approx \tfrac{|OUT_{Pi}|}{|OUT_{Ji}|}|OUT_{Jk}|$.  \revise{ This is
reasonable since the join input size to the $(i+1)$-th join is reduced
by a factor of $\tfrac{|OUT_{Pi}|}{|OUT_{Ji}|}$.} This assumption
simplifies the estimation of the final result size after inserting a
deduplication operation.
More precise estimation requires collecting more statistics, and may
incur much higher cost.

\end{enumerate}

We use $DP_i$ to denote the optimal time for computing 
$\Pi (R_1 \Join \cdots \Join R_i)$. 
We observe that if we already add deduplication to the $i$-th join,
the deduplicated result $OUT_{Pi}$ does not change if more
deduplication operations are added to the joins before the $i$-th
join.  Therefore, we have the following equations.
\begin{align*}
  DP_i= 
  \begin{split}
    \left \{
    \begin{array}{ll}
      0, & i=1 \\
      \min_{j=1}^{i-1}(DP_j + \sum_{h=j+1}^{i-1} JoinCost_h + JPCost_i), & i>1
    \end{array}
    \right.
  \end{split}
\end{align*}
The formula inside $min$ computes the case where a deduplication is
added to the $j$-th join and there is no deduplication from the
$(j+1)$-th to $(i-1)$-th join.  The total cost of this case has three
components: (i) $DP_j$, which is the optimal time for computing
$OUT_{Pj}$; (ii) the cost of the subsequent joins, i.e., the $(j+1)$-th
to $(i-1)$-th join; and (iii) the cost of the final join project
operation.  Note that when $j$=1, there is no deduplication before the
$i$-th join.  

Based on assumption (3), we estimate the input table size for
component (ii) and (iii).  Specifically, $JoinCost_h$ is the cost of
joining $\Pi(R_1 \Join \cdots \Join R_j) \Join \cdots \Join R_{h-1}$
and $R_h$.  The size of the former is estimated as
$\tfrac{|OUT_{Pj}|}{|OUT_{Jj}|}|OUT_{Jh-1}|$.  Moreover, $JPCost_i$ is
the cost of the Join-Project operation at the $i$-th join.  The sizes
of the two input tables to the Join-Project operation are
$\tfrac{|OUT_{Pj}|}{|OUT_{Jj}|}|OUT_{Ji-1}|$ and $|R_i|$.  The
intermediate join result in the Join-Project operation can be
estimated as $\tfrac{|OUT_{Pj}|}{|OUT_{Jj}|}|OUT_{Ji}|$ and the final
output size is $|OUT_{Pi}|$.  Given the sizes of the input tables, the
intermediate join result, and the final output, we can use the
formulas in Section~\ref{subsec:threshold} to estimate the cost of the
Join-Project operation.

%

For a \Removed{multi-way join query with projection} \revise{MJP query} on $n$ tables, the path to
get $DP_n$ in the DP process gives the optimal deduplication
placement.  The time complexity of this algorithm is $\Theta (n^3)$.
As $n$ is often not large, the overhead of this DP algorithm is
small.

\section{Performance Evaluation} \label{sec:exp}

In this section, we evaluate the performance of our proposed solutions
using both real-world and synthetic data sets.

\Removed{
There are three main goals:
\begin{enumerate}

  \item We compare DIM$^3$ with state-of-the-art solutions and with
RDBMSs to understand its performance benefits.

  \item We study the components of DIM$^3$ to verify the 
effectiveness of the design.

  \item We evaluate the proposed solutions for Join-Aggregate
and multi-way joins with projection.

\end{enumerate}
}

\subsection{Experimental Setup}
\label{subsec:setup}

\Paragraph{Machine Configuration} All experiments are performed on a
machine with Intel Core i7-9700 CPU (3.00GHz, Turbo Boost 4.70 GHz, 8
cores/8 threads, 12MB last level cache) and 32 GB RAM, running Ubuntu
18.04.5 LTS with Linux 5.4.0-81 kernel.  All code is written in C/C++
and compiled with \emph{g++} 7.5.0 using \emph{--std=c++11},
\emph{-O3}, and \emph{-mavx} flags.  \revise{MKL (Intel Math Kernel Library)}
is used in DHK.  By default, we run single-threaded experiments.  In
the parallel experiments, we use OpenMP for parallelization.

\Paragraph{Datasets}
We use six real-world datasets with different input size, $|OUT_J|$,
and $|OUT_P|$ in our experiments, as shown in
Table~\ref{tab:dataset-real}.  
\revise{Amazon~\cite{amazon} data set records the frequently co-purchased products on Amazon website. The Join-Project queries help to find potential products that can be co-purchased.}
%
%
Slashdot~\cite{Slashdot} is a technology-related news website.  The data set
contains friend/foe links between users of Slashdot.  The query computes the
indirectly connected pairs of users.
The data set and the query of HetRec~\cite{hetrec2011} follow the motivating
example in Section 1.
MV1 and MV2 are two MovieLens~\cite{Movielens} data sets.  They contain user
ratings for movies.
Similarly, Jokes~\cite{Jokes} contains user ratings for jokes.
The Join-Project queries on MovieLens and Jokes data sets find users that have
rated the same objects.
Friendster~\cite{Friend} is a social network dataset that contains the
retweet relationship between users.  We use this data set to study \Removed{multi-way
joins with projection} \revise{MJP}, which computes the multi-hop connections between users.

%
\revise{We also conduct experiments on the TPC-H data set with SF=10.
A representative query joins \emph{LineItem} and \emph{Orders}, then
projects on \emph{CUSTKEY} and \emph{SUPPKEY} to attain the purchase
relationship between customers and suppliers. However, since this join
is a primary-foreign key join, $|OUT_J|$ is similar to the size of
\emph{LineItem}.  Both DIM$^3$ and DHK choose the classical solution,
showing the same performance.  Hence, we do not study TPC-H further.}

In contrast, $|OUT_J|$ and $|OUT_P|$ in the other data sets are much
larger than their input sizes.  The higher the $pnz$, the denser the
dataset, and thus the more intermediate join results.
As we do not assume any special sort order in the input tables of
Join-Project, we randomly shuffle the data sets before the
experiments.

\begin{table}[t]
    \centering
    \small
    \addtolength{\tabcolsep}{-2.2pt}
    \caption{Real-world data sets used in experiments.}
    \label{tab:dataset-real}
    \vspace{-10pt}
    \begin{tabular}{lrrrrrr}
      \hline
      Data set &$|R|$ & $|S|$& $pnz(R)$& $pnz(S)$& $|OUT_J|$& $|OUT_P|$\\
      \hline
      \revise{Amazon}   & \revise{1.2M} & \revise{1.2M} & \revise{0.0018\%} & \revise{0.0018\%}  & \revise{14M} & \revise{11M} \\
      Slashdot & 905K & 905K & 0.02\%  & 0.02\%  & 118M    & 81M \\
      HetRec   & 487K & 438K & 0.02\%  & 0.03\%  & 125M    & 34M \\
      MV1      & 500K & 500K & 2.2\%   & 2.2\%   & 204M    & 29M \\
      MV2      & 1M   & 1M   & 4.5\%   & 4.5\%   & 816M    & 35M \\
      Jokes    & 617K & 617K & 25\%    & 25\%    & 10B     & 622M \\
      Friendster& \multicolumn{2}{c}{1.8B} & \multicolumn{2}{c}{$4.2\times10^{-7}$} & -- & -- \\
      \hline

    \end{tabular}\\
    Note: $pnz(M)$ is the percentage of non-zero elements in matrix $M$.

    \vspace{-0.2in}

\end{table}

\Paragraph{Solutions to Compare}
We compare two categories of solutions: stand-alone Join-Project
algorithms, and full-fledged RDBMSs.

We compare the following stand-alone implementations.
(1) \emph{Classical}: The classical solution performs the Radix
join~\cite{VLDB13Join} then a hash-based deduplication using a
flat\_hash\_map~\cite{Flathashmap}.
(2) \emph{MKL}: The state-of-the-art dense MM in Intel MKL is used to
evaluate Join-Project.  We implement the baseline mapping algorithm as
described in Section~\ref{sec:mapping}.
(3) \emph{DHK}: We obtained the DHK code from the authors of the DHK
paper~\cite{SIGMOD20}.  
\revise{Algorithm 3 in DHK chooses the classical solution when
$|OUT_J| \leq 20 \cdot N$, where $N=|R|=|S|$.  We find this feature is
missing in the DHK code, and implement the feature.}
The code assumes the input contains consecutive natural numbers.
Hence, we add the baseline mapping algorithm.  The code leaves the
final deduplication unimplemented.  Thus, we add a deduplication step
that checks every result \revise{$($}$x$,$z$\revise{$)$} from the
sparse part against the matrix $\mathbf{C}^{x\times z}$ computed in
the dense part.
(4) \emph{DIM$^3$}: Our proposed solution follows the description in
Section~\ref{sec:jp}.  We also study the individual components of
DIM$^3$. For the MM component, we compare DenseEC, SparseBMM, MM in
MKL, and a sub-cubic MM~\cite{FMM} based on the Strassen
algorithm~\cite{Strassen69}.

In addition to the stand-alone algorithms, we compare DIM$^3$ with
four full-fledged RDBMSs. (5) \emph{DBMSX}: one of the best performing
commercial RDBMSs.  
(6) \emph{PostgreSQL} version 13.3 and (7) \emph{MariaDB} version
10.5.11: two popular open-source RDBMSs.  
(8) \emph{MonetDB} version 11.39.17: a representative analytical main
memory RDBMS.
We set the configuration parameters of the RDBMSs to ensure that they
make full use of the memory.  For each experiment, we run the same
query on the target data set twice.  The first run warms up the RDBMSs
so that the input tables are loaded into main memory.  Then we measure
the performance of the second run.  

Our comparison between \emph{DIM$^3$} and RDBMSs is meaningful.  One
concern is that RDBMSs pay additional cost, including socket
communication, SQL parsing, query optimization.  (Note that we compute
a final count to avoid returning a large number of query results from
RDBMSs.) We quantify this additional cost by measuring the execution
time of the same queries on a very small data set.  The result is
1--3ms.  
In comparison, our reported run times on RDBMSs are from 7s to about 1
hour.  Hence, the extra cost of 1--3ms is negligible.  Our reported
run times indeed correspond to the cost of Join-Project query
processing in RDBMSs.

Unless otherwise noted, we focus on single-threaded performance in the
experiments.  The scalability results report multi-threaded
performance to show that our proposed algorithms are amenable to
parallelization.  Partial result caching is disable by default.  Every
experiment is run five times.  We report the average of the five runs.

\Removed{
Recall that in Section~\ref{sec:threshold} we compare time complexity with the sophisticated theoretical work~\cite{matrix1987} on MM, whose time complexity is $\mathcal{O}(n^{2.376})$. However, as we measured, the dense MM in MKL, as well as other packages like BLAS (Basic Linear Algebra Subprograms), is $\mathcal{O}(n^{3})$. This is because sub-cubic MM algorithms are not yet practical due to huge constant factors~\cite{LeGall12,Pegoraro20}. In other words, MM implemented by MKL, where many sophisticated optimizations like SIMD are made, practically runs faster than the sub-cubic algorithms in theoretical works. Therefore, the comparison is fair, because both in theory and practice, we compare DIM$^3$ with the better one.
}

\subsection{Evaluation for Join-Project Operations}
\label{subsec:result-jp}

\begin{figure}[t]

    \subfigcapskip=-2pt
    \subfigbottomskip=-2pt
    \subfigure[\revise{Comparison with stand-alone Join-Project algorithms.}]{
      \includegraphics[width=0.94\linewidth]{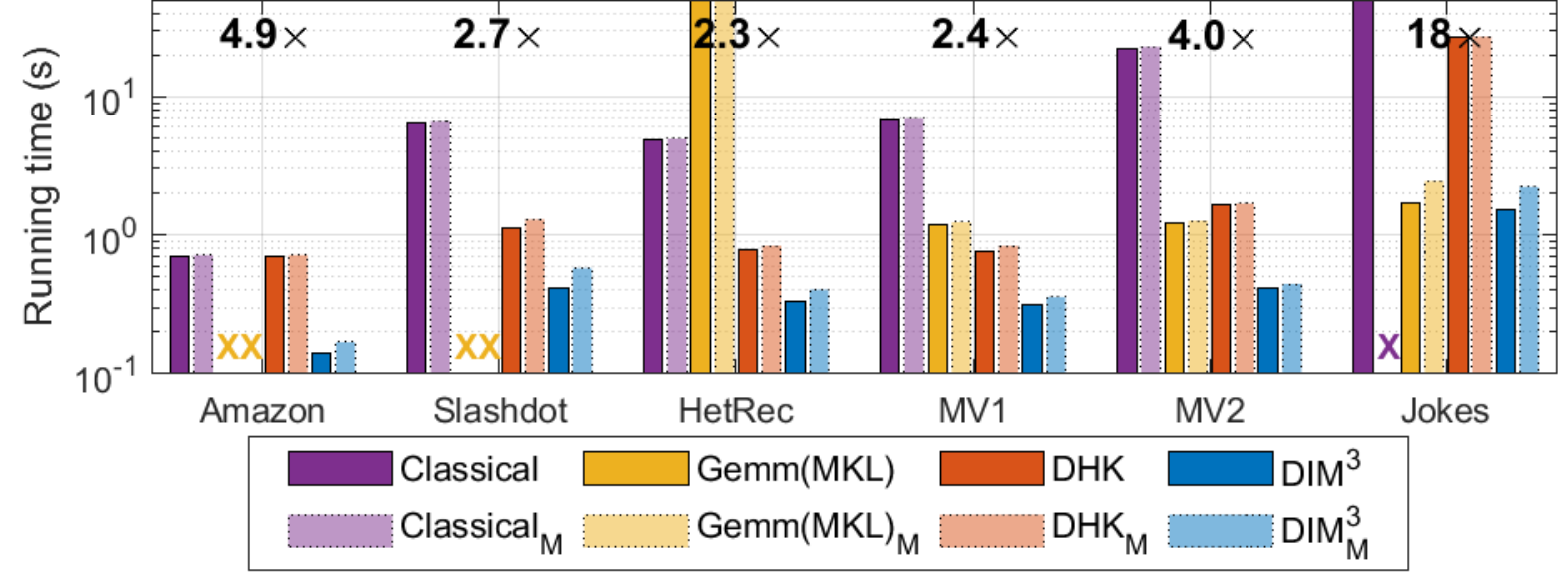}
      \label{fig:jp2}
    }
    \subfigure[\revise{Comparison with RDBMSs.}]{
      \includegraphics[width=0.94\linewidth]{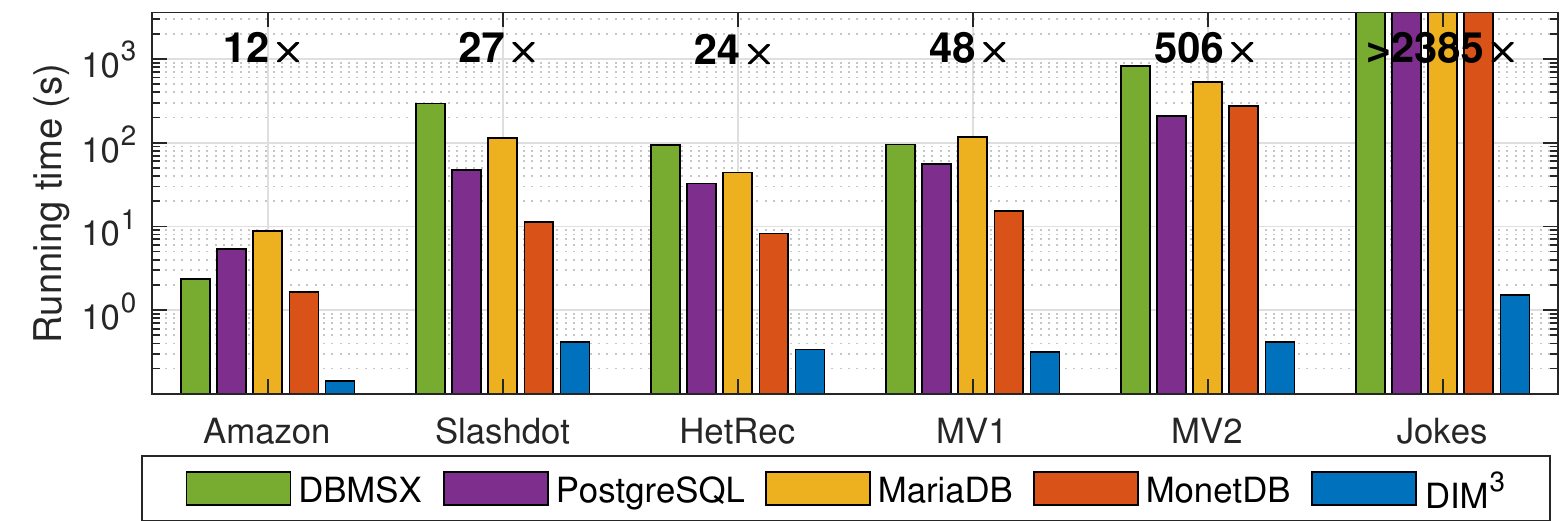}
      \label{fig:jp1}
    }

    \vspace{-0.1in}

    \caption{ \revise{Join-Project on real-world datasets. (By
default, the algorithms count then discard the result tuples.
Subscript 'M' denotes a version of an algorithm that materializes the
result tuples in memory.  'X' shows when MKL/Classical runs out of
memory and fails. We label speedups of DIM$^3$ over DHK in (a), and
over the best RDBMS solution in (b).)}}

    \label{fig:jp}
    \vspace{-0.25in}
\end{figure}

\Paragraph{Performance of Stand-alone Solutions on Real-World Data
Sets} Figure~\ref{fig:jp}(a) compares DIM$^3$ with stand-alone
Join-Project algorithms on real-world data sets.
%
%
\revise{ We see that DIM$^3$ achieves the best performance among all
stand-alone algorithms with or without result materialization.
Compared to \emph{DHK}, the state-of-the-art algorithm, DIM$^3$
achieves 2.3$\times$-18$\times$ improvements.  For all the six data
sets, DIM$^3$ chooses the hybrid strategy. In contrast, for Amazon,
DHK chooses the classical solution. The speedup of DIM$^3$ over DHK
for Amazon shows that our strategy selection function $f_1$ is more
accurate than the rule-of-thumb condition of DHK.  }

Moreover, DIM$^3$ has smallest memory footprints among all algorithms,
as shown in Table~\ref{tab:mem}.  We use '\emph{/usr/bin/time -v}' to
measure the peak memory usage (i.e., maximum resident set size).
Typically, the final Join-Project results will be consumed by
upper-level operators, and therefore we do not allocate space for
storing the final results.
Compared to \emph{Classical}, when the data sets are dense, DIM$^3$
saves \revise{80}\%--99\% memory because it chooses the hybrid strategy and
saves the space required by hash-based deduplication in
\emph{Classical}.
The memory usage of \emph{Gemm(MKL)} is mainly determined by the
matrix sizes.  MKL runs out of memory for the two sparsest data sets
(i.e., \revise{Amazon} and Slashdot).  When the datasets are dense, DIM$^3$
saves 68\%--99\% memory of \emph{Gemm(MKL)} because DenseEC saves the
space of the output matrix, and uses one bit per element rather than
4-byte integers in MKL.
As for \emph{DHK}, one main source of its memory usage is the input
and output matrices for dense MM invocation.  In comparison, DenseEC
significantly saves this memory.



\Paragraph{Comparison with RDBMSs on Real-World Data Sets}
Figure~\ref{fig:jp}(b) compares DIM$^3$ with RDBMSs.  We see that
DIM$^3$ outperforms all the RDBMSs.  For data sets with
$|OUT_J|\gg|R|+|S|$, DIM$^3$ achieves one to three orders of
magnitudes of speedups.

We examine the query plans generated by the RDBMSs and see that they
all essentially employ the \emph{Classical} solution, i.e., a join
followed by a deduplication of the intermediate results.
Among the RDBMSs, MonetDB is the best when memory is sufficient.
However, its performance drops sharply when the required space exceeds
the memory size (for MV2 and Jokes).
In the case of Jokes, all the RDBMSs fail to perform the deduplication
operation entirely in memory.  The queries take much longer time
because of the disk I/Os for storing and accessing the intermediate
results.  We stop the queries when they took longer than 1 hour.  The
figure reports the lower bound (i.e., 3600s) for RDBMSs on the Jokes
data set.  

As RDBMSs implement the classical solution, we focus on the comparison
with stand-alone solutions in the rest of this subsection.

\begin{table}[t]
    \centering
    \small
    \addtolength{\tabcolsep}{-2pt}

    \caption{Memory usage of stand-alone Join-Project algorithms.}
    \label{tab:mem}
    \vspace{-0.15in}

    \begin{tabular}{lcccccc}
      \hline
                & \revise{Amazon}   & Slashdot & HetRec & MV1   & MV2   & Jokes \\
      \hline
      DIM$^3$   & \revise{86MB}   & 64MB     & 32MB   & 39MB  & 72MB  & 47MB \\
      DHK       & \revise{426MB}   & 70MB     & 168MB  & 68MB  & 132MB & 2.4GB \\
      Gemm(MKL) & \revise{$>$32GB} & $>$32GB  & 7.1GB  & 211MB & 227MB & 2.4GB \\
      Classical & \revise{426MB}   & 3.0GB    & 1.5GB  & 789MB & 1.5GB & 24GB \\
      \hline
    \end{tabular}
    \vspace{-0.20in}
\end{table}

\begin{figure}[t]
    \centering
    \subfigcapskip=-2pt
    \subfigbottomskip=-2pt
    \subfigure[Performance with 8 threads.]{
      \includegraphics[width=0.95\linewidth]{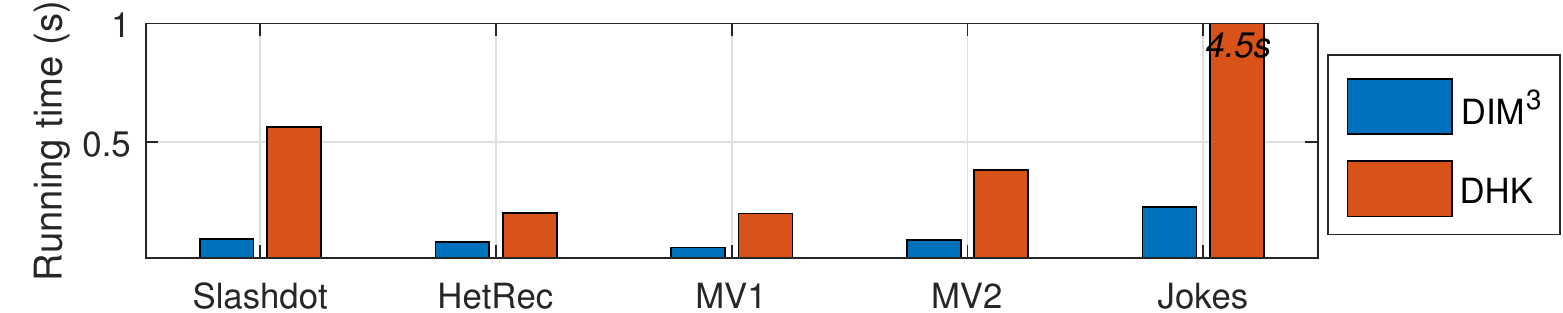}
    }
    \subfigure[Scalability on the HetRec data set.]{
      \includegraphics[width=0.44\linewidth]{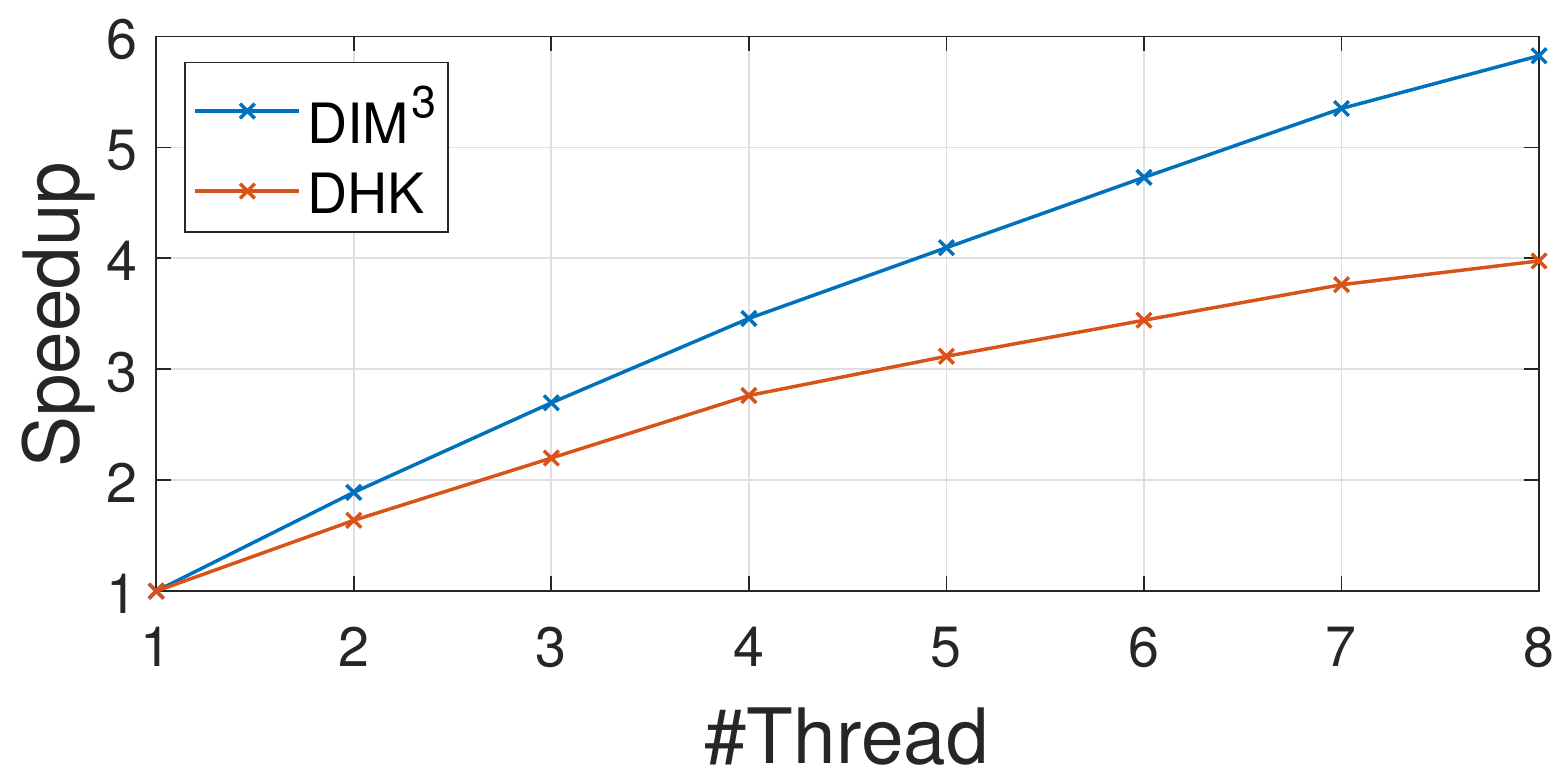}
    }
    \subfigure[Scalability on the MV1 data set.]{
      \includegraphics[width=0.44\linewidth]{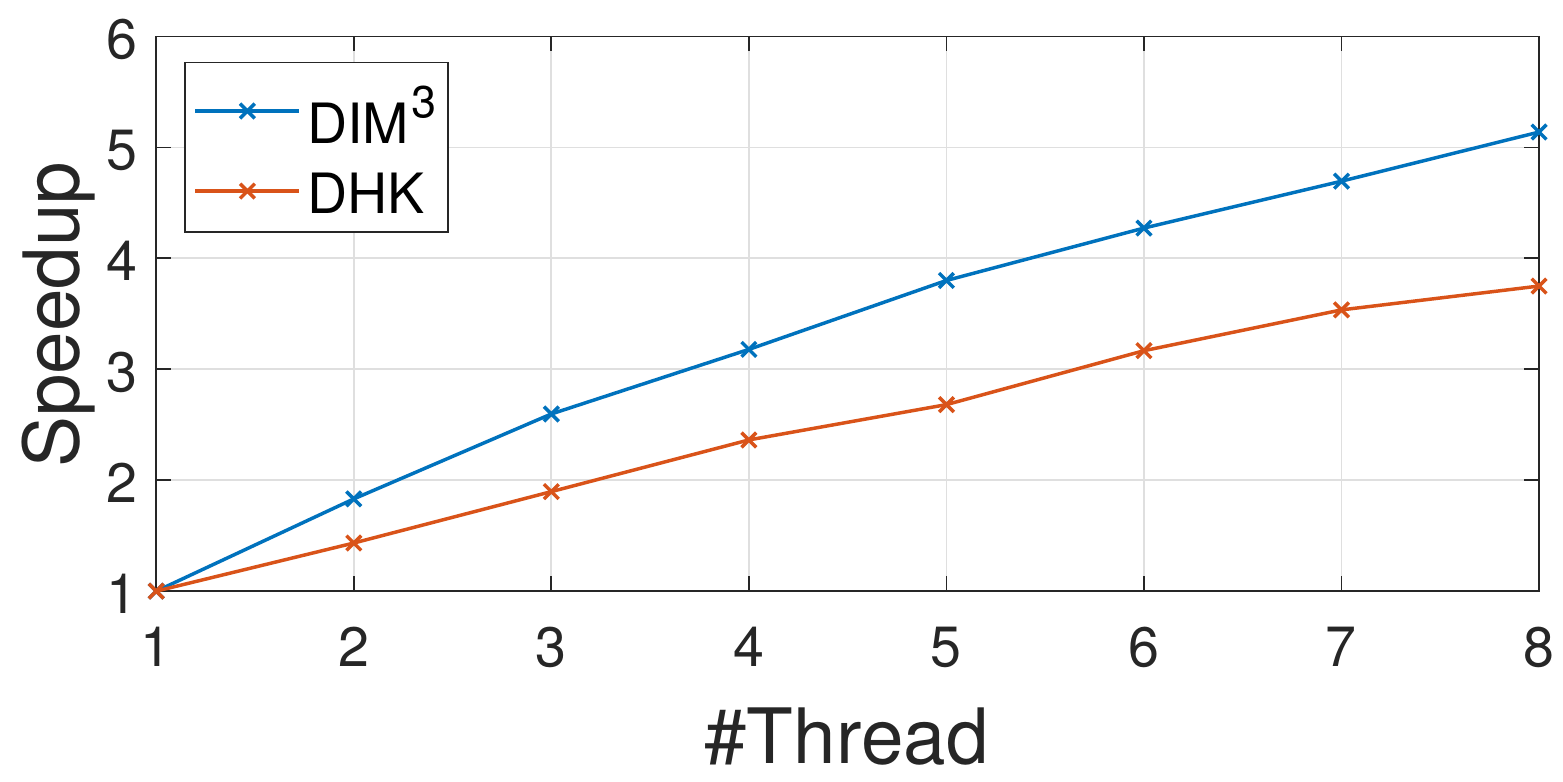}
    }

    \vspace{-0.08in}
    \caption{Multi-threaded performance.}
    \label{fig:jpparallel}
    \vspace{-0.21in}

\end{figure}

\Paragraph{Scalability} We implement the multi-threaded DIM$^3$ with
OpenMP by modifying around 20 lines of code. 
Specifically, we parallelize the outer \emph{for}-loop in SparseBMM
and DenseEC.  Since every outer loop iteration focuses on a specific
$x$, there is no dependence or contention among the execution of
different outer-loop iterations.
%
%
%
While more involved changes may better parallelize the algorithm, we
find that such simple modifications can already achieve promising
results.

We compare DIM$^3$ with \emph{DHK} using 8 threads in
Figure~\ref{fig:jpparallel}(a).  We see that DIM$^3$ achieves
2.6$\times$-20$\times$ speedups over DHK.
Figure~\ref{fig:jpparallel}(b)--(c) show the scalability of the two
algorithms.  The Y-axis reports the speedup compared to the
single-threaded execution of the same algorithm.  We see that DIM$^3$
shows good scalability as the number of threads increases from 1 to 8.

\begin{figure}[t]

  \centering

  \subfigcapskip=-2pt
  \subfigbottomskip=-2pt
  \subfigure[\revise{Varying $n$ ($|X|=|Y|=|Z|=n$).}]{
    \includegraphics[width=0.46\linewidth]{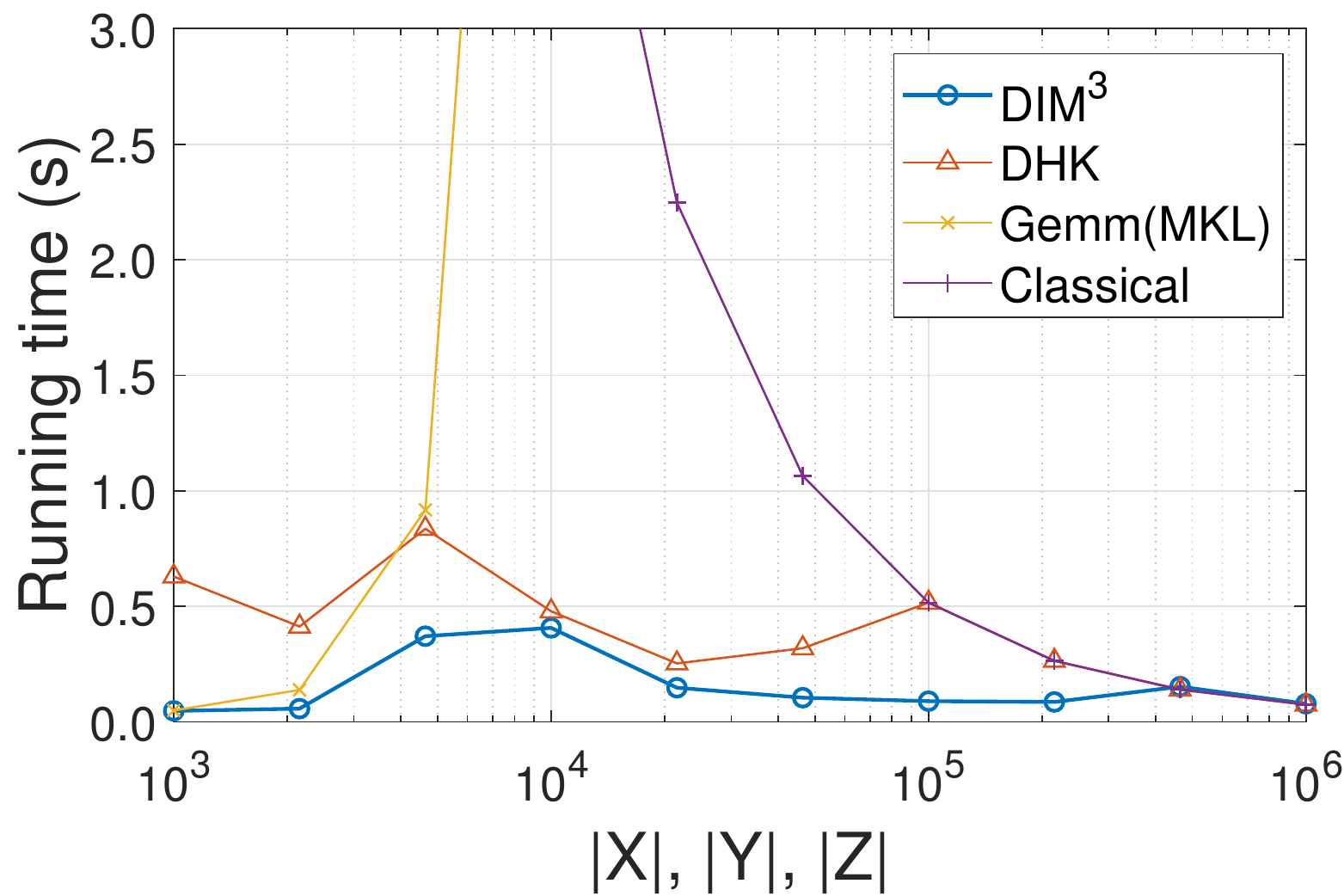}
    \label{fig:xyz}
  }
  \subfigure[Zipf on $R.x$ and $S.z$.]{
    \includegraphics[width=0.46\linewidth]{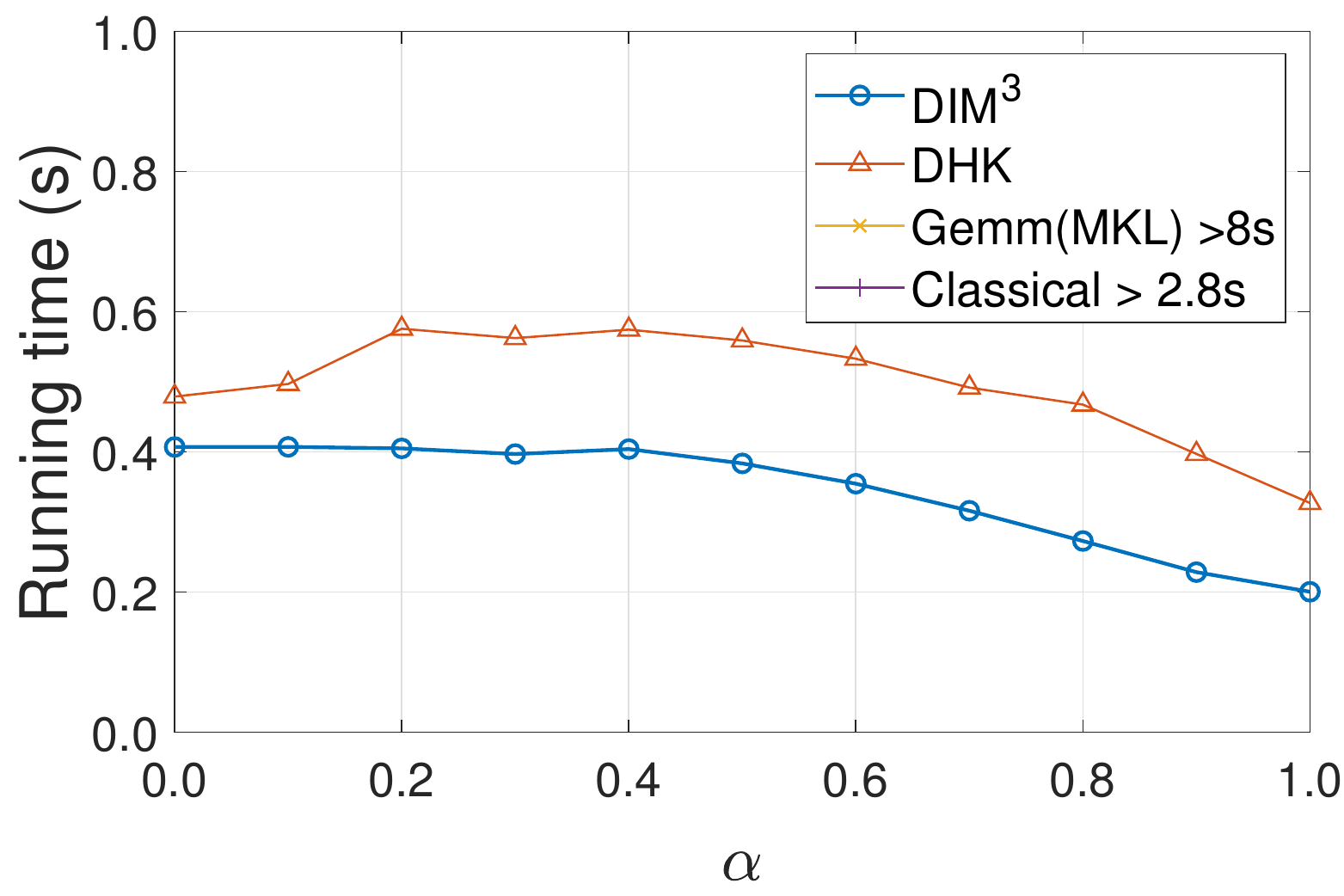}
    \label{fig:zipf1}
  }
  \subfigure[Zipf on $R.y$ and $S.y$.]{
    \includegraphics[width=0.46\linewidth]{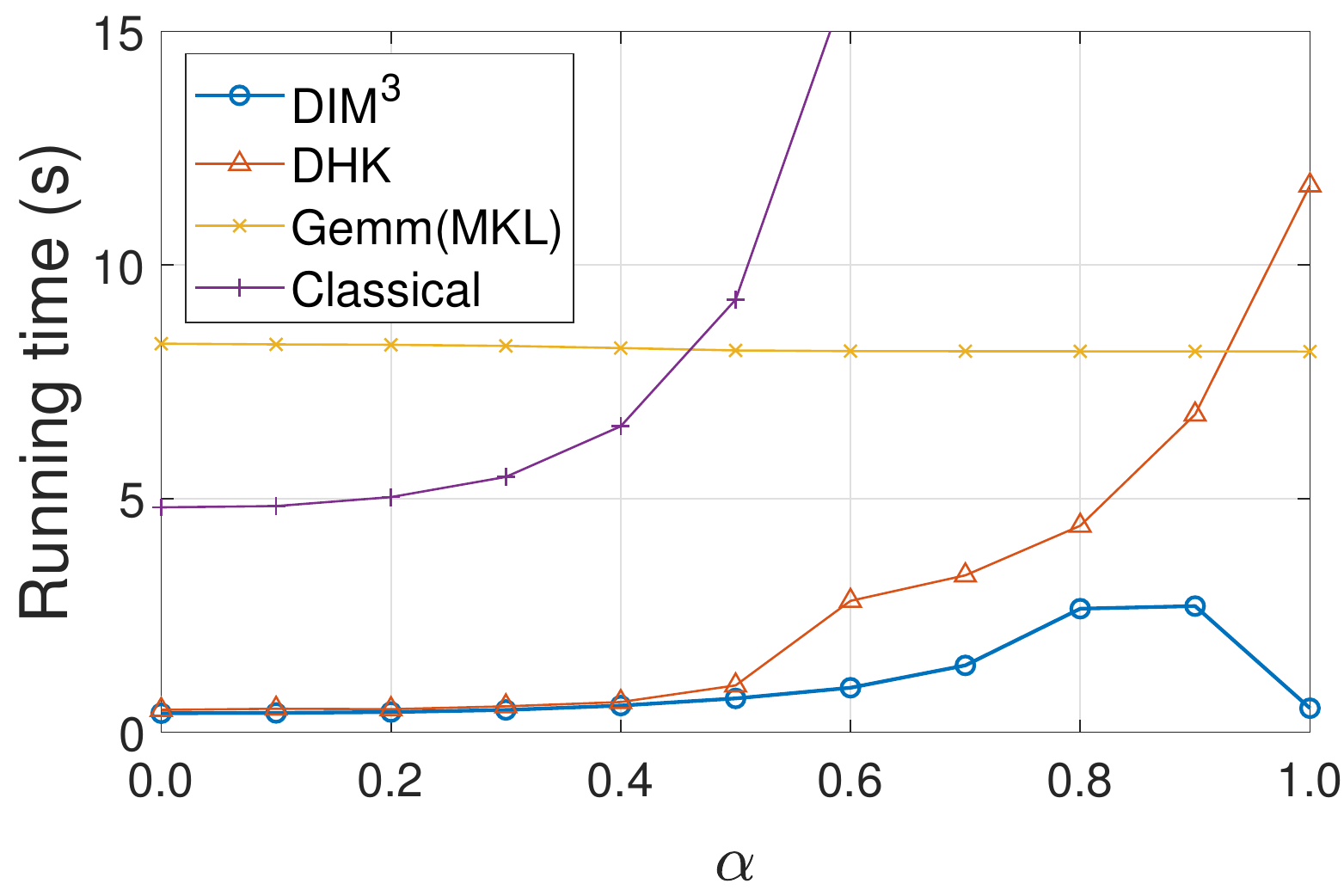}
    \label{fig:zipf2}
  }
  \subfigure[R-mat.]{
    \includegraphics[width=0.46\linewidth]{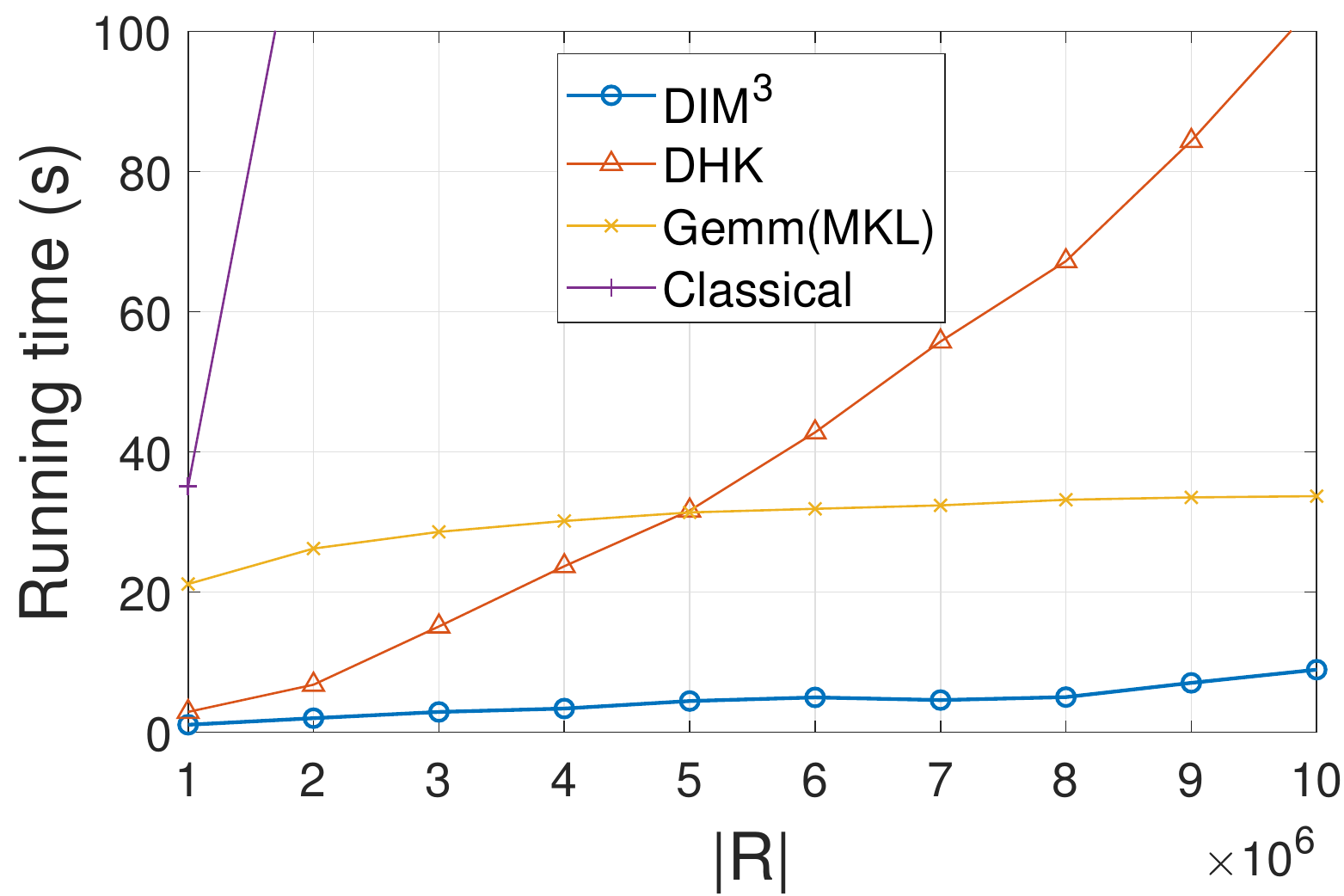}
    \label{fig:rmat}
  }
  \subfigure[\revise{Varying selectivity of $R.x$ and $S.z$.}]{
    \includegraphics[width=0.46\linewidth]{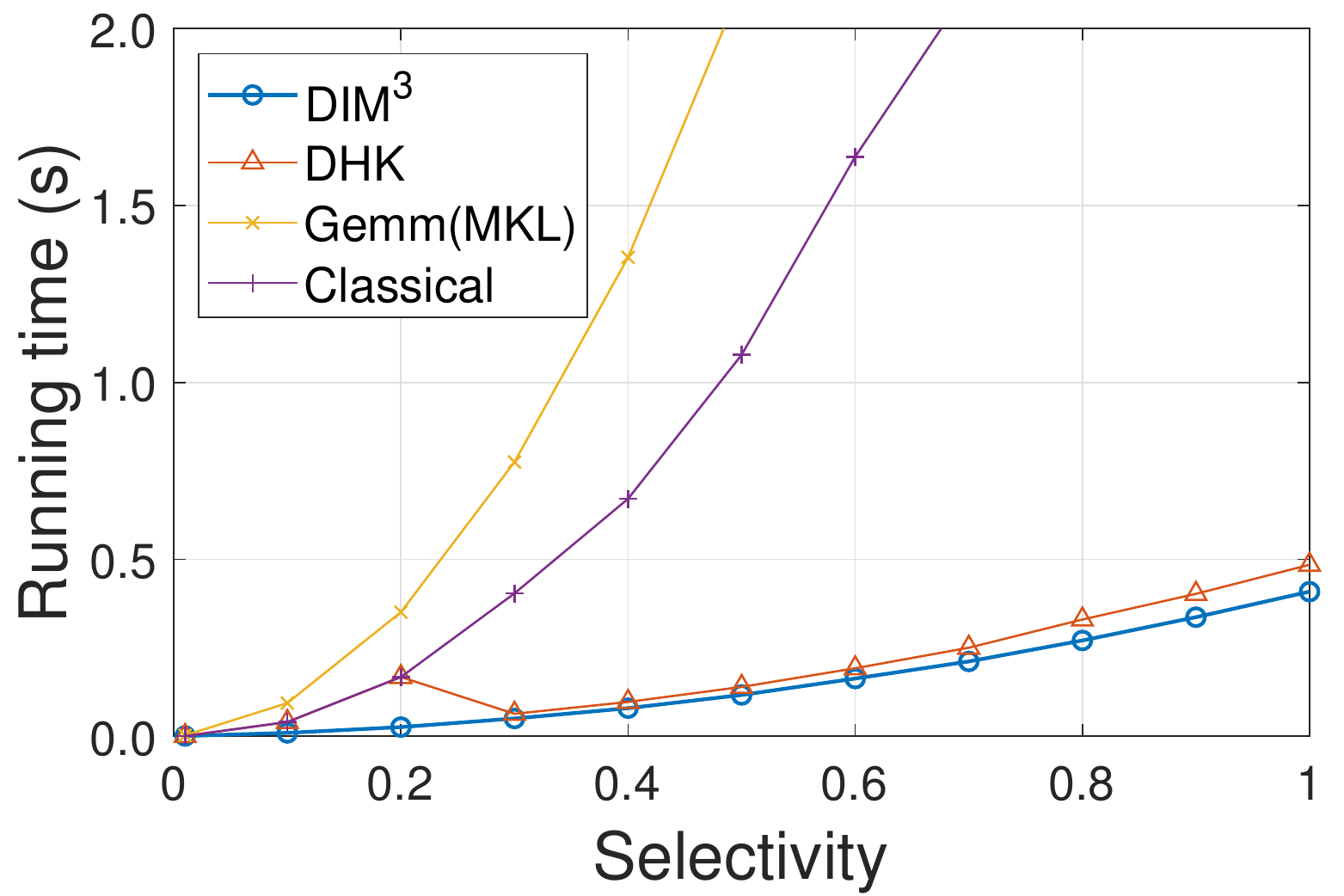}
    \label{fig:selectivity}
  }
  \subfigure[\revise{Zoom in to bottom left of (e).}]{
    \includegraphics[width=0.46\linewidth]{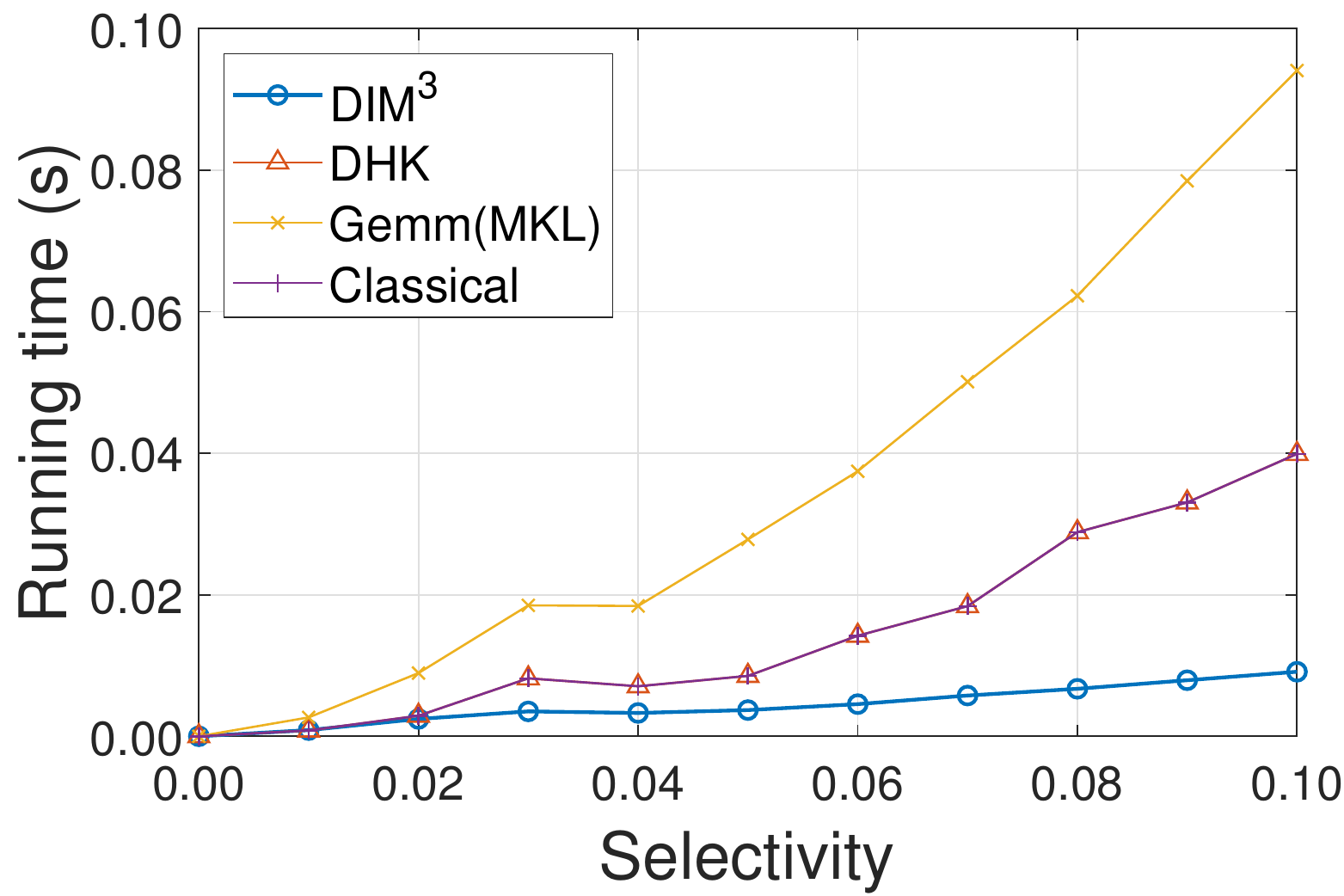}
    \label{fig:selectivity2}
  }

  \vspace{-0.08in}
  \caption{Sensitivity analysis of DIM$^3$.}
  \label{fig:analysis}
  \vspace{-0.2in}

\end{figure}

\Paragraph{Sensitivity Analysis with Synthetic Data Sets} We use
synthetic data sets to investigate the Join-Project performance for a
wide range of situations.  The default parameters in these experiments
are as follows.  All input columns are generated with uniform
distributions. $|R|=|S|=10^6$, and $|X|=|Y|=|Z|=n=10^4$.
We study the performance impact varying $n$, data skews, and
selectivity in Figure~\ref{fig:analysis}.  Overall, we see that
DIM$^3$ performs the best in all the cases.

\emph{(1) Varying $n$}.
Figure~\ref{fig:xyz} varies $n$, the number of distinct $x$/$y$/$z$,
while keeping the number of non-zero elements (i.e. $|R|$ and $|S|$)
fixed.  As $n$ increases, the input data sets become more and more
sparse, and $|OUT_J|$ decreases from $10^9$ to $10^6$. 
%
%
\emph{Classical} runs faster for more sparse data sets.
\emph{Gemm(MKL)} sees a sharp increase of run time as the matrix
dimensions (i.e., $n$) increase.
DIM$^3$ wisely selects evaluation strategies based on data density.
It employs DenseEC for all data when $n$<$10^4$, SparseBMM when $n \in
[10^4, 4\times 10^5]$, and the classical solution when $n$$>$$4\times
10^5$.  In this way, DIM$^3$ achieves the best performance in all
cases.

\revise{\emph{DHK} has good performance in the middle range in
Figure~\ref{fig:xyz} and switches to Classical in the higher range.
However, its performance is worse than \emph{Gemm(MKL)} in the lower
range.} To understand the reason, we look into the \emph{DHK} code and
find that \emph{DHK} restricts the degree thresholds (i.e., $\Delta_1$
and $\Delta_2$~\cite{SIGMOD20}) to have
$\Delta_2=\tfrac{|R|\Delta_1}{|OUT_P|}$ in order to reduce the
threshold search space.  Unfortunately, this narrows the range of
considered thresholds.  For the lower range when the data is very
dense, \emph{DHK} fails to put all data into the dense part, and thus
performs worse than \emph{Gemm(MKL)}.  In comparison, DIM$^3$ does not
suffer from this problems. DIM$^3$ makes the partition decision based
on $z$, which is simple and less error-prone to compute.


\emph{(2) Varying data skews}.
In Figure~\ref{fig:zipf1}, we generate $R.x$ and $S.z$ using the Zipf
distribution, and vary the parameter $\alpha$ from 0 to 1.  As $R.x$
and $S.z$ become more skewed, $|OUT_J|$ stays around $10^8$, while the
deduplicated result size $|OUT_P|$ decreases from $6\times10^7$ to
$1.4\times10^7$.
DIM$^3$ partitions the more skewed $z$'s into the dense part as
$\alpha$ increases, and achieves a speedup of $\sim$1.5$\times$
compared to \emph{DHK}.

%

In Figure~\ref{fig:zipf2}, we generate the join attribute, $R.y$ and
$S.y$, using Zipf distribution, and vary $\alpha$ from 0 to 1.  As
$R.y$ and $S.y$ become more skewed, the number of intermediate join
results ($|OUT_J|$) increases sharply from $10^8$ to $4.3\times10^9$,
and the data sets become increasingly dense.
%
%
\emph{Classical} runs slower as the data sets become denser.
\emph{Gemm(MKL)} sees a flat curve because the matrix dimensions are
the same.
\emph{DHK} suffers from the threshold search space problem for very
dense data sets.
DIM$^3$ achieves the best performance in all cases.  There is a dip in
the DIM$^3$ curve at $\alpha$=1 because early stopping effectively
reduces DenseEC computation in this case.


In Figure~\ref{fig:rmat}, we use R-mat~\cite{rmat} to generate skew
data so that the two columns in a table are correlated.  We use the
R-mat parameters in Graph500~\cite{graph500} (i.e., $a=0.57$,
$b=0.19$, $c=0.19$, and $d=0.05$).   We generate a directed graph with
$n$ vertices and $|R|$ edges.  The Join-Project finds 2-hop paths in
the graph.  As R-mat requires $n$ to be a power of 2, we set
$n=2^{14}$, which is close to the default $10^4$.  We vary $|R|$ from
$10^6$ to $10^7$.  From Figure~\ref{fig:rmat}, we see that DIM$^3$
performs significantly better than the other algorithms.  As $|R|$
increases, the data set becomes increasingly dense.  \emph{Classical},
\emph{Gemm(MKL)}, and \emph{DHK} curves show similar trends as in
Figure~\ref{fig:zipf2}.

%
%
%

\emph{(3) Varying selectivity}.
In Figure~\ref{fig:selectivity}, we consider the cases where there are
filtering predicates on both $R.x$ and $S.z$.  We vary their
selectivity at the same time from 0 to 1.
Figure~\ref{fig:selectivity2} zooms in to the bottom left of the
Figure~\ref{fig:selectivity}.
As the selectivity decreases, both the input sizes and the output size
decrease.  Hence, all the algorithms run faster.  From the figure, we
see that DIM$^3$ performs the best in all cases.   Note that when the
selectivity is very small (less than 0.03), DIM$^3$ switches to the
classical solution. 
\revise{When the selectivity is less or equal than 0.2, \emph{DHK} switches to the classical solution.}


\begin{figure}[t]

  \subfigcapskip=-2pt
  \subfigbottomskip=-2pt
  \subfigure[Mapping step varying $n$.]{
    \includegraphics[width=0.46\linewidth]{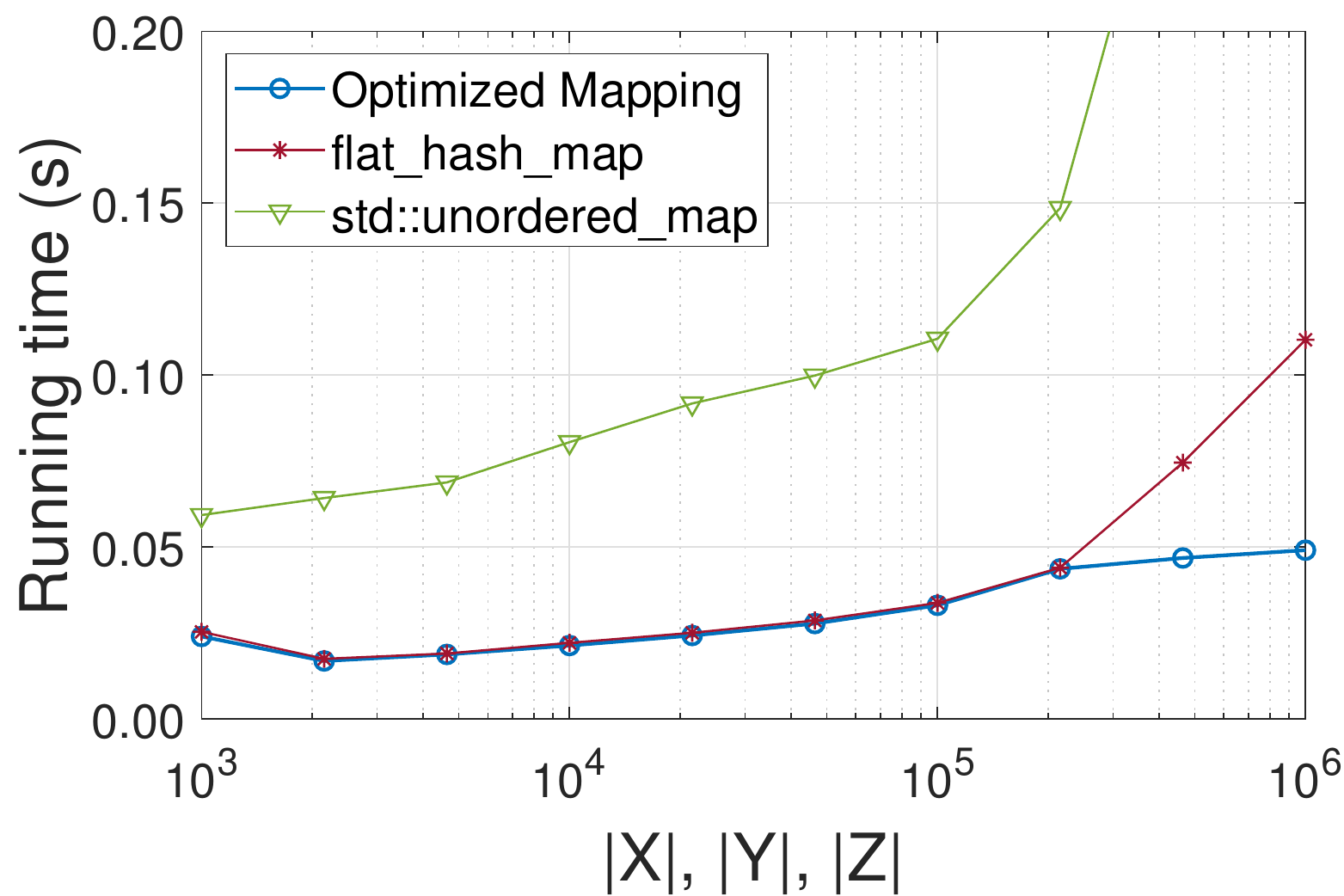}
    \label{fig:map1}
  }
  \subfigure[Mapping step varying $|R|=|S|$.]{
    \includegraphics[width=0.46\linewidth]{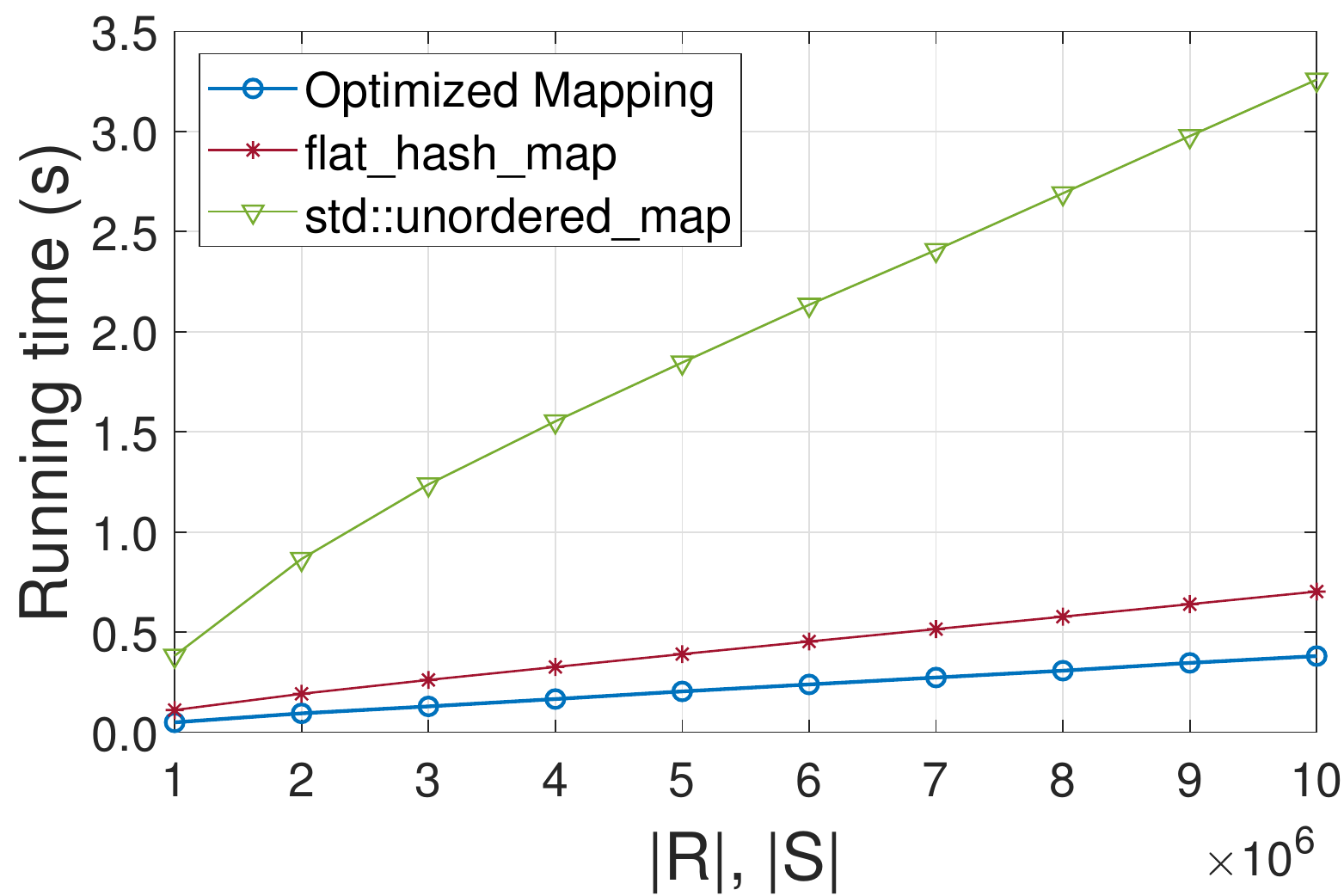}
    \label{fig:map2}
  }
  \subfigure[DenseEC varying $n$.]{
    \includegraphics[width=0.46\linewidth]{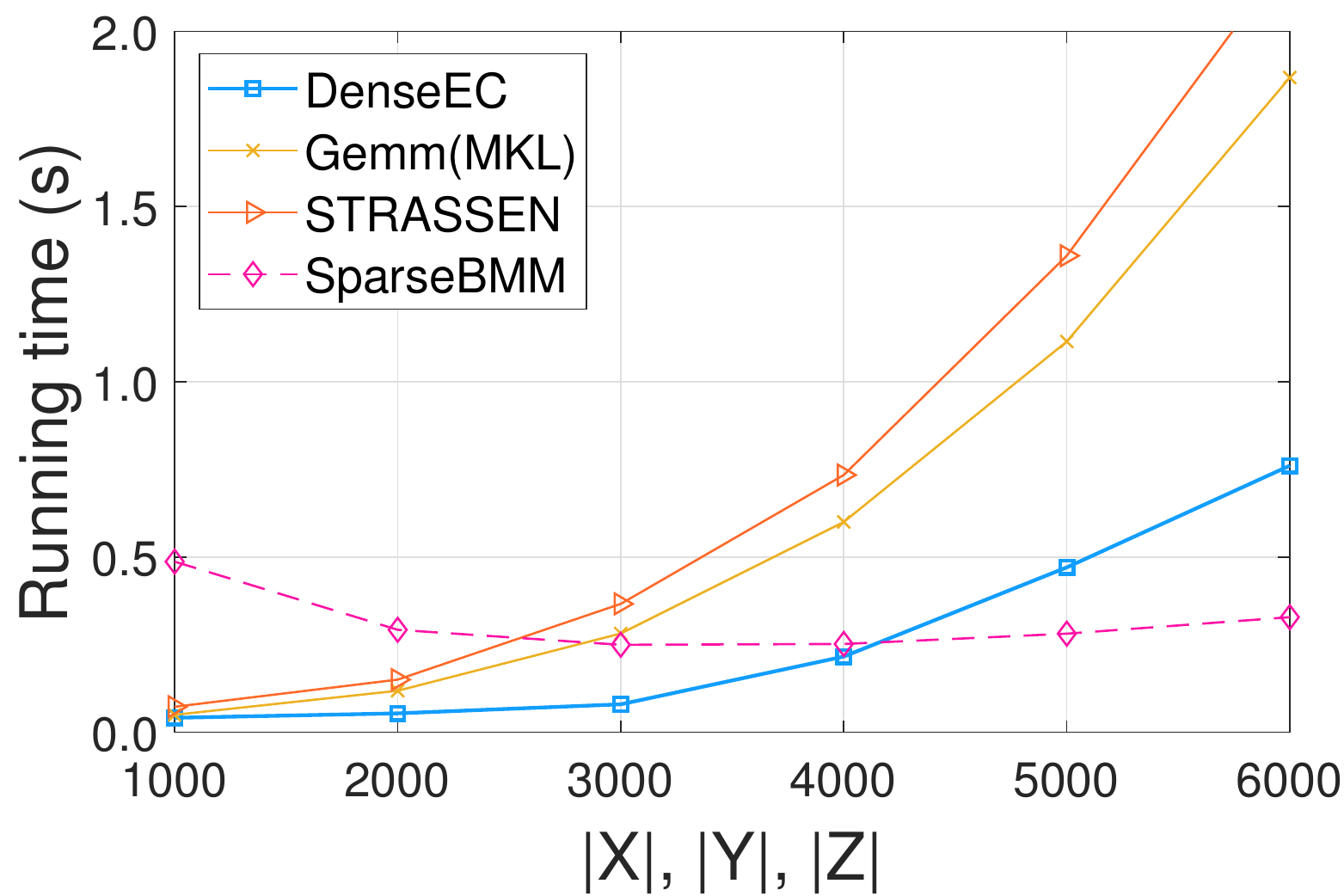}
    \label{fig:denseec}
  }
  \subfigure[SparseBMM varying $n$.]{
    \includegraphics[width=0.46\linewidth]{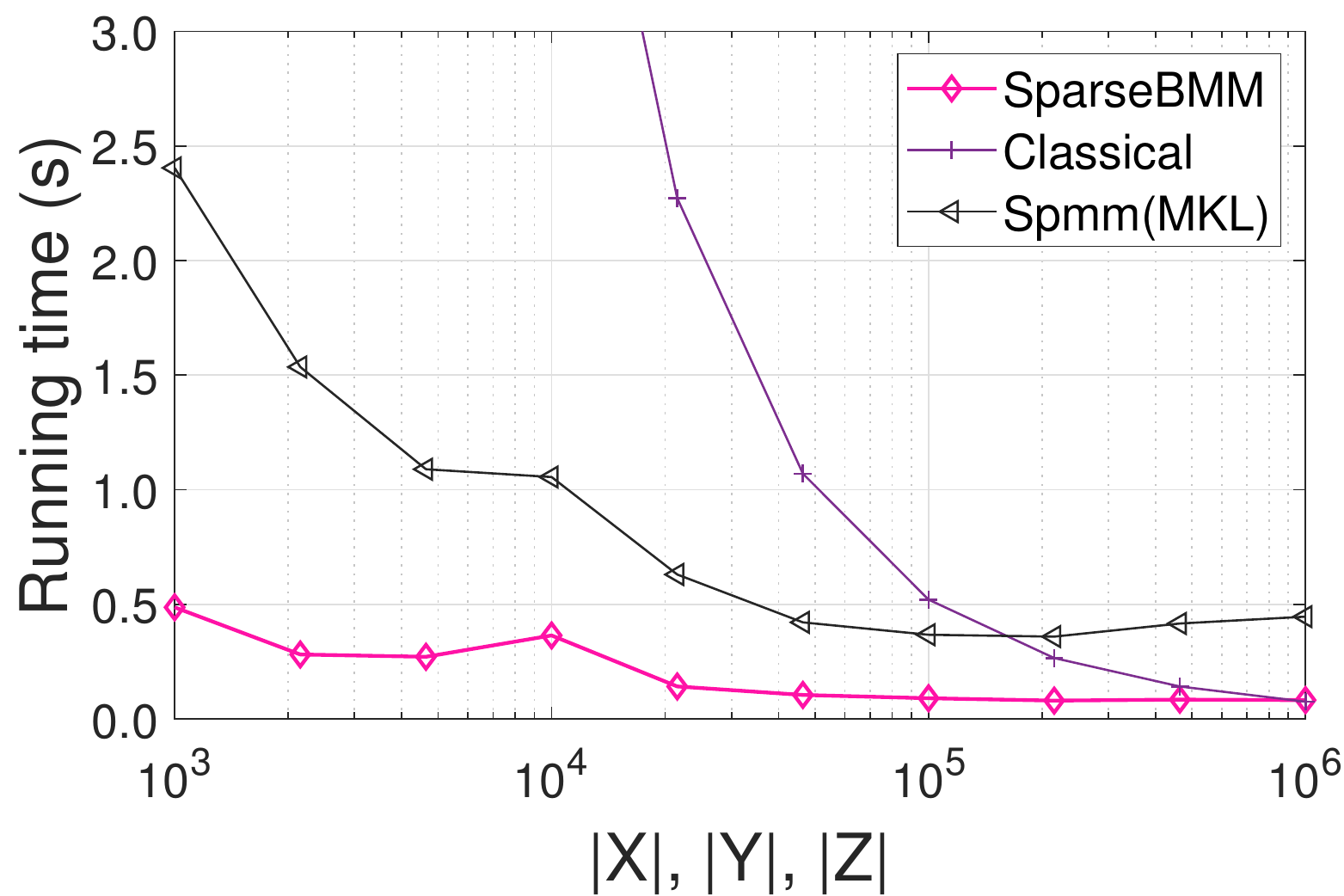}
    \label{fig:sparsebmm}
  }

  \vspace{-0.08in}
  \caption{Effectiveness of DIM$^3$ Components.}
  \label{fig:components}
  \vspace{-0.21in}

\end{figure}

\Paragraph{Effectiveness of DIM$^3$ Components} In the following, we
use synthetic data sets with the same default parameters as the above.

\emph{(1) Mapping step}.  We compare our optimized mapping solution
with the baseline using std::unordered\_map, and an improved baseline
using flat\_hash\_map. 
In Figure~\ref{fig:map1}, we fix $|R|$=$|S|$=$10^6$ while varying $n$
from $10^3$ to $10^6$.   The hash table size increases as $n$.  When
$n<4\times 10^5$, the hash table fits into the L3 cache, and thus we
do not perform cache partitioning.  Our solution has similar
performance compared to flat\_hash\_map.
When $n \geq 4\times 10^5$, we use cache partitioning to reduce
expensive CPU cache misses, thereby significantly out-performing
flat\_hash\_map.
As for std::unordered\_map, it is much slower because of chained
hashing.

In Figure~\ref{fig:map2}, we fix $n$=$10^6$ while varying $|R|$=$|S|$
from $10^6$ to $10^7$.  The size of the hash table is fixed.  The
number of hash visits increases.  Hence, we see the increasing trends
for all curves.   Overall, our optimized mapping solution performs the
best.

\revise{Moreover, we compare DIM$^3$ (with optimized mapping) and
DIM$^3$ with baseline std::unordered\_map for the Amazon data set.
The mapping step takes 54.9\% of the Join-Project run time for DIM$^3$
with baseline mapping.  Replacing the baseline mapping with optimized
mapping, DIM$^3$ achieves an improvement of 1.5x.  }

\emph{(2) DenseEC}. In Figure~\ref{fig:denseec}, we vary $n$ from 1000
to 6000, where the data sets are relatively dense and thus dense MM
makes sense.  We compare DenseEC with dense MM implementations.

We investigate sub-cubic fast MM algorithms.  While the best known
$\mathcal{O}(n^{2.373})$ algorithm~\cite{matrix2018} is considered
impractical because of its huge constant
factors~\cite{LeGall12,Pegoraro20}, recent work aims to make the
Strassen algorithm practical~\cite{FMM}.  We obtain and evaluate this
\emph{STRASSEN} implementation~\cite{FMM}.  Our initial results show
that it is slower than Intel MKL.  One potential reason is that the
original \emph{STRASSEN} supports only floating point MM, while the
MKL run uses integer MM.  Since the \emph{STRASSEN} code calls BLAS as
the underlying MM sub-routine, it is straight-forward to modify the
code to call MKL's integer MM.  The resulting integer \emph{STRASSEN}
indeed runs slightly faster, but it is still slower than MKL.
Figure~\ref{fig:denseec} shows the performance of the improved integer
\emph{STRASSEN}.

Overall, DenseEC out-performs dense MM implementations because
early stopping can effectively reduce computation cost.

\emph{(3) SparseBMM}. In Figure~\ref{fig:sparsebmm}, we vary $n$ from
$10^3$ to $10^6$.  The generated data sets become more and more
sparse, and therefore sparse MM and \emph{Classical} are more
suitable.  We compare SparseBMM with \emph{Classical} and MKL's sparse
MM.  From the figure, we see that SparseBMM performs the best.

We also plot the SparseBMM curve in Figure~\ref{fig:denseec}.  The
crossing point of SparseBMM and DenseEC is around 4000.  This shows
why DIM$^3$ chooses between DenseEC and SparseBMM with function $f_2$.


\subsection{Evaluation for Partial Result Caching}
\label{subsec:result-cache}

\begin{figure}[t]

    \subfigcapskip=-2pt
    \subfigbottomskip=-2pt
    \subfigure[Run time on real-world data sets.]{
      \includegraphics[width=0.46\linewidth]{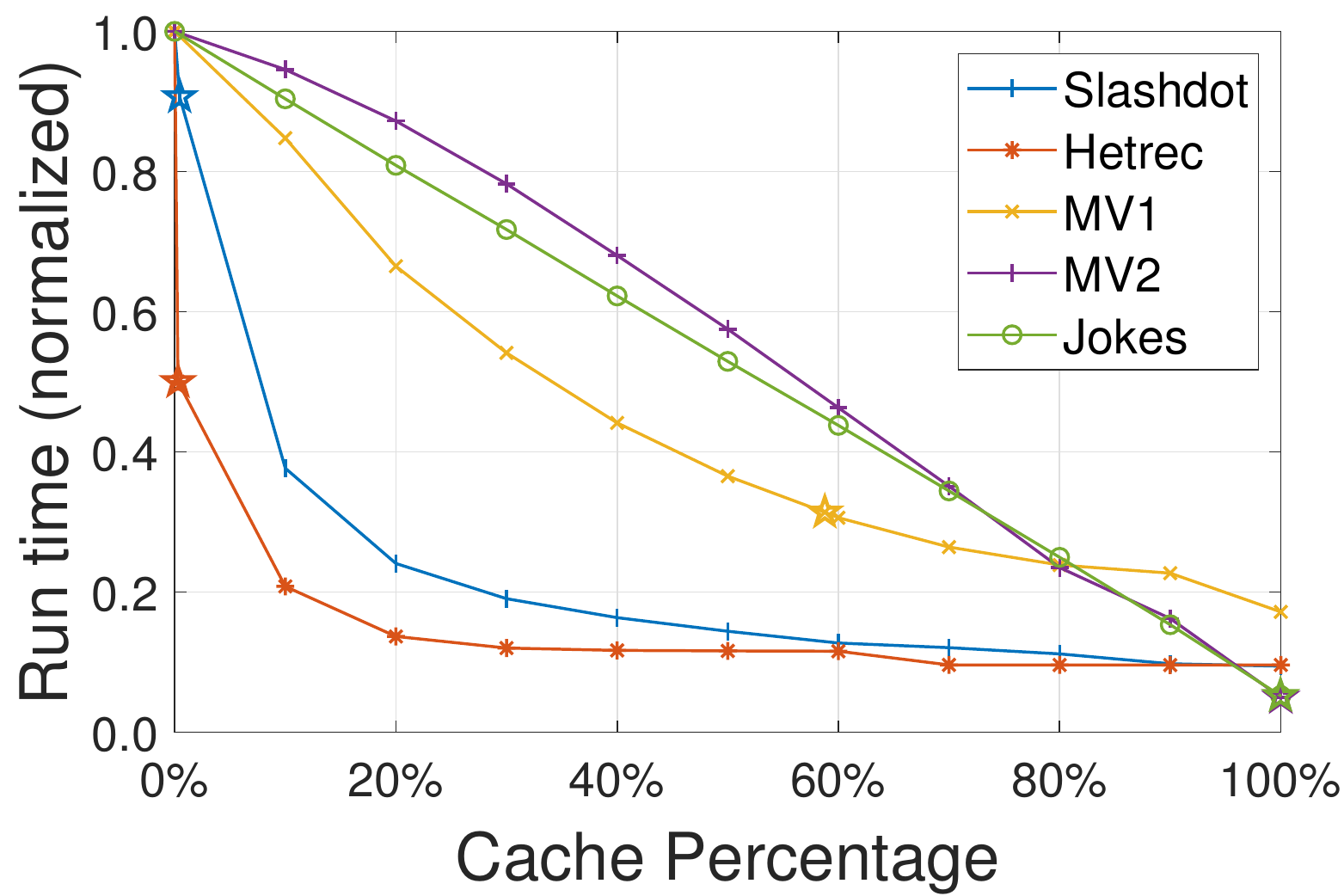}
      \label{fig:caching_rw1}
      }
    \subfigure[Space cost on real-world data sets.]{
      \includegraphics[width=0.46\linewidth]{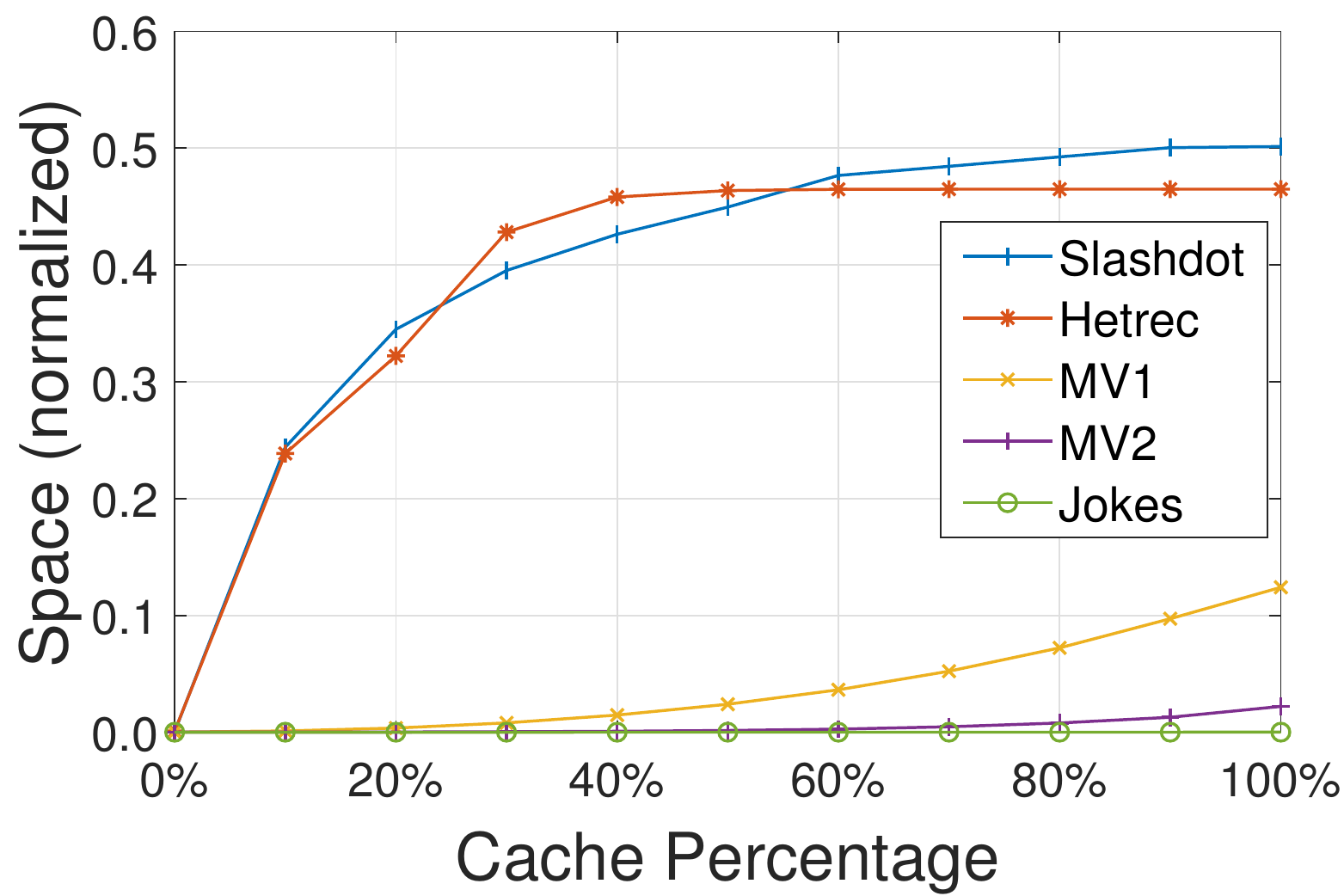}
      \label{fig:caching_rw2}
    }
    \subfigure[Run time on Zipf data.]{
      \includegraphics[width=0.46\linewidth]{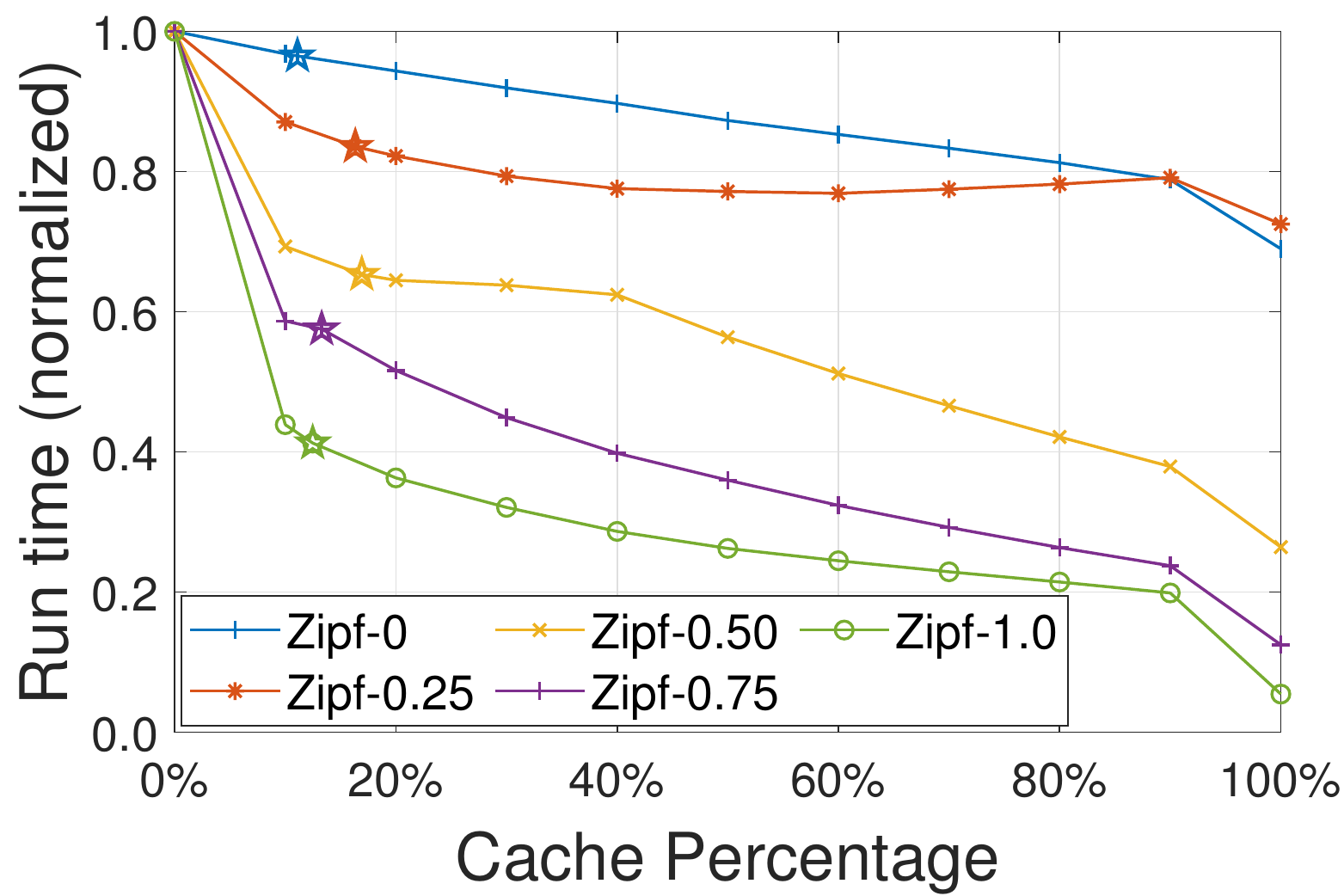}
      \label{fig:caching_zipf}
    }
    \subfigure[Run time on R-mat data.]{
      \includegraphics[width=0.46\linewidth]{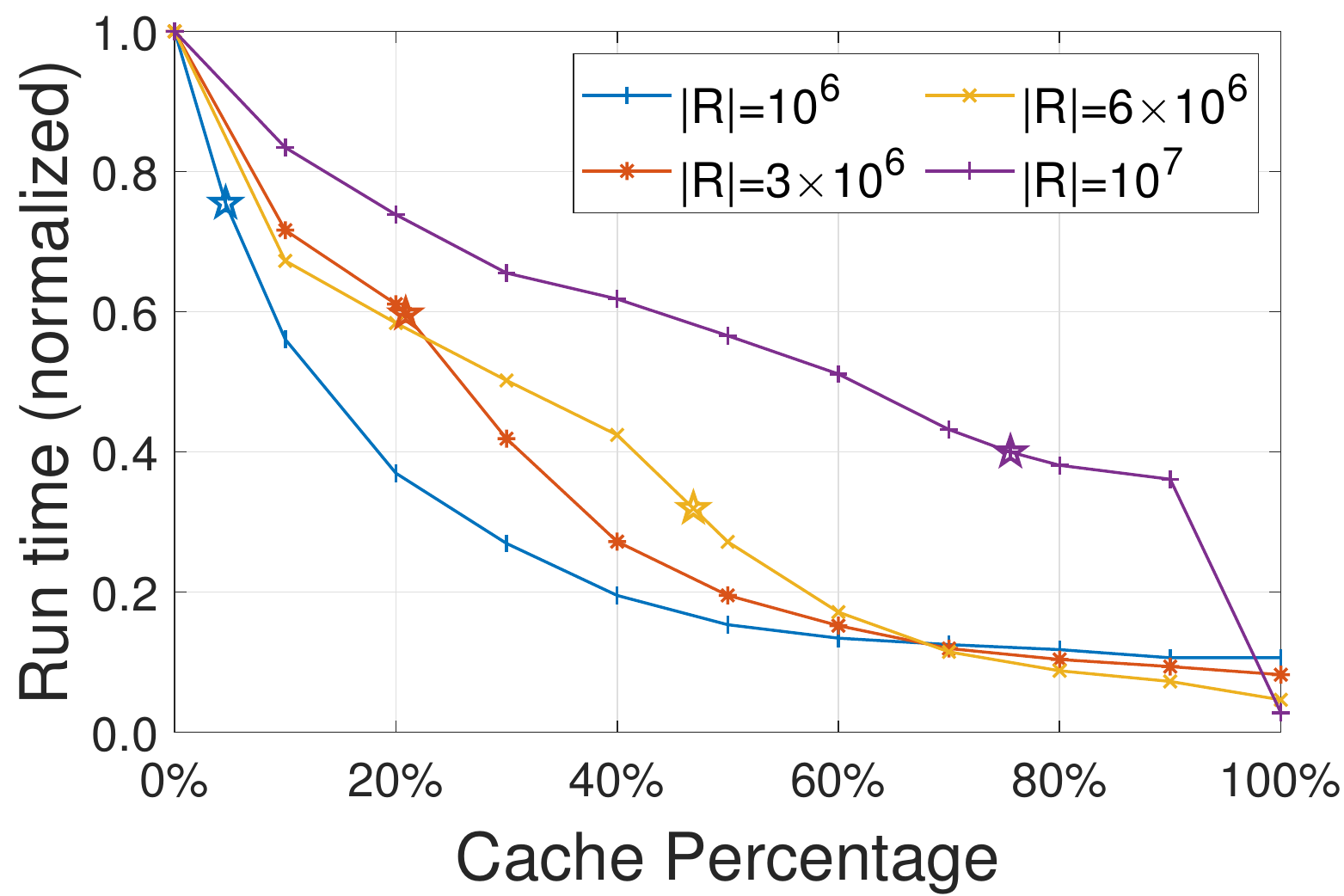}
      \label{fig:caching_rmat}
    }

    \vspace{-0.08in}
    \caption{Partial result caching. ($\star$ shows the point where
cached result size $=$ input data size.)}
    \label{fig:caching}
    \vspace{-0.2in}

\end{figure}

We evaluate partial result caching for DIM$^3$ using both real-world
and synthetic data sets, as shown in Figure~\ref{fig:caching}.  The
synthetic data sets use the same default parameters as in
Section~\ref{subsec:result-jp}.  Our solution retrieves cached partial
results and performs JoinProject between $R$ and $\{$\revise{$($}$z,y$\revise{$)$}$|$\revise{$($}$z,y$\revise{$)$}$\in S
\land z \notin Z_{cached}\}$.
The X-axis varies the percentage of $z$ values that are cached.  The
Y-axis reports either the run time normalized to DIM$^3$ without
caching, or the cached result size normalized to the size of the total
JoinProject results.


Figure~\ref{fig:caching_rw1} and~\ref{fig:caching_rw2} show the run
time and the cached result size on real-world data sets, respectively.
For Slashdot and HetRec, the computation time distributions for $z$
values are very skewed.   
\revise{The space cost is relatively high because most results are
cached by their original values.}
Caching a relatively small percentage (e.g., 20\%) of $z$ values can
significantly speedup DIM$^3$.
On the other hand, the denser data sets (i.e., MV1, MV2, and Jokes)
see less skewed computation time distributions for $z$ values.  As the
number of results for the same $z$ is often large, our technique to
cache the complement set of result vectors can drastically reduce the
cached result size.  For example, the space at 100\% (i.e., for
caching all results) is only 12\%, 6\%, and less than 1\% of the total
result sizes for MV1, MV2, and Jokes, respectively.  Thus, we can
use a small amount of cache space to achieve significant
speedups.  Overall, if we cache either 20\% of $z$ values or up to the
input data size, partial result caching achieves 3.3x--20x
improvements over DIM$^3$ without caching.


Figure~\ref{fig:caching_zipf} shows the run time on synthetic data
sets where $R.x$ and $S.z$ follow Zipf distribution.  When Zipf
parameter $\alpha=0$, the data is uniformly distributed, and the
benefits of caching is low.  As $\alpha$ increases from 0.25 to 1, the
computation time distribution for $z$ values becomes more skewed and
caching is more helpful.  Caching only 10\% of $z$ values reduces the
run time by 13\% to 57\%.

Figure~\ref{fig:caching_rmat} shows the run time on R-mat data varying
the input table size $|R|$ from $10^6$ to $10^7$.  If we limit the
cache space to the size of the input table, partial result caching
reduces the run time of DIM$^3$ by 25\%--67\%.


\nottechreport{\vspace{-3pt}}
\subsection{Evaluation for Join-OP Query Types}
\label{subsec:join-op-exp}

\begin{figure}[t]
    \centering

    \subfigcapskip=-2pt
    \subfigbottomskip=-2pt
    \subfigure[\revise{Join-Aggregate. ($_{x,z}\mathcal{G}_{avg(|R.v-S.v|)}(R(x,y,v) \Join_y S(z,y,v))$)}]{
      \includegraphics[width=0.95\linewidth]{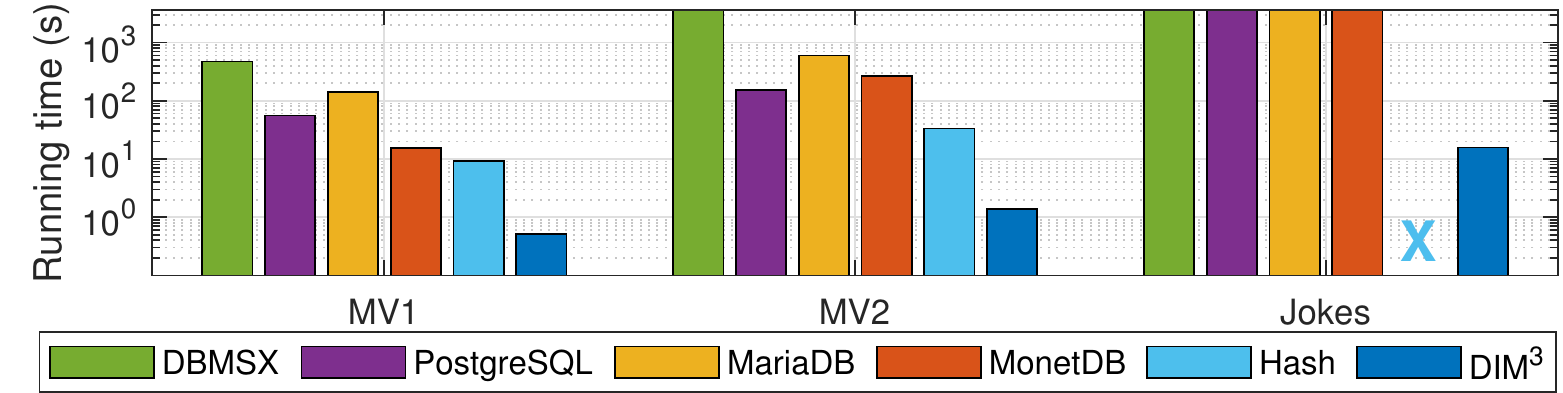}
      \label{fig:agg}
    }
    \subfigure[\revise{MJP.}]{
      \includegraphics[width=0.95\linewidth]{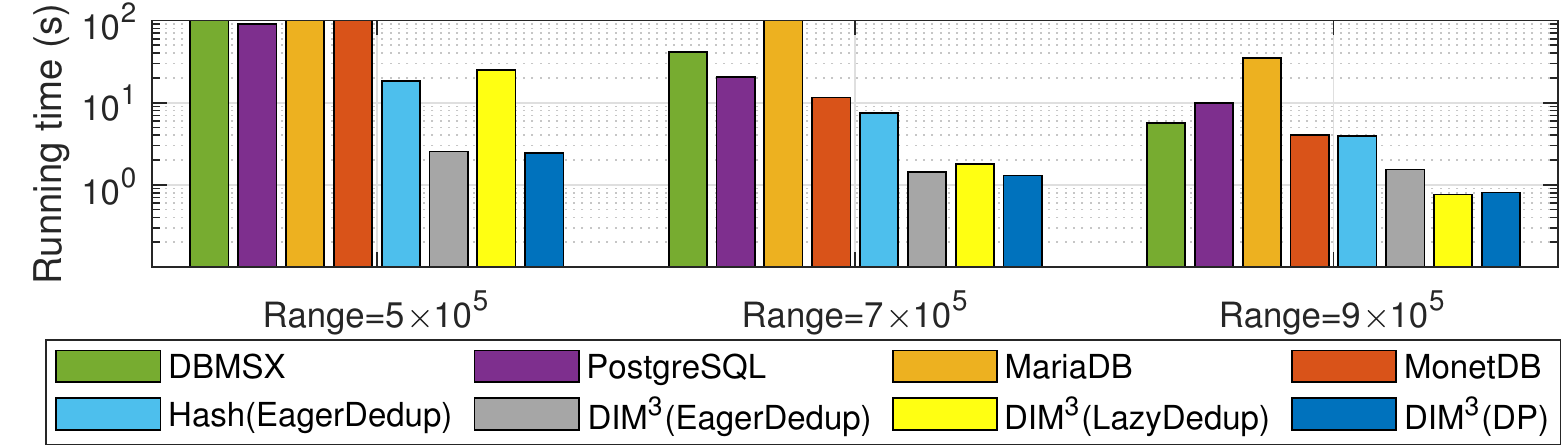}
      \label{fig:cq}
    }

    \vspace{-0.1in}
    \caption{\revise{Evaluation for Join-$op$ query
types.(\emph{Hash}:
stand-alone implementation using flat\_hash\_map for
join, aggregation, and deduplication. 'X': \emph{Hash} runs out of memory and fails.)}}

    \label{fig:result-joinop}
    \nottechreport{\vspace{-0.2in}}
    \techreport{\vspace{-0.1in}}

\end{figure}

\nottechreport{\vspace{-3pt}}
\Paragraph{Join-Aggregate} Figure~\ref{fig:agg} compares the Join-Aggregate
performance of DIM$^3$ with RDBMSs \revise{and the stand-alone hash-based
solution} on MV1, MV2, and Jokes.  We focus on the more complex case where the
group-by attributes are from both tables.  As described in
Section~\ref{sec:ja}, if the group-by attributes are from one table, we can
rewrite the query and employ the classical solution for efficient evaluation.
We compute aggregates on the ratings for the MovieLens and Jokes data sets.
(We omit Slashdot and HetRec in this set of experiments because they do not
have natural value columns to express the aggregation.) 
We see that DIM$^3$ performs the best among all solutions.
\revise{Compared with the hash-based solution, DIM$^3$ obtains 
a speedup of 18--24$\times$.}
Compared with the fastest RDBMS solution, DIM$^3$ obtains 
a speedup of as least 30$\times$. 
%




\Paragraph{\Removed{Multi-Way Joins with Projection} \revise{MJP}} We
evaluate the effectiveness of our proposed DP algorithm. We compare
three DIM$^3$ variants with RDBMSs: 1) DIM$^3$\emph{(DP)} runs DP to
find the optimal query plan; 2) DIM$^3$\emph{(EagerDedup)} places a
deduplication after every join; and 3) DIM$^3$\emph{(LazyDedup)}
performs only one deduplication after all the joins.

In this set of experiments, we use the Friendster data
set~\cite{Friend} and compute multi-hop connections between users.
Since the original Friendster data set is large and very sparse, as
shown in Table~\ref{tab:dataset-real}, we construct subsets of the
data set using a parameter $range$.  Given a $range$, we filter out
any tuples whose attributes are both larger than $range$.  From the
remaining tuples, we randomly extracted 10 tables, each with $10^6$
tuples, then perform a 10-way join with projection.  
We set $range$= $5\times10^5$, $7\times10^5$ or $9\times10^5$.  Note
that the larger the $range$, the sparser the input tables.  For the
three sub data sets, the final output size $|OUT_P|$ is 46M, 18M, and
8M, and DIM$^3$\emph{(DP)} decides to place 4, 3, and 2
deduplication operations, respectively.

As shown in Figure~\ref{fig:cq}, all DIM$^3$ variants run faster than
RDBMSs because of the better Join-Project performance of DIM$^3$.
\revise{Compared with the hash-based solution with EagerDedup
strategy, DIM$^3$(DP) obtains a speedup of 4.9--7.4$\times$.}
When $range$= $5\times10^5$, DIM$^3$\emph{(DP)} gains a 10x speedup
compared to \emph{LazyDedup}.
When $range$= $7\times10^5$, \emph{DP} are 11\% and 39\% better than
\emph{EagerDedup} and \emph{LazyDedup}, respectively.
When $range$= $9\times10^5$, \emph{DP} is 1.9x as fast as
\emph{EagerDedup}.
However, \emph{DP} is slightly (i.e., 5\%) slower than
\emph{LazyDedup}.  In this case, \emph{DP} places a deduplication
after the first join.  The hope is that the amount of computation in
subsequent joins will be proportionally reduced, but the actual
reduction is not as significant.
Overall, we see that DIM$^3$\emph{(DP)} performs the best among the
three variants.  This confirms the effectiveness of the DP
algorithm for finding the optimal plan for evaluating \Removed{multi-way joins
with projection} \revise{MJP}.

\nottechreport{\vspace{-2pt}}
\revise{\section{Conclusion and Future Work}}

In this paper, we propose DIM$^3$ that combines intersection-free partitioning,
optimized mapping, and DenseEC and SparseBMM algorithms to improve
the state-of-the-art DHK solution.  Moreover, we investigate partial result
caching and extend DIM$^3$ to efficiently compute Join-$op$ queries.  Our
results show that DIM$^3$ is a promising solution for the widely used
Join-Project operation.

\revise{Several promising future directions remain to be explored.}
\revise{The first is out-of-core computation.}
\Removed{Out-of-core computation is an interesting topic to investigate in future work.}
When the input table $R$ and $S$ are too large to fit in the allocated memory,
one way is to perform I/O partitioning for $R$ and $S$ according to $R.x$ and
$S.z$, respectively.  Then, we load each pair of $R.x$ partition and $S.z$
partition into memory and employ DIM$^3$ to compute the results.
An alternative way is to partition the tables according to the join key $y$.
However, a final deduplication step is necessary because the intermediate
results from different partitions may contain duplicates.  

\revise{The second direction is to study caching for multiple queries
when partial result caching is enabled.  In addition to traditional
considerations, such as query access statistics, the cache space
allocated to a Join-Project query becomes a tunable parameter.  This
adds a new dimension in the design of the caching strategy.}

\revise{Last but not least, it is interesting to exploit new hardware to
accelerate database operations. For example, GPUs (Graphics Processing
Units)~\cite{bress2014gpu} and TPUs (Tensor Core Units)~\cite{hu2021tcudb} can
significantly accelerate matrix multiplication in Join-Project.  NVM
(Non-Volatile Memory) provides an interesting combination of persistence,
capacity, and performance, which can be used to speed up queries with huge
memory consumption~\cite{van2018managing}.}


\begin{acks}
        This work is partially supported by Natural Science Foundation of China (62172390). Shimin Chen is the corresponding author.
\end{acks}

\bibliographystyle{ACM-Reference-Format}
\bibliography{myref}


\begin{thebibliography}{44}


\ifx \showCODEN    \undefined \def \showCODEN     #1{\unskip}     \fi
\ifx \showDOI      \undefined \def \showDOI       #1{#1}\fi
\ifx \showISBNx    \undefined \def \showISBNx     #1{\unskip}     \fi
\ifx \showISBNxiii \undefined \def \showISBNxiii  #1{\unskip}     \fi
\ifx \showISSN     \undefined \def \showISSN      #1{\unskip}     \fi
\ifx \showLCCN     \undefined \def \showLCCN      #1{\unskip}     \fi
\ifx \shownote     \undefined \def \shownote      #1{#1}          \fi
\ifx \showarticletitle \undefined \def \showarticletitle #1{#1}   \fi
\ifx \showURL      \undefined \def \showURL       {\relax}        \fi
\providecommand\bibfield[2]{#2}
\providecommand\bibinfo[2]{#2}
\providecommand\natexlab[1]{#1}
\providecommand\showeprint[2][]{arXiv:#2}

\bibitem[\protect\citeauthoryear{Amossen and Pagh}{Amossen and Pagh}{2009}]%
        {ICDT09}
\bibfield{author}{\bibinfo{person}{Rasmus~Resen Amossen} {and}
  \bibinfo{person}{Rasmus Pagh}.} \bibinfo{year}{2009}\natexlab{}.
\newblock \showarticletitle{Faster join-projects and sparse matrix
  multiplications}. In \bibinfo{booktitle}{\emph{Proceedings of the 12th
  International Conference on Database Theory}}. \bibinfo{pages}{121--126}.
\newblock


\bibitem[\protect\citeauthoryear{Balkesen, Alonso, Teubner, and
  {\"O}zsu}{Balkesen et~al\mbox{.}}{2013}]%
        {VLDB13Join}
\bibfield{author}{\bibinfo{person}{Cagri Balkesen}, \bibinfo{person}{Gustavo
  Alonso}, \bibinfo{person}{Jens Teubner}, {and} \bibinfo{person}{M~Tamer
  {\"O}zsu}.} \bibinfo{year}{2013}\natexlab{}.
\newblock \showarticletitle{Multi-core, main-memory joins: Sort vs. hash
  revisited}.
\newblock \bibinfo{journal}{\emph{Proceedings of the VLDB Endowment}}
  \bibinfo{volume}{7}, \bibinfo{number}{1} (\bibinfo{year}{2013}),
  \bibinfo{pages}{85--96}.
\newblock


\bibitem[\protect\citeauthoryear{Benson and Ballard}{Benson and
  Ballard}{2015}]%
        {FMM}
\bibfield{author}{\bibinfo{person}{Austin~R Benson} {and} \bibinfo{person}{Grey
  Ballard}.} \bibinfo{year}{2015}\natexlab{}.
\newblock \showarticletitle{A framework for practical parallel fast matrix
  multiplication}.
\newblock \bibinfo{journal}{\emph{ACM SIGPLAN Notices}} \bibinfo{volume}{50},
  \bibinfo{number}{8} (\bibinfo{year}{2015}), \bibinfo{pages}{42--53}.
\newblock


\bibitem[\protect\citeauthoryear{Blakeley, Larson, and Tompa}{Blakeley
  et~al\mbox{.}}{1986}]%
        {BlakeleyLT86}
\bibfield{author}{\bibinfo{person}{Jos{\'{e}}~A. Blakeley},
  \bibinfo{person}{Per{-}{\AA}ke Larson}, {and} \bibinfo{person}{Frank~Wm.
  Tompa}.} \bibinfo{year}{1986}\natexlab{}.
\newblock \showarticletitle{Efficiently Updating Materialized Views}. In
  \bibinfo{booktitle}{\emph{Proceedings of the 1986 {ACM} {SIGMOD}
  International Conference on Management of Data, Washington, DC, USA, May
  28-30, 1986}}. \bibinfo{publisher}{{ACM} Press}, \bibinfo{pages}{61--71}.
\newblock


\bibitem[\protect\citeauthoryear{Bre{\ss}, Heimel, Siegmund, Bellatreche, and
  Saake}{Bre{\ss} et~al\mbox{.}}{2014}]%
        {bress2014gpu}
\bibfield{author}{\bibinfo{person}{Sebastian Bre{\ss}}, \bibinfo{person}{Max
  Heimel}, \bibinfo{person}{Norbert Siegmund}, \bibinfo{person}{Ladjel
  Bellatreche}, {and} \bibinfo{person}{Gunter Saake}.}
  \bibinfo{year}{2014}\natexlab{}.
\newblock \showarticletitle{Gpu-accelerated database systems: Survey and open
  challenges}.
\newblock In \bibinfo{booktitle}{\emph{Transactions on Large-Scale Data-and
  Knowledge-Centered Systems XV}}. \bibinfo{publisher}{Springer},
  \bibinfo{pages}{1--35}.
\newblock


\bibitem[\protect\citeauthoryear{Cai, Balazinska, and Suciu}{Cai
  et~al\mbox{.}}{2019}]%
        {cai2019pessimistic}
\bibfield{author}{\bibinfo{person}{Walter Cai}, \bibinfo{person}{Magdalena
  Balazinska}, {and} \bibinfo{person}{Dan Suciu}.}
  \bibinfo{year}{2019}\natexlab{}.
\newblock \showarticletitle{Pessimistic cardinality estimation: Tighter upper
  bounds for intermediate join cardinalities}. In
  \bibinfo{booktitle}{\emph{Proceedings of the 2019 International Conference on
  Management of Data}}. \bibinfo{pages}{18--35}.
\newblock


\bibitem[\protect\citeauthoryear{Cantador, Brusilovsky, and Kuflik}{Cantador
  et~al\mbox{.}}{2011}]%
        {hetrec2011}
\bibfield{author}{\bibinfo{person}{Iv{\'a}n Cantador}, \bibinfo{person}{Peter
  Brusilovsky}, {and} \bibinfo{person}{Tsvi Kuflik}.}
  \bibinfo{year}{2011}\natexlab{}.
\newblock \showarticletitle{Second workshop on information heterogeneity and
  fusion in recommender systems (HetRec2011)}. In
  \bibinfo{booktitle}{\emph{Proceedings of the fifth ACM conference on
  Recommender systems}}. \bibinfo{pages}{387--388}.
\newblock


\bibitem[\protect\citeauthoryear{Chakrabarti, Zhan, and Faloutsos}{Chakrabarti
  et~al\mbox{.}}{2004}]%
        {rmat}
\bibfield{author}{\bibinfo{person}{Deepayan Chakrabarti},
  \bibinfo{person}{Yiping Zhan}, {and} \bibinfo{person}{Christos Faloutsos}.}
  \bibinfo{year}{2004}\natexlab{}.
\newblock \showarticletitle{R-MAT: A recursive model for graph mining}. In
  \bibinfo{booktitle}{\emph{Proceedings of the 2004 SIAM International
  Conference on Data Mining}}. SIAM, \bibinfo{pages}{442--446}.
\newblock


\bibitem[\protect\citeauthoryear{Dalton, Olson, and Bell}{Dalton
  et~al\mbox{.}}{2015}]%
        {dalton2015optimizing}
\bibfield{author}{\bibinfo{person}{Steven Dalton}, \bibinfo{person}{Luke
  Olson}, {and} \bibinfo{person}{Nathan Bell}.}
  \bibinfo{year}{2015}\natexlab{}.
\newblock \showarticletitle{Optimizing sparse matrix—matrix multiplication
  for the gpu}.
\newblock \bibinfo{journal}{\emph{ACM Transactions on Mathematical Software
  (TOMS)}} \bibinfo{volume}{41}, \bibinfo{number}{4} (\bibinfo{year}{2015}),
  \bibinfo{pages}{1--20}.
\newblock


\bibitem[\protect\citeauthoryear{Deep, Hu, and Koutris}{Deep
  et~al\mbox{.}}{2020}]%
        {SIGMOD20}
\bibfield{author}{\bibinfo{person}{Shaleen Deep}, \bibinfo{person}{Xiao Hu},
  {and} \bibinfo{person}{Paraschos Koutris}.} \bibinfo{year}{2020}\natexlab{}.
\newblock \showarticletitle{Fast join project query evaluation using matrix
  multiplication}. In \bibinfo{booktitle}{\emph{Proceedings of the 2020 ACM
  SIGMOD International Conference on Management of Data}}.
  \bibinfo{pages}{1213--1223}.
\newblock


\bibitem[\protect\citeauthoryear{Deep, Hu, and Koutris}{Deep
  et~al\mbox{.}}{2021}]%
        {ICDT21Enum}
\bibfield{author}{\bibinfo{person}{Shaleen Deep}, \bibinfo{person}{Xiao Hu},
  {and} \bibinfo{person}{Paraschos Koutris}.} \bibinfo{year}{2021}\natexlab{}.
\newblock \showarticletitle{Enumeration Algorithms for Conjunctive Queries with
  Projection}. In \bibinfo{booktitle}{\emph{24th International Conference on
  Database Theory, {ICDT} 2021, March 23-26, 2021, Nicosia, Cyprus}}
  \emph{(\bibinfo{series}{LIPIcs})}, \bibfield{editor}{\bibinfo{person}{Ke~Yi}
  {and} \bibinfo{person}{Zhewei Wei}} (Eds.), Vol.~\bibinfo{volume}{186}.
  \bibinfo{publisher}{Schloss Dagstuhl - Leibniz-Zentrum f{\"{u}}r Informatik},
  \bibinfo{pages}{14:1--14:17}.
\newblock


\bibitem[\protect\citeauthoryear{Gall and Urrutia}{Gall and Urrutia}{2018}]%
        {matrix2018}
\bibfield{author}{\bibinfo{person}{Fran{\c{c}}ois~Le Gall} {and}
  \bibinfo{person}{Florent Urrutia}.} \bibinfo{year}{2018}\natexlab{}.
\newblock \showarticletitle{Improved rectangular matrix multiplication using
  powers of the Coppersmith-Winograd tensor}. In
  \bibinfo{booktitle}{\emph{Proceedings of the Twenty-Ninth Annual ACM-SIAM
  Symposium on Discrete Algorithms}}. SIAM, \bibinfo{pages}{1029--1046}.
\newblock


\bibitem[\protect\citeauthoryear{Goldberg, Roeder, Gupta, and Perkins}{Goldberg
  et~al\mbox{.}}{2001}]%
        {Jokes}
\bibfield{author}{\bibinfo{person}{Ken Goldberg}, \bibinfo{person}{Theresa
  Roeder}, \bibinfo{person}{Dhruv Gupta}, {and} \bibinfo{person}{Chris
  Perkins}.} \bibinfo{year}{2001}\natexlab{}.
\newblock \showarticletitle{Eigentaste: A constant time collaborative filtering
  algorithm}.
\newblock \bibinfo{journal}{\emph{information retrieval}} \bibinfo{volume}{4},
  \bibinfo{number}{2} (\bibinfo{year}{2001}), \bibinfo{pages}{133--151}.
\newblock


\bibitem[\protect\citeauthoryear{Graefe and Kuno}{Graefe and Kuno}{2011}]%
        {Btree2011}
\bibfield{author}{\bibinfo{person}{Goetz Graefe} {and} \bibinfo{person}{Harumi
  Kuno}.} \bibinfo{year}{2011}\natexlab{}.
\newblock \showarticletitle{Modern B-tree techniques}. In
  \bibinfo{booktitle}{\emph{2011 IEEE 27th International Conference on Data
  Engineering}}. IEEE, \bibinfo{pages}{1370--1373}.
\newblock


\bibitem[\protect\citeauthoryear{Harper and Konstan}{Harper and
  Konstan}{2015}]%
        {Movielens}
\bibfield{author}{\bibinfo{person}{F~Maxwell Harper} {and}
  \bibinfo{person}{Joseph~A Konstan}.} \bibinfo{year}{2015}\natexlab{}.
\newblock \showarticletitle{The movielens datasets: History and context}.
\newblock \bibinfo{journal}{\emph{Acm transactions on interactive intelligent
  systems (tiis)}} \bibinfo{volume}{5}, \bibinfo{number}{4}
  (\bibinfo{year}{2015}), \bibinfo{pages}{1--19}.
\newblock


\bibitem[\protect\citeauthoryear{Hertzschuch, Hartmann, Habich, and
  Lehner}{Hertzschuch et~al\mbox{.}}{2021}]%
        {hertzschuch2021simplicity}
\bibfield{author}{\bibinfo{person}{Axel Hertzschuch}, \bibinfo{person}{Claudio
  Hartmann}, \bibinfo{person}{Dirk Habich}, {and} \bibinfo{person}{Wolfgang
  Lehner}.} \bibinfo{year}{2021}\natexlab{}.
\newblock \showarticletitle{Simplicity Done Right for Join Ordering.}. In
  \bibinfo{booktitle}{\emph{CIDR}}.
\newblock


\bibitem[\protect\citeauthoryear{Ho and Park}{Ho and Park}{2016}]%
        {Btree2016}
\bibfield{author}{\bibinfo{person}{VanPhi Ho} {and} \bibinfo{person}{Dong-Joo
  Park}.} \bibinfo{year}{2016}\natexlab{}.
\newblock \showarticletitle{A survey of the-state-of-the-art b-tree index on
  flash memory}.
\newblock \bibinfo{journal}{\emph{International Journal of Software Engineering
  and Its Applications}} \bibinfo{volume}{10}, \bibinfo{number}{4}
  (\bibinfo{year}{2016}), \bibinfo{pages}{173--188}.
\newblock


\bibitem[\protect\citeauthoryear{Hu and Yi}{Hu and Yi}{2020}]%
        {PODS20Parallel}
\bibfield{author}{\bibinfo{person}{Xiao Hu} {and} \bibinfo{person}{Ke Yi}.}
  \bibinfo{year}{2020}\natexlab{}.
\newblock \showarticletitle{Parallel Algorithms for Sparse Matrix
  Multiplication and Join-Aggregate Queries}. In
  \bibinfo{booktitle}{\emph{Proceedings of the 39th ACM SIGMOD-SIGACT-SIGAI
  Symposium on Principles of Database Systems}}. \bibinfo{pages}{411--425}.
\newblock


\bibitem[\protect\citeauthoryear{Hu, Li, and Tseng}{Hu et~al\mbox{.}}{2021}]%
        {hu2021tcudb}
\bibfield{author}{\bibinfo{person}{Yu-Ching Hu}, \bibinfo{person}{Yuliang Li},
  {and} \bibinfo{person}{Hung-Wei Tseng}.} \bibinfo{year}{2021}\natexlab{}.
\newblock \showarticletitle{TCUDB: Accelerating Database with Tensor
  Processors}.
\newblock \bibinfo{journal}{\emph{arXiv preprint arXiv:2112.07552}}
  (\bibinfo{year}{2021}).
\newblock


\bibitem[\protect\citeauthoryear{Ioannidis and Kang}{Ioannidis and
  Kang}{1991}]%
        {ioannidis1991left}
\bibfield{author}{\bibinfo{person}{Yannis~E Ioannidis} {and}
  \bibinfo{person}{Younkyung~Cha Kang}.} \bibinfo{year}{1991}\natexlab{}.
\newblock \showarticletitle{Left-deep vs. bushy trees: An analysis of strategy
  spaces and its implications for query optimization}. In
  \bibinfo{booktitle}{\emph{Proceedings of the 1991 ACM SIGMOD international
  conference on Management of data}}. \bibinfo{pages}{168--177}.
\newblock


\bibitem[\protect\citeauthoryear{Kanda, Morita, and Fuketa}{Kanda
  et~al\mbox{.}}{2017}]%
        {kanda2017practical}
\bibfield{author}{\bibinfo{person}{Shunsuke Kanda}, \bibinfo{person}{Kazuhiro
  Morita}, {and} \bibinfo{person}{Masao Fuketa}.}
  \bibinfo{year}{2017}\natexlab{}.
\newblock \showarticletitle{Practical string dictionary compression using
  string dictionary encoding}. In \bibinfo{booktitle}{\emph{2017 International
  Conference on Big Data Innovations and Applications (Innovate-Data)}}. IEEE,
  \bibinfo{pages}{1--8}.
\newblock


\bibitem[\protect\citeauthoryear{Kepner and Gilbert}{Kepner and
  Gilbert}{2011}]%
        {Book11Spmm}
\bibfield{author}{\bibinfo{person}{Jeremy Kepner} {and} \bibinfo{person}{John
  Gilbert}.} \bibinfo{year}{2011}\natexlab{}.
\newblock \bibinfo{booktitle}{\emph{Graph algorithms in the language of linear
  algebra}}.
\newblock \bibinfo{publisher}{SIAM}.
\newblock


\bibitem[\protect\citeauthoryear{Kim, Kaldewey, Lee, Sedlar, Nguyen, Satish,
  Chhugani, Di~Blas, and Dubey}{Kim et~al\mbox{.}}{2009}]%
        {VLDB09Join}
\bibfield{author}{\bibinfo{person}{Changkyu Kim}, \bibinfo{person}{Tim
  Kaldewey}, \bibinfo{person}{Victor~W Lee}, \bibinfo{person}{Eric Sedlar},
  \bibinfo{person}{Anthony~D Nguyen}, \bibinfo{person}{Nadathur Satish},
  \bibinfo{person}{Jatin Chhugani}, \bibinfo{person}{Andrea Di~Blas}, {and}
  \bibinfo{person}{Pradeep Dubey}.} \bibinfo{year}{2009}\natexlab{}.
\newblock \showarticletitle{Sort vs. hash revisited: Fast join implementation
  on modern multi-core CPUs}.
\newblock \bibinfo{journal}{\emph{Proceedings of the VLDB Endowment}}
  \bibinfo{volume}{2}, \bibinfo{number}{2} (\bibinfo{year}{2009}),
  \bibinfo{pages}{1378--1389}.
\newblock


\bibitem[\protect\citeauthoryear{Kocberber, Grot, Picorel, Falsafi, Lim, and
  Ranganathan}{Kocberber et~al\mbox{.}}{2013}]%
        {kocberber2013meet}
\bibfield{author}{\bibinfo{person}{Onur Kocberber}, \bibinfo{person}{Boris
  Grot}, \bibinfo{person}{Javier Picorel}, \bibinfo{person}{Babak Falsafi},
  \bibinfo{person}{Kevin Lim}, {and} \bibinfo{person}{Parthasarathy
  Ranganathan}.} \bibinfo{year}{2013}\natexlab{}.
\newblock \showarticletitle{Meet the walkers accelerating index traversals for
  in-memory databases}. In \bibinfo{booktitle}{\emph{2013 46th Annual IEEE/ACM
  International Symposium on Microarchitecture (MICRO)}}. IEEE,
  \bibinfo{pages}{468--479}.
\newblock


\bibitem[\protect\citeauthoryear{Kraska, Beutel, Chi, Dean, and
  Polyzotis}{Kraska et~al\mbox{.}}{2018}]%
        {kraska2018case}
\bibfield{author}{\bibinfo{person}{Tim Kraska}, \bibinfo{person}{Alex Beutel},
  \bibinfo{person}{Ed~H Chi}, \bibinfo{person}{Jeffrey Dean}, {and}
  \bibinfo{person}{Neoklis Polyzotis}.} \bibinfo{year}{2018}\natexlab{}.
\newblock \showarticletitle{The case for learned index structures}. In
  \bibinfo{booktitle}{\emph{Proceedings of the 2018 International Conference on
  Management of Data}}. \bibinfo{pages}{489--504}.
\newblock


\bibitem[\protect\citeauthoryear{Kurzak, Alvaro, and Dongarra}{Kurzak
  et~al\mbox{.}}{2009}]%
        {kurzak2009optimizing}
\bibfield{author}{\bibinfo{person}{Jakub Kurzak}, \bibinfo{person}{Wesley
  Alvaro}, {and} \bibinfo{person}{Jack Dongarra}.}
  \bibinfo{year}{2009}\natexlab{}.
\newblock \showarticletitle{Optimizing matrix multiplication for a short-vector
  SIMD architecture--CELL processor}.
\newblock \bibinfo{journal}{\emph{Parallel Comput.}} \bibinfo{volume}{35},
  \bibinfo{number}{3} (\bibinfo{year}{2009}), \bibinfo{pages}{138--150}.
\newblock


\bibitem[\protect\citeauthoryear{Larson, Goldstein, and Zhou}{Larson
  et~al\mbox{.}}{2004}]%
        {LarsonGZ04}
\bibfield{author}{\bibinfo{person}{Per{-}{\AA}ke Larson},
  \bibinfo{person}{Jonathan Goldstein}, {and} \bibinfo{person}{Jingren Zhou}.}
  \bibinfo{year}{2004}\natexlab{}.
\newblock \showarticletitle{MTCache: Transparent Mid-Tier Database Caching in
  {SQL} Server}. In \bibinfo{booktitle}{\emph{Proceedings of the 20th
  International Conference on Data Engineering, {ICDE} 2004, 30 March - 2 April
  2004, Boston, MA, {USA}}}. \bibinfo{publisher}{{IEEE} Computer Society},
  \bibinfo{pages}{177--188}.
\newblock


\bibitem[\protect\citeauthoryear{Larson and Yang}{Larson and Yang}{1985}]%
        {LarsonY85}
\bibfield{author}{\bibinfo{person}{Per{-}{\AA}ke Larson} {and}
  \bibinfo{person}{H.~Z. Yang}.} \bibinfo{year}{1985}\natexlab{}.
\newblock \showarticletitle{Computing Queries from Derived Relations}. In
  \bibinfo{booktitle}{\emph{VLDB'85, Proceedings of 11th International
  Conference on Very Large Data Bases, August 21-23, 1985, Stockholm, Sweden}},
  \bibfield{editor}{\bibinfo{person}{Alain Pirotte} {and}
  \bibinfo{person}{Yannis Vassiliou}} (Eds.). \bibinfo{publisher}{Morgan
  Kaufmann}, \bibinfo{pages}{259--269}.
\newblock


\bibitem[\protect\citeauthoryear{Le~Gall}{Le~Gall}{2012}]%
        {LeGall12}
\bibfield{author}{\bibinfo{person}{Fran{\c{c}}ois Le~Gall}.}
  \bibinfo{year}{2012}\natexlab{}.
\newblock \showarticletitle{Faster algorithms for rectangular matrix
  multiplication}. In \bibinfo{booktitle}{\emph{2012 IEEE 53rd annual symposium
  on foundations of computer science}}. IEEE, \bibinfo{pages}{514--523}.
\newblock


\bibitem[\protect\citeauthoryear{Leis, Radke, Gubichev, Kemper, and
  Neumann}{Leis et~al\mbox{.}}{2017}]%
        {leis2017cardinality}
\bibfield{author}{\bibinfo{person}{Viktor Leis}, \bibinfo{person}{Bernhard
  Radke}, \bibinfo{person}{Andrey Gubichev}, \bibinfo{person}{Alfons Kemper},
  {and} \bibinfo{person}{Thomas Neumann}.} \bibinfo{year}{2017}\natexlab{}.
\newblock \showarticletitle{Cardinality Estimation Done Right: Index-Based Join
  Sampling.}. In \bibinfo{booktitle}{\emph{Cidr}}.
\newblock


\bibitem[\protect\citeauthoryear{Leskovec, Adamic, and Huberman}{Leskovec
  et~al\mbox{.}}{2007}]%
        {amazon}
\bibfield{author}{\bibinfo{person}{Jure Leskovec}, \bibinfo{person}{Lada~A.
  Adamic}, {and} \bibinfo{person}{Bernardo~A. Huberman}.}
  \bibinfo{year}{2007}\natexlab{}.
\newblock \showarticletitle{The dynamics of viral marketing}.
\newblock \bibinfo{journal}{\emph{{ACM} Trans. Web}} \bibinfo{volume}{1},
  \bibinfo{number}{1} (\bibinfo{year}{2007}), \bibinfo{pages}{5}.
\newblock
\urldef\tempurl%
\url{https://doi.org/10.1145/1232722.1232727}
\showDOI{\tempurl}


\bibitem[\protect\citeauthoryear{Leskovec, Lang, Dasgupta, and
  Mahoney}{Leskovec et~al\mbox{.}}{2009}]%
        {Slashdot}
\bibfield{author}{\bibinfo{person}{Jure Leskovec}, \bibinfo{person}{Kevin~J
  Lang}, \bibinfo{person}{Anirban Dasgupta}, {and} \bibinfo{person}{Michael~W
  Mahoney}.} \bibinfo{year}{2009}\natexlab{}.
\newblock \showarticletitle{Community structure in large networks: Natural
  cluster sizes and the absence of large well-defined clusters}.
\newblock \bibinfo{journal}{\emph{Internet Mathematics}} \bibinfo{volume}{6},
  \bibinfo{number}{1} (\bibinfo{year}{2009}), \bibinfo{pages}{29--123}.
\newblock


\bibitem[\protect\citeauthoryear{Mishra and Eich}{Mishra and Eich}{1992}]%
        {join1992}
\bibfield{author}{\bibinfo{person}{Priti Mishra} {and}
  \bibinfo{person}{Margaret~H Eich}.} \bibinfo{year}{1992}\natexlab{}.
\newblock \showarticletitle{Join processing in relational databases}.
\newblock \bibinfo{journal}{\emph{ACM Computing Surveys (CSUR)}}
  \bibinfo{volume}{24}, \bibinfo{number}{1} (\bibinfo{year}{1992}),
  \bibinfo{pages}{63--113}.
\newblock


\bibitem[\protect\citeauthoryear{Murphy, Wheeler, Barrett, and Ang}{Murphy
  et~al\mbox{.}}{2010}]%
        {graph500}
\bibfield{author}{\bibinfo{person}{Richard~C Murphy}, \bibinfo{person}{Kyle~B
  Wheeler}, \bibinfo{person}{Brian~W Barrett}, {and} \bibinfo{person}{James~A
  Ang}.} \bibinfo{year}{2010}\natexlab{}.
\newblock \showarticletitle{Introducing the graph 500}.
\newblock \bibinfo{journal}{\emph{Cray Users Group (CUG)}}
  \bibinfo{volume}{19} (\bibinfo{year}{2010}), \bibinfo{pages}{45--74}.
\newblock


\bibitem[\protect\citeauthoryear{Pegoraro, Uysal, and van~der Aalst}{Pegoraro
  et~al\mbox{.}}{2020}]%
        {Pegoraro20}
\bibfield{author}{\bibinfo{person}{Marco Pegoraro},
  \bibinfo{person}{Merih~Seran Uysal}, {and} \bibinfo{person}{Wil~MP van~der
  Aalst}.} \bibinfo{year}{2020}\natexlab{}.
\newblock \showarticletitle{Efficient time and space representation of
  uncertain event data}.
\newblock \bibinfo{journal}{\emph{Algorithms}} \bibinfo{volume}{13},
  \bibinfo{number}{11} (\bibinfo{year}{2020}), \bibinfo{pages}{285}.
\newblock


\bibitem[\protect\citeauthoryear{Qiu, Wang, Yi, Li, Wu, and Zhan}{Qiu
  et~al\mbox{.}}{2021}]%
        {qiu2021weighted}
\bibfield{author}{\bibinfo{person}{Yuan Qiu}, \bibinfo{person}{Yilei Wang},
  \bibinfo{person}{Ke Yi}, \bibinfo{person}{Feifei Li}, \bibinfo{person}{Bin
  Wu}, {and} \bibinfo{person}{Chaoqun Zhan}.} \bibinfo{year}{2021}\natexlab{}.
\newblock \showarticletitle{Weighted Distinct Sampling: Cardinality Estimation
  for SPJ Queries}. In \bibinfo{booktitle}{\emph{Proceedings of the 2021
  International Conference on Management of Data}}.
  \bibinfo{pages}{1465--1477}.
\newblock


\bibitem[\protect\citeauthoryear{Saule, Kaya, and {\c{C}}ataly{\"u}rek}{Saule
  et~al\mbox{.}}{2013}]%
        {saule2013performance}
\bibfield{author}{\bibinfo{person}{Erik Saule}, \bibinfo{person}{Kamer Kaya},
  {and} \bibinfo{person}{{\"U}mit~V {\c{C}}ataly{\"u}rek}.}
  \bibinfo{year}{2013}\natexlab{}.
\newblock \showarticletitle{Performance evaluation of sparse matrix
  multiplication kernels on intel xeon phi}. In
  \bibinfo{booktitle}{\emph{International Conference on Parallel Processing and
  Applied Mathematics}}. Springer, \bibinfo{pages}{559--570}.
\newblock


\bibitem[\protect\citeauthoryear{Schneider and DeWitt}{Schneider and
  DeWitt}{1990}]%
        {schneider1990tradeoffs}
\bibfield{author}{\bibinfo{person}{Donovan~A Schneider} {and}
  \bibinfo{person}{David~J DeWitt}.} \bibinfo{year}{1990}\natexlab{}.
\newblock \bibinfo{booktitle}{\emph{Tradeoffs in processing complex join
  queries via hashing in multiprocessor database machines}}.
\newblock \bibinfo{publisher}{University of Wisconsin-Madison. Computer
  Sciences Department}.
\newblock


\bibitem[\protect\citeauthoryear{Skarupke}{Skarupke}{2017}]%
        {Flathashmap}
\bibfield{author}{\bibinfo{person}{Malte Skarupke}.}
  \bibinfo{year}{2017}\natexlab{}.
\newblock \bibinfo{title}{I Wrote The Fastest Hashtable}.
\newblock
  \bibinfo{howpublished}{https://probablydance.com/2017/02/26/i-wrote-the-fastest-hashtable/}.
\newblock


\bibitem[\protect\citeauthoryear{Strassen}{Strassen}{1969}]%
        {Strassen69}
\bibfield{author}{\bibinfo{person}{Volker Strassen}.}
  \bibinfo{year}{1969}\natexlab{}.
\newblock \showarticletitle{Gaussian elimination is not optimal}.
\newblock \bibinfo{journal}{\emph{Numer. Math.}}  \bibinfo{volume}{13}
  (\bibinfo{year}{1969}), \bibinfo{pages}{354--356}.
\newblock


\bibitem[\protect\citeauthoryear{van Renen, Leis, Kemper, Neumann, Hashida, Oe,
  Doi, Harada, and Sato}{van Renen et~al\mbox{.}}{2018}]%
        {van2018managing}
\bibfield{author}{\bibinfo{person}{Alexander van Renen},
  \bibinfo{person}{Viktor Leis}, \bibinfo{person}{Alfons Kemper},
  \bibinfo{person}{Thomas Neumann}, \bibinfo{person}{Takushi Hashida},
  \bibinfo{person}{Kazuichi Oe}, \bibinfo{person}{Yoshiyasu Doi},
  \bibinfo{person}{Lilian Harada}, {and} \bibinfo{person}{Mitsuru Sato}.}
  \bibinfo{year}{2018}\natexlab{}.
\newblock \showarticletitle{Managing non-volatile memory in database systems}.
  In \bibinfo{booktitle}{\emph{Proceedings of the 2018 International Conference
  on Management of Data}}. \bibinfo{pages}{1541--1555}.
\newblock


\bibitem[\protect\citeauthoryear{Wang, Zhang, Shen, Zhang, Lu, Wu, and
  Wang}{Wang et~al\mbox{.}}{2014}]%
        {MKL}
\bibfield{author}{\bibinfo{person}{Endong Wang}, \bibinfo{person}{Qing Zhang},
  \bibinfo{person}{Bo Shen}, \bibinfo{person}{Guangyong Zhang},
  \bibinfo{person}{Xiaowei Lu}, \bibinfo{person}{Qing Wu}, {and}
  \bibinfo{person}{Yajuan Wang}.} \bibinfo{year}{2014}\natexlab{}.
\newblock \showarticletitle{Intel math kernel library}.
\newblock In \bibinfo{booktitle}{\emph{High-Performance Computing on the
  Intel{\textregistered} Xeon Phi™}}. \bibinfo{publisher}{Springer},
  \bibinfo{pages}{167--188}.
\newblock


\bibitem[\protect\citeauthoryear{Yang and Leskovec}{Yang and Leskovec}{2015a}]%
        {DBLP}
\bibfield{author}{\bibinfo{person}{Jaewon Yang} {and} \bibinfo{person}{Jure
  Leskovec}.} \bibinfo{year}{2015}\natexlab{a}.
\newblock \showarticletitle{Defining and evaluating network communities based
  on ground-truth}.
\newblock \bibinfo{journal}{\emph{Knowledge and Information Systems}}
  \bibinfo{volume}{42}, \bibinfo{number}{1} (\bibinfo{year}{2015}),
  \bibinfo{pages}{181--213}.
\newblock


\bibitem[\protect\citeauthoryear{Yang and Leskovec}{Yang and Leskovec}{2015b}]%
        {Friend}
\bibfield{author}{\bibinfo{person}{Jaewon Yang} {and} \bibinfo{person}{Jure
  Leskovec}.} \bibinfo{year}{2015}\natexlab{b}.
\newblock \showarticletitle{Defining and evaluating network communities based
  on ground-truth}.
\newblock \bibinfo{journal}{\emph{Knowledge and Information Systems}}
  \bibinfo{volume}{42}, \bibinfo{number}{1} (\bibinfo{year}{2015}),
  \bibinfo{pages}{181--213}.
\newblock


\end{thebibliography}

\nottechreport{
\clearpage
\input{response}
}

\end{document}